\pdfoutput=1

\documentclass[amsfonts, aps, amssymb, amsmath, reprint, showkeys, nofootinbib, twoside,superscriptaddress,floatfix,prx]{revtex4-2}
\usepackage[letterpaper, total={7in, 9in}]{geometry}
\usepackage[english]{babel}
\usepackage[utf8]{inputenc}
\usepackage[T1]{fontenc}
\usepackage[colorinlistoftodos, color=green!40, prependcaption]{todonotes}
\usepackage{mathrsfs}
\usepackage[pdftex, pdftitle={Article}, pdfauthor={Author}]{hyperref} 
\usepackage{xcolor}
\usepackage{comment}
\usepackage{siunitx}
\usepackage{textcomp}

\def\_#1{\textsubscript{#1}}
\def\^#1{\textsuperscript{#1}}
\usepackage[page,toc,titletoc,title]{appendix}

\usepackage{hyperref}

\newcommand{\beginsupplement}{
        \setcounter{table}{0}
        \renewcommand{\thetable}{S\arabic{table}}
        \setcounter{figure}{0}
        \renewcommand{\thefigure}{S\arabic{figure}}
        \setcounter{equation}{0}
        \renewcommand{\theequation}{S\arabic{equation}}
        \setcounter{section}{0}
        \renewcommand{\thesection}{S\arabic{section}}
        }

\begin{document}

\title{A scanning probe microscopy approach for identifying defects in aluminum oxide}

\author{Leah Tom}
\author{Zachary J.\ Krebs}
\affiliation{Department of Physics, University of Wisconsin-Madison, Madison, WI 53706, USA}
\author{Joel B.\ Varley}
\affiliation{Lawrence Livermore National Laboratory, Livermore, California 94550, USA}
\author{E. S. Joseph}
\author{Wyatt A.\ Behn}
\author{M.\ A.\ Eriksson}
\affiliation{Department of Physics, University of Wisconsin-Madison, Madison, WI 53706, USA}
\author{Keith G.\ Ray}
\author{Vincenzo Lordi}
\affiliation{Lawrence Livermore National Laboratory, Livermore, California 94550, USA}
\author{S.\ N.\ Coppersmith}
\affiliation{School of Physics, University of New South Wales, Sydney NSW 2052, Australia}
\author{Victor W.\ Brar}
\author{Mark Friesen}
\affiliation{Department of Physics, University of Wisconsin-Madison, Madison, WI 53706, USA}
\email[Correspondence email address: ]{}

\date{\today} 

\maketitle

\noindent
\textbf{The coherence of quantum dot qubits fabricated in semiconductors is often limited by charge noise from defects in gate dielectrics, which are material- and process-dependent.  
Characterizing these defects is an important step towards reducing their impact and improving qubit coherence.  
The identification of individual defects requires atomic-scale spatial resolution, however, and sufficient spectral sensitivity to determine their electronic structure. 
Electrostatic force microscopy (EFM) provides highly resolved maps of the surface potential of dielectrics, and importantly, is also sensitive to single-electron charging processes that reflect the spectral structure of underlying defects. 
In this work, we use cryogenic EFM to characterize aluminum oxide grown by atomic layer deposition (ALD) on bulk silicon.   
These measurements reveal defects close to the surface that exchange electrons with the EFM tip as they transition through different charge states. 
Detailed electrostatic modeling opens the door to powerful techniques for mapping tip-backgate charging voltages onto defect transition energies, allowing defects such as aluminum vacancies, and carbon, oxygen, or hydrogen impurities to be identified, by comparing to density functional theory (DFT).
These results point towards EFM as a powerful tool for exploring defect structures in solid-state qubits. }

\begin{figure*}[t]
    \centering
\includegraphics[width=0.9\textwidth]{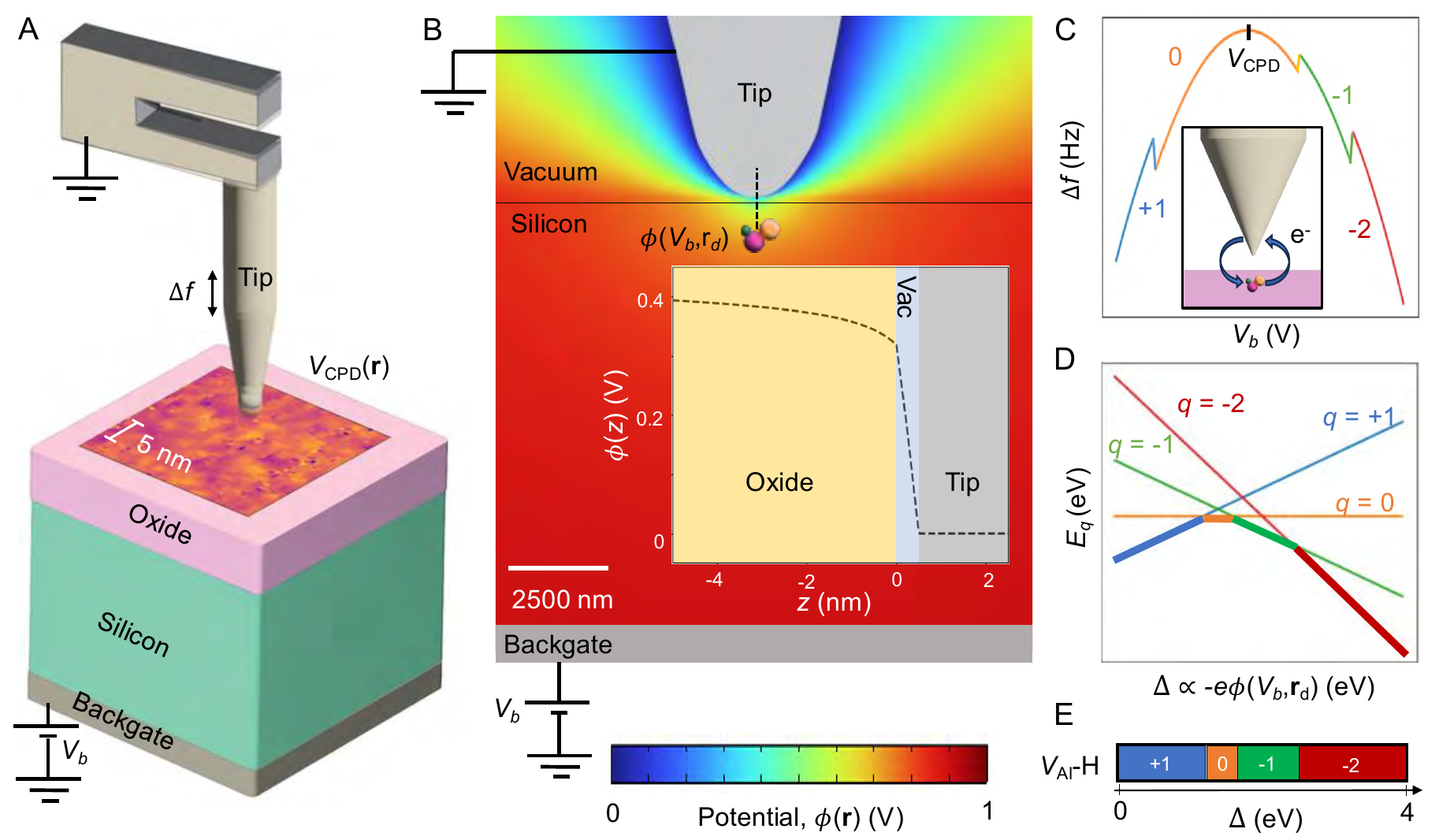}
    \caption{Experimental setup. 
    (\emph{A}) Schematic of a typical electrostatic force microscope (EFM) experiment: an oscillating tip is raster-scanned in vacuum across a sample comprised of Al$_2$O$_3$ grown atop a silicon substrate with a backgate.
    Interactions between the tip and charged defects in the oxide cause a change in the tip resonant frequency $\Delta f$ as a function of the tip-backgate bias voltage $V_b$.
    The color-scale plot shows a spatial map of the measured contact potential (CPD) difference, $V_b=V_\text{CPD}$, that maximizes $\Delta f$ at a given location. 
    Variations in $V_\text{CPD}$ are caused by charged defects near the sample surface.
   (\emph{B}) Electrostatic simulations are used to compute the electrostatic potential at the location of the defect $\phi(V_{b},{\mathbf r}_d)$.
   (Here, $V_{b}$ = 1~V.) 
   The inset shows a linecut directly below the tip, corresponding to the dashed line in the main figure. Note that the oxide layer is not labeled here, because its much smaller thickness (60~nm) is not visible on the larger scale of the image.
   (\emph{C}) Schematic variations of the resonant frequency as a function of bias (an $f$-$V$ curve).
    Jumps indicate discrete charging events, as electrons tunnel between the tip and the defect.
    (Defect charge states are indicated.)
    (\emph{D}) A corresponding energy diagram, with matching colors, as a function of the chemical potential of the defect. 
    The lowest-energy states are indicated by thick lines. 
    (\emph{E}) A corresponding charge-occupancy-chemical-potential diagram, which we use to identify the chemical species of the defect by comparing to density functional theory (DFT) calculations.}
    \label{fig:setup_experiment}
\end{figure*}

\vspace{0.2in} \noindent
The fidelities of gate operations in quantum dot qubits are often limited by charge noise~\cite{Yoneda2018fidelitychargenoise,Seedhouse2023spatio}.
A principal noise sources is charge traps in gate dielectrics~\cite{Connors2019_chargenoise}, including crystalline vacancies, interstitials, impurities, and dangling bonds~\cite{Choi2013_nativepointdefectAl2O3,Holder2013_HydrogenDefectsAlumina,Weber2011_NativeDefectsMOSdevices,Goes2008}. 
Identifying and understanding these defects is an important step towards mitigating their effects.
Various characterization techniques have therefore been used to probe the physical, chemical and electronic structure of the defects~\cite{tuomisto2019characterisation}, including deep-level transient spectroscopy~\cite{lang_dlts},  positron annihilation spectroscopy~\cite{defect_identification_semiconductors_PAS}, time-dependent defect spectroscopy~\cite{TDDS_Grasser}, electron spin resonance~\cite{ludwig1962electron,poindexter1983characterization}, X-ray photoelectron spectroscopy~\cite{mclellan2023chemical,ishikawa2011direct}, spectroscopic ellipsometry ~\cite{price2012}, and charge-pumping spectroscopy~\cite{Zahid2010, brugler1969charge, hori2019charge, tsuchiya2020detection}. 
While these approaches are able to chemically identify the defects and measure their energy levels, they typically rely on spatial averaging to characterize large ensembles and therefore do not provide details about the spatial locations or depths of defects below the surface. 
Although in some cases it may be possible to resolve single defects, when they are well isolated~\cite{hori2019charge, tsuchiya2020detection}, these techniques do not have sufficient resolution to distinguish the behavior of closely spaced defects.

Scanning probe methods have been developed to address these problems by simultaneously imaging and probing the electronic properties of atomic-scale objects at non-conducting surfaces, including impurities, atoms, molecules, and quantum dots. 
In particular, electrostatic force microscopes (EFM) are used to measure changes in the bias-dependent electrostatic force on a microscope tip due to charge rearrangement in the sample induced by electrostatic fields from the tip itself.  
Moreover, the tip bias provides a direct knob for manipulating the reorganized charges and their corresponding electron states~\cite{ludeke2001imaging, suganuma2002probing, bussmann2005single, dana2005electrostatic, stomp2005detection, zhu2005frequency, ludeke_structural_2004, Patera2019, kocic_periodic_2015, Gross2009measuring}. 
This technique can be applied in a variety of ways to study these systems.  
For example, discrete charging events may be observed as sudden jumps in the bias-dependent EFM resonant frequency shift~\cite{ludeke2001imaging,dana2005electrostatic,bussmann2006single,suganuma2002probing}.
This has been used to probe individual molecules~\cite{steurer2015probe,fatayer2018reorganization} and impurity states on insulating surfaces~\cite{bussmann2006single,konig2011defects,johnson2009atomic,ludeke2001imaging,dana2005electrostatic,bussmann2005single}, weakly-coupled quantum dots~\cite{stomp2005detection,zhu2005frequency}, and metallic nanoparticles~\cite{azuma2006single,zheng2010electronic}.  
Alternatively, the bias-dependence of the dissipation of an EFM cantilever can be used to probe charge motion in a sample, either due to electron tunneling or charge reorganization associated with certain defect states.
The latter technique has been used to probe subsurface interface states~\cite{cowie2024spatially,cowie2024spatially2} and quantum dots~\cite{Cockins2010}. 
Two other quantities in EFM experiments that contain information about charging events in dielectrics include the effective tip-sample separation and the corresponding bias-dependent capacitance~\cite{naitou2007investigation,johnson2009atomic}.   
But while these various signals and methods can provide defect maps, they do not typically yield spectroscopic information about the underlying electronic states, as required for chemical analysis.  
For example, converting tip bias measurements (which carry information about charge reorganization) to defect energies requires detailed knowledge about the tip-sample geometry and its associated electrostatics. 
While several models have been employed to perform these conversions, they often over-simplify the tip geometry, and are most effective when the system details known \textit{a priori}.


This work expands on previous techniques by developing EFM as a general tool for probing and identifying distributions of defect species on dielectric surfaces, which are not known in advance.
We perform measurements of an Al$_{2}$O$_{3}$ sample grown by atomic layer deposition (ALD), on which we observe a high density of charged defects and many single-electron charging events.
These are detected as sudden jumps in the EFM resonance frequency as a function of tip bias and, in many cases, for multiple biases at the same tip location.  
The charging events are position dependent, and by repeating our measurements as a function of lateral and vertical tip position, we are able to estimate the tip shape and defect depth.  
Using this information, we then perform electrostatic simulations to estimate the defect energy levels.  
The results are finally compared to a survey of energy levels of candidate defect species obtained from density functional theory (DFT).  
We find that this approach provides reasonable guesses about the defect identities, despite the high density of defects and disorder at the sample surface.  

\begin{figure*}[t]
    \centering
    \includegraphics[width=0.95\textwidth]{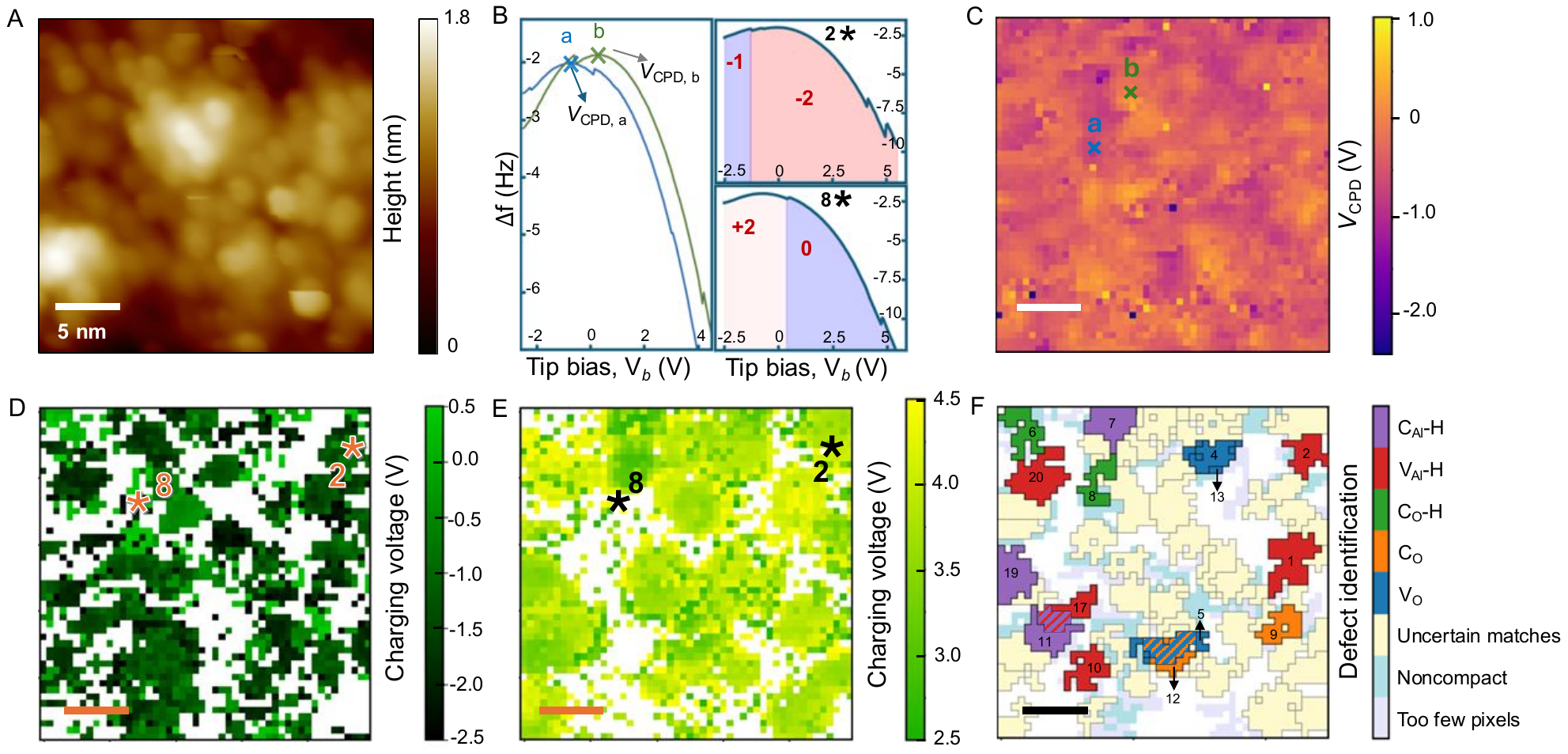}
    \caption{
    EFM results.
    Note that all scale bars in this figure are 5~nm.
    (\emph{A}) Topography map of the oxide-on-silicon sample studied in this work, obtained via nc-AFM.
    (\emph{B}) $f$-$V$ curves are obtained for every pixel in a CPD map; several examples are shown here.
    Special features include maxima, corresponding to $V_b=V_\text{CPD,a(b)}$ (left-hand side), and charging transitions, corresponding to $V_b=V_\text{transition}$ (right-hand side), with charging states identified by different colored regions
    (see \emph{SI Appendix}, section S11).
    Here, a, b, 2, and 8 refer to distinct pixels, whose locations are shown in the maps of panels \emph{C}-\emph{E}.
    In particular, pixels 2 and 8 coincide with the centers of Defects 2 and 8, as identified in \emph{F}.
    Following procedures described in the main text, 
    only one charging transition was considered for Defects 2 and 8.
    (\emph{C}) A $V_\text{CPD}$ map describing the same scan region as \emph{A}.
    (\emph{D,E}) Maps of $V_\text{transition}$ for the same scan region, within the voltage range indicated by color bars.
    (White pixels indicate that no transitions are observed in this range.)
    (\emph{F}) Colored regions indicate the pixels assigned to different defects, as determined by the clustering algorithm, for 15 different defects. 
    (Five other defects were identified using high-resolution scans.)
    The color bar indicates our best guesses for defect species, according to the ``single-defect analysis'' described below, based on comparison with DFT.
    Hatched regions indicate overlapping defects, whose charging transitions can still be uniquely identified. 
    Defects whose chemical species could not be identified are also shown in color, and fall into three main categories:
    (i) charging transitions in the forward and backward bias sweeps that cannot be matched (`Uncertain matches'); 
    (ii) defects that are too irregular in shape (`Noncompact'); or
    (iii) that have too few pixels assigned to them by the combined clustering and Gaussian-mixture algorithms (`Too few pixels'). }
    \label{fig:transitions_defects_maps}
\end{figure*}

\vspace{.1in}\noindent
\textbf{\large{Results}}

\vspace{.1in}\noindent
\textbf{EFM measurements.} 
A schematic illustration of the experimental setup and technique is shown in Fig.~\ref{fig:setup_experiment}. 
The sample is formed of 60~nm of Al\_2O\_3 grown atop a $381~\SI{}{\micro\meter}$ silicon wafer, as described in Methods and Materials.
A quartz-cantilever (qPlus) EFM tip is used to probe the sample surface in non-contact, constant-frequency mode at a temperature of 4.5~K and ultra-high vacuum (UHV) conditions. 
Below, we refer to the grid points of this scan as ``pixels.''
For many of the images obtained here, the pixels are spaced by 0.5~nm, although in several cases the grid spacings can be as small as 0.1~nm. 
A topography map obtained by non-contact atomic force microscopy (nc-AFM) is shown 
Fig.~\ref{fig:transitions_defects_maps}\emph{A} for a $25\times25$~nm$^2$ portion of sample, revealing an average roughness of 1.5~nm.  

We also perform EFM measurements at each pixel location.
Here, the feedback loop is turned off and the change in the tip resonance frequency $\Delta f$ is recorded as a function of the tip-sample bias $V_b$, as shown  for two different pixels on the left-hand side of Fig.~\ref{fig:transitions_defects_maps}\emph{B}.
Such curves are referred to as ``$f$-$V$'' curves and are central to our analysis. 
As illustrated schematically in Fig.~\ref{fig:setup_experiment}\emph{C}, the curves are roughly parabolic in shape, with a maximum occurring at the voltage $V_\text{CPD}$ for which the contact potential difference (CPD) between the tip and sample is zero~\cite{kitamura2005mapping, Gross2009measuring, neff_insights_2015,Rosenwaks2004}.
CPD maps are a common product of Kelvin probe force microscopy (KPFM) experiments, where they characterize the work function difference between the tip and backgate, modulated by local fluctuations of the electrostatic potential.
Figures~\ref{fig:setup_experiment}\emph{A} and \ref{fig:transitions_defects_maps}\emph{C} show CPD maps obtained for the same region of sample as Fig.~\ref{fig:transitions_defects_maps}\emph{A}.
The points labeled $a$ and $b$ correspond to the $f$-$V$ curves in Fig.~\ref{fig:transitions_defects_maps}\emph{B}.
Although these points are separated by only $\sim$5~nm, their $V_\text{CPD}$ values are distinct, highlighting the sensitivity of EFM to local electrostatic environments.

The $f$-$V$ curves also display prominent jumps, representing single-electron charging events between the tip and a defect near the sample surface~\cite{ludeke2001imaging, bussmann2005single, dana2005electrostatic, stomp2005detection, zhu2005frequency, azuma2006single, bussmann2006single}. 
The charging process and its effect on scanning probe measurements is illustrated in Fig.~\ref{fig:setup_experiment}\emph{B-D}. 
Here, the tip bias $V_b$ causes strong gradients of the electrostatic potential $\phi$ near the vacuum/oxide interface, with corresponding changes in the defect's chemical potential.
Under appropriate conditions, charging or discharging can occur, if the tip and defect are close enough to allow tunneling. 
(In \emph{SI Appendix}, Fig.~\ref{fig:highResData_chargingVoltages}, we confirm tunneling as the charging mechanism by tracking hysteresis of charging voltages as a function of tip-sample separation.)
The sudden change of charge modifies the local electrostatic potential, as manifested by a jump in the resonant frequency of the EFM cantilever.
For a wide enough range of $V_b$, multiple charging events are likely to be observed at a single pixel, which originate from one or more defects. 
For example, Figs.~\ref{fig:transitions_defects_maps}\emph{D} and \emph{E} show very different charging maps, for two different bias windows at the same location.
Below, the combination of local charging transitions and CPD-based spectroscopy techniques provide powerful tools for characterizing defects and identifying their chemical species, as shown in Fig.~\ref{fig:transitions_defects_maps}\emph{F}.

\begin{figure*}[t]
    \centering
    \includegraphics[width=0.9\textwidth]{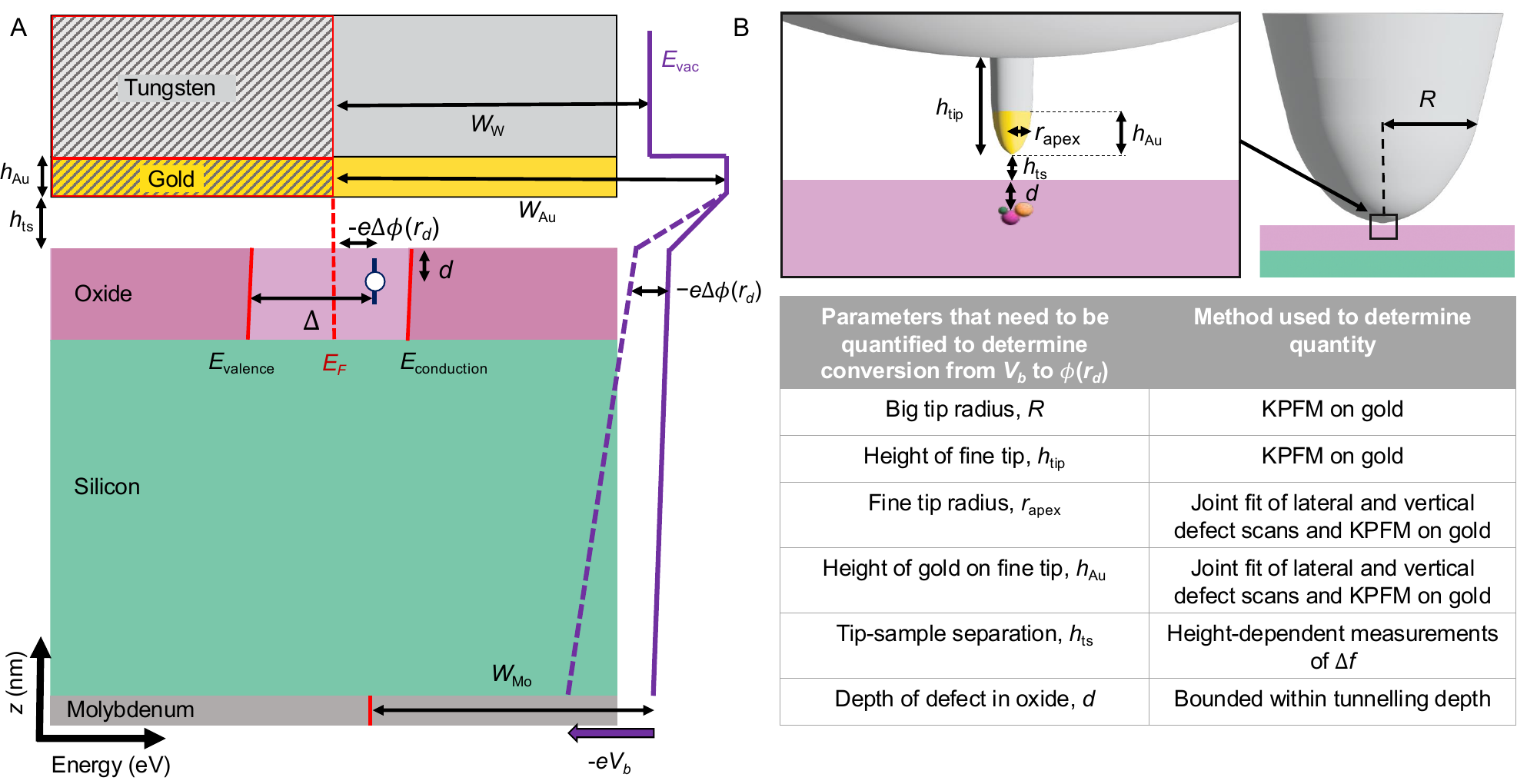}
    \caption{
    Device and sample parameters.
    The experimental set-up, including (\emph{A}) vertical profiles of the tip and sample, and (\emph{B}) the lateral tip geometry, in which a fine tip (left) is attached to the apex of a big tip (right).
    The work function of gold ($W_\text{Au}$) is well known, while the work functions of tungsten and molybdenum ($W_\text{W}$ and $W_\text{Mo}$) are determined in \emph{SI Appendix}, section~\ref{AuMeasurements:simulations}.
    Unknown parameters are listed in the table and illustrated above it.
    These parameters are determined as described in Fig.~\ref{fig:tip_characterization}  through a series of experiments, including EFM scans as a function of tip height, above either the bare oxide surface or a gold-coated sample.
    The local electrostatic potential depends on the tip height, the voltage bias $V_b$ between tip and back gate, and the presence of charged defects in the sample.
    The reference vacuum energy level (solid purple line) is obtained at the special bias $V_b=\overline V_\text{CPD}$ for which interactions between the tip and sample vanish.
    (We define the reference with respect to the globally averaged CPD potential.)
    Local shifts in the electrostatic potential $\Delta \phi$ (dashed purple line) are computed via electrostatics simulations.
    A charging transition occurs when a defect's chemical potential crosses the Fermi energy of the tip $E_F$, as a function of $V_b$.
    DFT calculations also provide the chemical potential for these transitions, $\Delta$, relative to the valance-band edge.
    Combining experimental measurments, DFT calculations, and electrostatics simulations allow us to identify the chemical species of a defect.    
}
    \label{fig:tip_shape_fitting}
\end{figure*}

\vspace{.1in}\noindent
\textbf{Tip-sample simulations and characterization.}
Charging transitions can occur when a defect's chemical potential falls below the Fermi level set by the conducting tip.
The chemical potential is modulated by the electrostatic potential at the defect's location $\phi({\mathbf r}_d)$, which depends on the tip shape, material work functions, and of course, the tip-backgate bias.  
A key focus of this work is to develop accurate methods for characterizing these various parameters, as illustrated in Fig.~\ref{fig:tip_shape_fitting}.  
Previous work has employed parallel-plate capacitor models to calculate the tip-defect bias $\phi ({\mathbf r}_d)$~\cite{bussmann2006single,cowie2024spatially,cowie2024spatially2,zheng2010electronic}.
However, such models break down when the tip-backgate separation is greater than the tip radius; for qubit devices and for our sample, these models are inappropriate.
A conical tip model provides more-accurate results for long-range tip-sample interactions~\cite{dana2005electrostatic}.
However, the fine structure of the tip point, which allows for better imaging resolution~\cite{Behn2021}, can significantly modify the short-range interactions.
In this work, we adopt a realistic model of a parabolically terminated conical tungsten tip with a nanometer-scale protruding apex, as illustrated in Fig.~\ref{fig:tip_shape_fitting}\emph{B}, whose endpoint has been coated with Au by poking into a gold pad.
The tip parameters that play a role in our simulations and in defect characterization are also listed in the figure.
The procedures used to determine these parameters are described in Methods, with details given in the \emph{SI Appendices}.

The oxide-on-silicon sample studied here accommodates a high density of charged defects near the sample surface~\cite{Wolfe2024defects}.
Indeed, our EFM measurements suggest that the tip interacts with many defects simultaneously, making individual identifications challenging.
Each defect can also undergo multiple charging transitions, further complicating their identification.
In this work, we apply a combination of techniques to systematically isolate individual defects.
We then assign charge-transition voltages $V_\text{transition}$ to the defects.
This procedure is complicated by the fact that the charging voltage is hysteretic with respect to forward and backward bias sweeps, and depends on the tip-defect separation.
Finally, the transition voltages are converted to chemical potentials $\Delta$, like those illustrated in Fig.~\ref{fig:setup_experiment}\emph{D}.
The series of procedures used to extract defect chemical potentials from EFM measurements are summarized in Methods, with details given in \emph{SI Appendix}.

\begin{figure*}[t]
    \centering
    \includegraphics[width=0.7\textwidth]{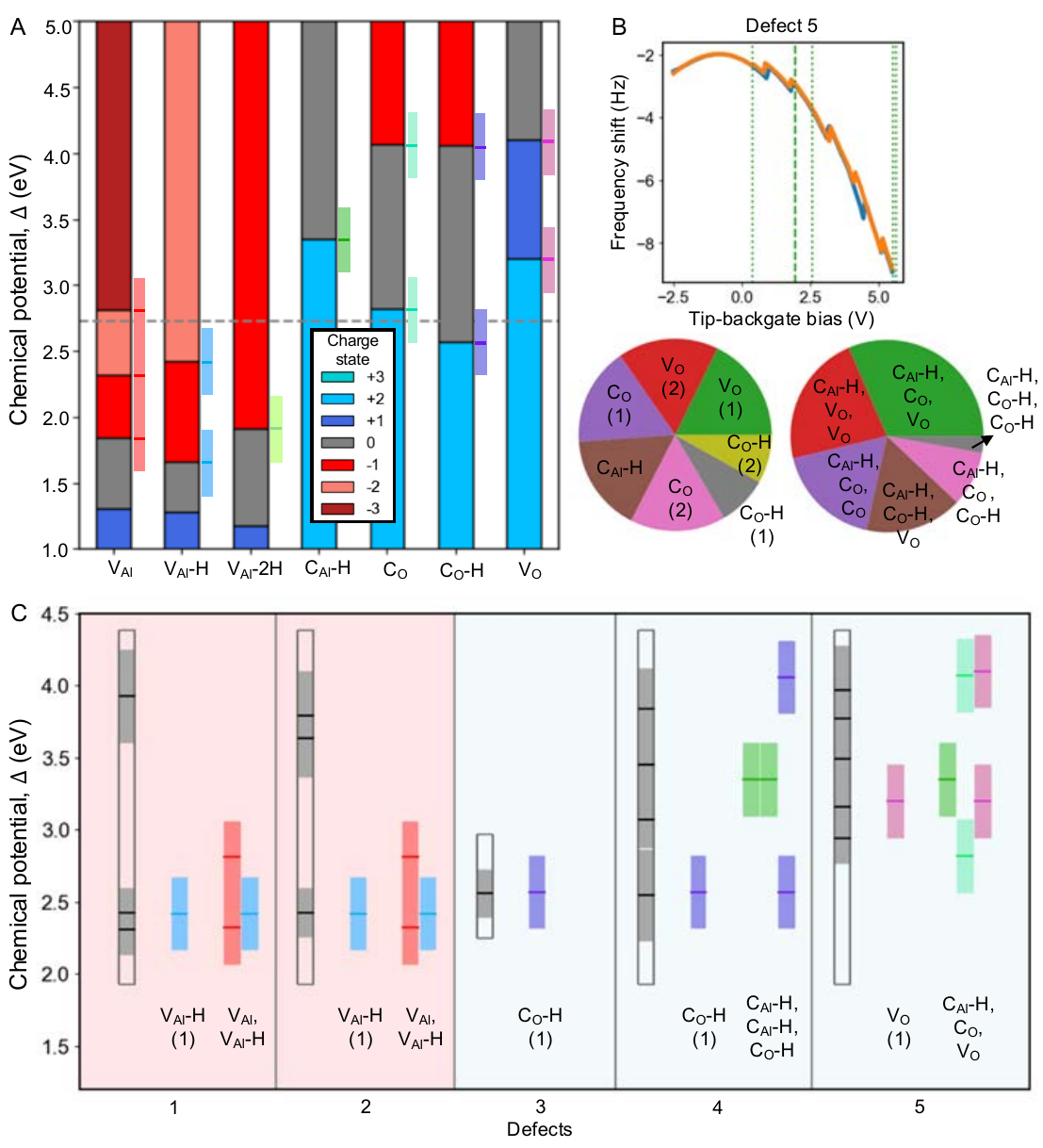}
    \caption{Identifying the chemical species of defects in Al$_2$O$_3$ by comparing EFM measurements and DFT analyses.
    (\emph{A}) Wide bars show DFT predictions for the chemical potentials ($\Delta$) of multiple charging transitions, for seven different defect species, with charge states indicated by colors (inset).
    Here, $\Delta$ is referenced to the valence-band edge of Al$_2$O$_3$ and the gray dashed line represents the Fermi level for the case where the vacuum potentials of the tungsten portion of the EFM tip and the molybdenum backgate are aligned ($V_b=V_{\Delta\Phi}$, see main text).
    The small bars to the right of the wide bars define the color coding used to identify the defects, below, and the length of the bar indicates the theoretical energy uncertainty $\sigma_i$ for a given transition $i$.
    (\emph{B}) Top: the measured $f$-$V$ curve for Defect~5.
    The vertical dashed line shows our best match (V$_\text{O}$) to a theoretically predicted transition, based on the single-defect analysis.
    The vertical dotted lines (including one behind the dashed line) show our best theoretical matches (C$_\text{Al}$-H, C$_\text{O}$, and V$_\text{O}$), based on the multi-defect analysis.
    Bottom: pie charts showing the relative assignment probabilities for different defect species, based on the single-defect analysis (left) and the multi-defect analysis (right).
    Here, the notation (1) indicates that one transition was ignored in the analysis, because it was an edge case, while (2) indicates that all transitions were included in the analysis.
    (\emph{C}) For Defects 1-5, we show our best matches between theoretical predictions (left column) and experimentally measured charging transitions, based on the single-defect analysis (middle column) and the multi-defect analysis (right column).
    Here, black lines represent the experimental transition voltages, converted to chemical potentials (see main text), the gray bars represent the experimental uncertainties for the transition ($\rho_i$), the black boxes represent the experimental measurement range, a red background indicates a negatively charged defect, and a blue background indicates a positively charged defect.
    Colored boxes are reproduced from \emph{A}, with boxes removed if they fall outside the experimental range.
    The defects are ordered from left to right in decreasing order of assignment probabilities, based on the single-defect analysis. }
    \label{fig:dft_energy_levels}
\end{figure*}

\vspace{.1in}\noindent
\textbf{Defect identification.}
Defects are identified chemically by comparing our estimates of $\Delta$ from  EFM measurements to those from DFT calculations.
The latter are obtained for a set of seven candidate point defects in Al$_2$O$_3$, including native defects like aluminum and oxygen vacancies (V$_\text{Al}$, V$_\text{O}$), substitutional impurities (C$_\text{O}$), and 
hydrogenic complexes involving these defects (V$_\text{Al}$-H, V$_\text{Al}$-2H, C$_\text{Al}$-H, C$_\text{O}$-H).
(See Methods for details.)
Each defect species can host multiple charge-state configurations, as illustrated in Figs.~\ref {fig:setup_experiment}\emph{D} and \emph{E}. 
Figure~\ref{fig:dft_energy_levels}\emph{A} shows the results of DFT calculations, including the chemical potentials $\Delta$ at which charging occurs (relative to the valence-band edge), with the corresponding charge states labeled.
For each defect species, the thin colored bar next to the main bar illustrates the defect color-coding scheme, used later in the figure, and its size reflects the error bar of the DFT calculations (see \emph{SI Appendix}, section~\ref{sec:errors}).

We employ a three-step procedure to assign the most likely chemical species for each of the experimentally measured defects. 
First, we determine the sign of the charged defect by comparing the $V_\text{CPD}$ value of its centermost pixel to the CPD of a charge-neutral sample (see \emph{SI Appendix}, section~\ref{sec:sign}); this allows us to narrow down the set of candidate defects.
We note that defects classified as neutral at the CPD condition ($V_b=V_\text{CPD}$) are excluded from further consideration because, when we account for experimental uncertainties, their true charge sign could be masked by a nearby defect with the opposite sign.
Here, ``neutral'' is defined as having a $V_\text{CPD}$ falling between the hysteretic, sample-averaged CPD values in forward vs backward voltage scans (i.e., $-0.33<V_\text{CPD}<-0.17$~V), since this range represents an effective error bar for our $V_\text{CPD}$ estimates.
Second, we exclude any defects for which theoretically predicted charging transitions are not observed within the experimental bias window. 
However, we do allow for the possibility of observing transitions from up to three different defects at a given pixel, since the defects may be in close proximity.
Finally, from the remaining defects, we determine the most likely candidate using a maximum likelihood technique~\cite{Perneger2021}. 
Here, we define a likelihood functional
\begin{equation}
    \mathscr{L}(E_1^{(\theta)},\dots,E_k^{(\theta)})
    = \prod_{i=1}^{k}\frac{1}{\sqrt{2\pi(\sigma_i^2+\rho_i^2)}}
    e^{-\frac{1}{2}\frac{(\Delta_i-E_i^{(\theta)})^2}{\sigma_i^2+\rho_i^2} } ,
    \label{eq:likelihood}
\end{equation}
where $\theta$ is an index labeling the defect species, $k$ is the number of charging transitions predicted for that species, $E_i^{(\theta)}$ is the theoretically computed chemical potential of the $i^\text{th}$ charging transition (note that $k$ and $E_i^{(\theta)}$ depend implicitly on $\theta$), and $\Delta_i$ is the experimentally determined chemical potential for the $i^\text{th}$ transition.
The functional also includes the experimental (theoretical) one-sigma error bars for $\Delta_i$ ($E_i^{(\theta)}$), designated $\rho_i$ ($\sigma_i$), which we combine in quadrature.
Experimental errors in our calculations include contributions from the unknown depth of the defect, hysteresis in defect charging and discharging voltages, surface roughness, uncertainties in tip geometry parameters, and the center location of the defect. 
(A further discussion of errors is given in \emph{SI Appendix}, section~\ref{sec:errors}.)

\begin{table*}
\centering
\begin{tabular}{ c c c c c c c }
\multicolumn{7}{l}{\textbf{\large{Table 1. Defect assignments and probabilities}}} \\ [0.5ex]
\hline 
 & Best assignment & Probability of & Probability & Best assignment & Probability of & Probability \\ 
  Defect No. & (single-defect) & assignment & ratio & (multi-defect) & assignment & ratio \\ 
\hline 
 1 & V$_\text{Al}$-H & 0.9998 & 2.25 & V$_\text{Al}$, V$_\text{Al}$-H & 0.02306 & 2.52 \\
 2 & V$_\text{Al}$-H & 0.9998 & 4.36 & V$_\text{Al}$, V$_\text{Al}$-H & 0.0001621 & 10.7 \\
 3 & C$_\text{O}$-H & 0.9996 & 1.45 & - & - & - \\
 4 & C$_\text{O}$-H & 0.9984 & 1.04 & C$_\text{Al}$-H, C$_\text{Al}$-H, C$_\text{O}$-H & 0.5409 & 1.28 \\
 5 & V$_\text{O}$ & 0.9924 & 1.06 & C$_\text{Al}$-H, C$_\text{O}$, V$_\text{O}$ & 0.5664 & 1.42 \\
 6 & C$_\text{O}$-H & 0.9917 & 1.07 & C$_\text{O}$-H, C$_\text{O}$-H & 0.9544 & 1.03 \\
 7 & C$_\text{Al}$-H & 0.9887 & 1.04 & V$_\text{O}$ & 0.6228 & 1.03 \\
 8 & C$_\text{O}$-H & 0.9845 & 1.13 & C$_\text{O}$-H & 0.6987 & 1.16 \\
 9 & C$_\text{O}$ & 0.9828 & 1.00 & C$_\text{Al}$-H, C$_\text{O}$ & 0.8565 & 1.65 \\
 10 & V$_\text{Al}$-H & 0.9802 & 1.76 & V$_\text{Al}$, V$_\text{Al}$-H & 0.009246 & 1.72 \\
 11 & C$_\text{Al}$-H & 0.9785 & 1.01 & C$_\text{Al}$-H, C$_\text{Al}$-H, C$_\text{O}$-H & 0.4337 & 2.60 \\
 12 & C$_\text{O}$ & 0.9760 & 1.00 & C$_\text{Al}$-H, C$_\text{O}$, V$_\text{O}$ & 0.5477 & 1.50 \\
 13 & V$_\text{O}$ & 0.9498 & 1.04 & C$_\text{Al}$-H, V$_\text{O}$ & 0.7353 & 1.38 \\
 14 & V$_\text{Al}$-H & 0.9232 & 1.21 & - & - & - \\
 15 & C$_\text{O}$-H & 0.9206 & 1.78 & - & - & - \\
 16 & V$_\text{Al}$-H & 0.9046 & 1.22 & - & - & - \\
 17 & V$_\text{Al}$-H & 0.8934 & 1.80 & V$_\text{Al}$, V$_\text{Al}$-H & 2.519e-05 & 5.13 \\
 18 & C$_\text{Al}$-H & 0.8769 & 1.34 & - & - & - \\
 19 & C$_\text{Al}$-H & 0.8762 & 1.38 & V$_\text{O}$ & 0.4725 & 1.16 \\
 20 & V$_\text{Al}$-H & 0.8675 & 1.63 & V$_\text{Al}$ & 0.02823 & 32.0 \\
 \hline
\end{tabular}
\caption{The 20 defects identified in this work are listed in decreasing order of their highest \emph{probability assignment}, defined as the ratio of the likelihood [Eq.~(\ref{eq:likelihood})] to its maximum value (see main text and \emph{SI Appendix}, section~\ref{sec:likelihood}). 
The probability \emph{ratio} is defined as the ratio of the best and second-best probability assignments. 
Single-defect assignments require all DFT-predicted charging transitions to be accounted for (except for potential edge cases), while ignoring non-predicted transitions. 
Multi-defect assignments must account for all the experimentally observed charging transitions (again, making exceptions for edge cases).
In cases where only one transition occurs in the measurement window, the two methods yield the same result, and we leave the multi-defect assignments blank. }
\label{table:table_likelihoods}
\end{table*}

The likelihood functional of Eq.~(\ref{eq:likelihood}) is normalized; however, its units depend on the number of charging transitions $k$.
To provide a fair comparison between different defect species, with different $k$, we therefore define two dimensionless quantities, derived from Eq.~(\ref{eq:likelihood}).
First, we define a \emph{probability of assignment}, which represents a normalized likelihood.
In this case, the likelihood functional is computed from Eq.~(\ref{eq:likelihood}), and then recomputed to give its maximum value, by setting $\Delta_i=E_i^{(\theta)}$ for each transition $i$.
The (dimensionless) assignment probability is then given by the ratio of these two quantities.
We also consider the \emph{probability ratio} between the highest and second-highest probability assignments, which provides a measure of confidence in the best assignment.

We also consider two defect assignment procedures.
The first provides a comparison between the theoretically predicted and experimentally measured charging transitions, when just a \emph{single defect} species is considered.
Here, we insist that all the predicted transitions for the defect should be observed if they fall within the experimental voltage range
(see \emph{SI Appendix}, section~\ref{sec:likelihood}).
We allow for the possibility that other charging transitions, associated with nearby defects, may be present in the data, and we do not fit those transitions. 
We also allow for edge cases, in which the theoretically predicted transitions occur within a error bar $\rho_i$ of the edge of the measurement range.
We then estimate the assignment probabilities with and without these edge transitions.
In the second assignment procedure, we simultaneously consider \emph{multiple defects}.
In this case, we insist that all experimentally observed transitions should be matched to theoretical predictions.
We note it is possible to satisfy this constraint with just a single defect (indeed, this is a common result, as shown below); however, we still refer to this method as ``multi-defect,'' for simplicity.
Finally, we note that edge cases are also allowed in this method.

Figures~\ref{fig:dft_energy_levels}\emph{B} and \emph{C} showcase some outcomes of this analysis.
The top panel of Fig.~\ref{fig:dft_energy_levels}\emph{B} shows the $f$-$V$ curve for the center pixel of Defect~5.
The vertical dashed line shows the theoretical prediction that most closely matches the charging transition seen in the data (V$_\text{O}$, based on the single-defect analysis). 
The pie chart on the left shows the full set of assignments for this analysis and their relative probabilities.
Here, the designation (1) indicates an analysis where one of the transitions is an ignored edge case, while the designation (2) indicates that all edge cases are included in the analysis.
The vertical dotted lines in the top panel show theoretical predictions most closely matching the full set of charging transitions in the multi-defect analysis. 
(One of the dotted lines coincides with the dashed line.)
Here, the most-probable solution corresponds to the defect set $\{$C$_\text{Al}$-H, C$_\text{O}$, V$_\text{O}\}$.
The pie chart on the right shows the full set of assignments for the multi-defect analysis.

Figure~\ref{fig:dft_energy_levels}\emph{C} describes the most-probable defect assignments for Defects 1-7, shown as bar graphs.
Here, pink shading indicates negatively charged defects (1 and 2), while blue shading indicates positively charged defects (3-5).
The experimental charging transitions and their corresponding error bars ($\rho_i$) are shown as black lines/gray bars.
The second column in each box shows the theoretical predictions most closely matching the experimental transitions, based on the single-defect analysis, using the color coding scheme from Fig.~\ref{fig:dft_energy_levels}\emph{A}.
The third column in each box shows the theoretical predictions most closely matching the experimental transitions, based on the multi-defect analysis.

Table~1 summarizes our main results.
Although EFM maps suggest an abundance of defects, our analysis focuses on the clearest and least ambiguous examples, which probably occur near the sample surface.
We thus arrive at a set of 20 defects.
For each of these defects, we provide our best guesses for their chemical assignments, with corresponding probabilities and probability ratios.
We report results for both the single-defect (columns 2-4) and multi-defect (columns 5-7) assignment methods.
In the latter case, the assignments may involve more than one object; however, we retain the terminology ``defects'' here, for simplicity.
In the table, the defects are labeled and ordered according to the probability estimates from the single-defect analysis.

The table shows some interesting results.
We first notice that the V$_\text{Al}$-H complex is well represented among the various defect assignments, often yielding high probabilities and probability ratios (e.g., Defects 1 and 2), where the latter indicates that the second-best assignment is somewhat less likely than the V$_\text{Al}$-H assignment.
These results are corroborated by the multi-defect analysis, which also favors V$_\text{Al}$-H assignments.
Typically however, the corresponding probabilities are \emph{low} in this analysis, especially when V$_\text{Al}$-H is paired with V$_\text{Al}$, indicating that V$_\text{Al}$ does not match the data very well, despite being our best estimate.
Taken together, these results confirm the prevalence of V$_\text{Al}$-H impurities, but they also suggest that our DFT analysis may have missed a defect species that is often found in the proximity of V$_\text{Al}$-H.
Thus, although V$_\text{Al}$ vacancies are listed in the table, there is no strong evidence to support their presence.
Other well-supported defect species include C$_\text{O}$-H, V$_\text{O}$, C$_\text{Al}$-H, and V$_\text{Al}$-H, emphasizing the prominance of hydrogenic impurities in materials, which can arise due to surface contamination, or defect nucleation at OH ligands in the oxide~\cite{Zhang2011}.
It is also interesting to note that, in several cases, defects listed in the table under the single-defect analysis do not appear on the corresponding list for the multi-defect analysis.
This is a consequence of the stricter constraints used in the multi-defect analysis, requiring that all measured transitions must be accounted for in the theory, and cannot be ignored.

Finally, we close by highlighting an interesting, alternative approach for addressing the question of whether multiple transitions, observed at a single pixel, can be assigned to a singe defect or to multiple defects, based on more direct evidence.
In \emph{SI Appendix}, section~\ref{sec:matching_multiple}, we describe a powerful approach based on hysteresis correlations between neighboring pixels, which accurately pinpoints the center location of a given transition.
This provides an interesting direction for future work.

\vspace{.1in}\noindent
\textbf{\large{Conclusions}}

\vspace{.1in}\noindent
In summary, we have demonstrated how EFM can be used as a chemically specific tool for identifying defects in non-metallic substrates. Building upon previous work in which single-defect charging processes can be observed in EFM, we show that, when combined with detailed tip modeling, such charging processes can still reveal spectral signatures within a denser array of defects. 
By comparing these signatures to theoretically predicted chemical potentials for charging events, it is possible to make informed guesses about their exact chemical structures. 
The technique described here offers a complementary approach for surface analysis, which can provide more accurate identifications of surface impurities and lead to more-specific techniques for surface preparation. 
We note that the amorphous surface studied in this work contains a high density of unknown defect species, representing sub-optimal conditions for our technique. 
Despite such challenges, we have shown that EFM can still provide convincing guesses about the chemical identities of the defect species. 
Future work using EFM to characterize defects in flat, crystalline materials with fewer defects, such as those being explored for next-generation quantum devices, would provide an excellent opportunity to utilize EFM with greater accuracy and reduced uncertainty, facilitating the identification, elimination, or mitigation of specific defects.

\vspace{.1in}\noindent
\textbf{\large{Materials and Methods}}

\begin{figure*}[t]
    \centering
    \includegraphics[width=0.9\textwidth]{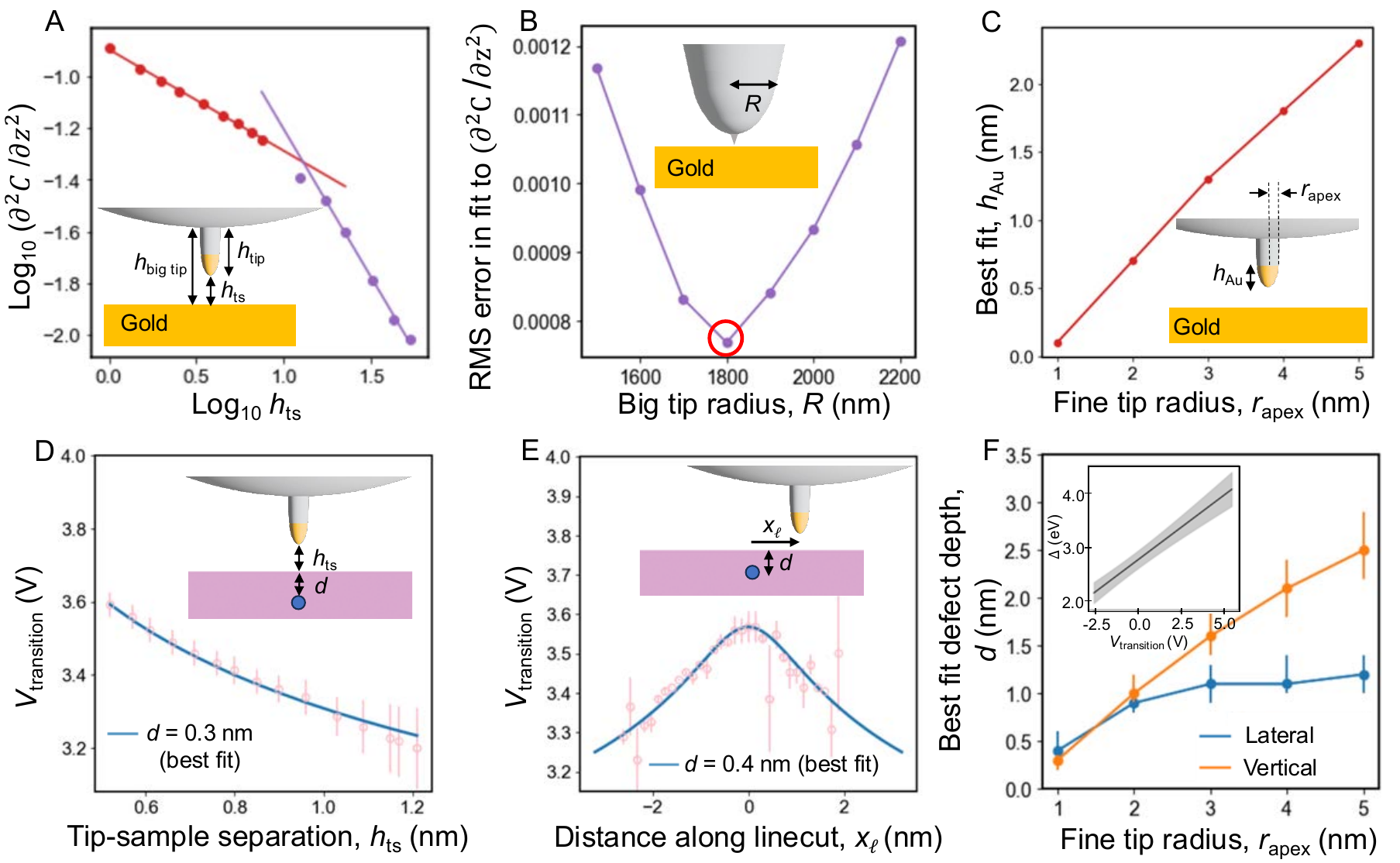}
    \caption{An overview of the methods used to characterize the geometrical parameters of the two-section EFM tip.
    (Full details are presented in \emph{SI Appendix}, section~\ref{sec:Tip_Geometry}.)
    (\emph{A})-(\emph{C}) Measurements above a gold sample.
    (\emph{A}) The capacitance curvature $\partial^2C/\partial z^2$ is measured as a function of tip-sample separation $h_\text{ts}$.
    Two types of behavior are observed: for small $h_\text{ts}$, tip-sample interactions are affected by both the coarse and fine-tip sections (red data), while for $h_\text{ts}>12.5$~nm, interactions are dominated by just the coarse (``big'') tungsten portion of the tip (purple data).
    (\emph{B}) First, COMSOL simulations are used to characterize the tungsten portion of the tip.
    For the purple data in (\emph{A}), simulations are performed as a function of the tip-sample separation and the big-tip radius $R$.
    Minimizing the RMS difference between $\partial^2C/\partial z^2$ values from the experiments and simulations (red circle) provides a two-parameter fit, yielding $R=1800$~nm.
    Additional knowledge of the tip-sample separation $h_\text{ts}$ from \emph{SI Appendix}, section~\ref{sec:tip-sample_separation}, gives the optimal value for the fine-tip height $h_\text{tip}=h_\text{big tip}-h_\text{ts}=23$~nm.
    (\emph{C}) Fine-tip parameters are obtained similarly, by comparing experimental data to COMSOL simulations, which now include the full two-section tip geometry.
    A fitting procedure gives the optimal relation between the length of the gold coating on the fine tip $h_\text{Au}$ and the fine-tip radius $r_\text{apex}$.
    However, since these two parameters have a similar effect on $\partial^2C/\partial z^2$, we are not able to determine a globally optimized pair of values $(h_\text{Au},r_\text{apex})$ from
    experiments above a gold sample.
    (\emph{D})-(\emph{E}) Measurements above an oxide-on-Si sample.
    Here, we choose an isolated defect with a well-defined charging transition and track this transition voltage $V_b=V_\text{transition}$ as the tip is scanned both vertically and laterally.
    (\emph{D}) Data points (pink markers) show $V_\text{transition}$ as a function of $h_\text{ts}$ when the tip is placed directly above the defect and scanned vertically.
    Simulation results (blue curve) are obtained for a fixed set of parameters $(h_\text{Au},r_\text{apex},d)$, where $d$ is the unknown depth of the defect below the sample surface, and the correlated parameters $h_\text{Au}$ and $r_\text{apex}$ are taken from (\emph{C}).
    $d$ is used as a fitting parameter, here, to provide the best match between simulations and experiments.
    The figure shows simulations results for the case $r_\text{apex}=1$~nm.
    (\emph{E}) The same procedure is used for lateral scans passing directly over the defect.
    Simulations results are again shown for $r_\text{apex}=1$~nm.
    (\emph{F}) Results of the fitting procedure in (\emph{D}, orange) and (\emph{E}, blue) showing the best-fit results for $d$.
    Since the optimal $d$ value should be consistent between these two cases, we settle on $d=(0.65\pm 0.30)$~nm as our final optimized result, with corresponding values for $h_\text{Au}$ and $r_\text{apex}$.
    Based on these parameter characterizations, the inset shows the resulting conversion between an experimentally measured charging-transition voltage $V_\text{transition}$ for a given defect, to its corresponding chemical potential $\Delta$, as discussed in \emph{SI Appendix}, section~\ref{sec:converting_Vtransion}. 
    Here, the shading indicates the uncertainty.}
    \label{fig:tip_characterization}
\end{figure*}

\vspace{.1in}\noindent
\textbf{Sample preparation.}
The sample used in this study was fabricated on a p-type Si~(100) wafer with a resistivity of 10–20~$\Omega\cdot$cm. 
Before fabrication, the wafer was cleaned using a preliminary piranha dip, followed by an RCA process, and a final HF dip, with the HF diluted with deionized water to a ratio of 1:50, to remove the native oxide.
A 60~nm Al$_2$O$_3$ dielectric layer was then deposited by thermal atomic layer deposition (ALD) at a substrate temperature of 250~$^\circ$C using room-temperature precursors trimethylaluminum (TMA) and H$_2$0.

\vspace{.1in}\noindent
\textbf{Tip characterization.}
We adopt a realistic model of a parabolically terminated conical tungsten tip with a nanometer-scale protruding apex, as illustrated in Fig.~\ref{fig:tip_shape_fitting}\emph{B}, whose endpoint has been coated with Au by poking into a gold pad.
Several tip parameters can be deduced from height-dependent measurements of the CPD and the tip-sample capacitance when the tip is positioned above a Au(111) surface, as described in the caption of Fig.~\ref{fig:tip_characterization}\emph{A-C}.
(For details, see \emph{SI Appendix}, section S2.)
The crossover between short- and long-ranged interactions is demarcated by red and purple in Fig.~\ref{fig:tip_characterization}\emph{A}.

To determine the remaining tip parameters, we also perform vertical and lateral scans above a defect undergoing charging transitions in the oxide-on-Si sample studied here.
Such measurements are described in the caption of Fig.~\ref{fig:tip_characterization}\emph{D-F} and \emph{SI Appendix}, section~S2.
The combined measurements yield the following tip parameters:
big-tip radius $R=(1800\pm 200)$~nm, fine-tip height $h_\text{tip}=(23 \pm 2)$~nm, fine-tip radius $r_\text{apex}=(1.5\pm 0.5)$~nm, and fine-tip gold height $h_\text{Au}=(0.5\pm 0.3)$~nm.
The subsurface depth of the defect used for this analysis is obtained as a byproduct, giving $d=(0.65\pm 0.35)$~nm.
This corroborates our intuition that defect charging transitions resulting from the tunneling of electrons to or from the EFM tip can only occur when the defects are very near the surface.
When applying the procedures described here for chemical analysis, we do not employ such an arduous method for determining $d$ in every defect.
Instead, we take the defect to be located at the sample surface ($d=0$) and assign a generous error bar of size $\Delta d=2.5$~nm~\cite{Jared}. 

\vspace{.1in}\noindent
\textbf{Converting transition voltages to defect energies.}
We apply a combination of techniques to systematically isolate individual defects, as a first step to characterizing their transition energies.
We now summarize one of these procedures, leaving the details to \emph{SI Appendix}, section~\ref{sec:Identifying_Defects}.
We first identify the charging transitions at a given pixel by analyzing slope discontinuities and deviations from parabolicity in the CPD curves.
(See \emph{SI Appendix}, section~\ref{sec:locating_transitions}.)
We then apply a Connected Component Analysis~\cite{opencv_library,LindaG_Shapiro1996} to the resulting transition maps to identify groups of adjacent pixels whose transition voltages fall within the same narrow voltage window.
Although the width of this voltage window is fixed, its center voltage value is swept continuously, to allow the capture of all possible transitions.
However, we then need to track the transitions across these shifting windows; to do this, we apply a Spectral Clustering Algorithm~\cite{towardsdatascienceSpectralClustering} to associate groups of pixels with spectral overlaps, largely eliminating duplicate defect identifications.
Unfortunately, this method is unable to distinguish transitions occurring in closely spaced defects of the same species.
We therefore apply a Gaussian Mixture Model Algorithm~\cite{scikit_learn}, which enforces a ``compactness'' (or roundness) metric that splits apart clusters of pixels that are large and irregularly shaped into smaller, more-regular clusters.
This completes the identification of individual defects and their charging events.

We next assign charge-transition voltages to the defects.
This procedure is complicated by the fact that the charging process is hysteretic: charging occurs at different tip-bias voltages during forward and backward voltage sweeps.
Moreover, different types of hysteresis are observed in the data (e.g., see \emph{SI Appendix}, Fig.~\ref{fig:highResData_chargingVoltages}), suggesting interesting hidden physics that will be explored in future work.
Here, we simply note that the hysteresis is minimized at the point of closest tip-defect separation, where the coupling is strongest.
To locate this point, we first note that the tunneling process is stochastic, yielding noisy transition maps like those shown in Figs.~\ref{fig:transitions_defects_maps}\emph{D} and \emph{E}, in which the defect centers can be difficult to identify.
To mitigate this problem, we apply an averaging procedure that convolves the data with a gaussian function (see \emph{SI Appendix}, section~\ref{sec:gaussian_blur}); the center of the gaussian is identified as the center of the defect.
To minimize the effects of hysteresis, we compare the forward and backward voltage sweeps at this center point.
The $f$-$V$ curve at this pixel may exhibit many charging transitions, including transitions that originate on neighboring defects.
We must therefore match the corresponding transitions in the forward and backward voltage sweeps; this is done visually, based on similarities in their jump amplitudes and behavior.
(See \emph{SI Appendix}, section~\ref{sec:matching_transitions}.)
Transitions with ambiguous matches are assumed to arise from different defects and are discarded.
By definition, the \emph{intrinsic} charging voltage contains no hysteresis, and we approximate it here as the average of the forward and backward transition voltages for the pixel at the center of a given defect.

In \emph{SI Appendix}, section~\ref{sec:Identifying_Defects2}, we also apply a second method to identify defects and locate their center positions, making use of the fact that the transition-voltage hysteresis is minimized when the EFM tip is centered directly above a defect.
Combining these two procedures, we identify a total of 20 distinct defects in our data set, as indicated in Fig.~2\emph{F}.

As a final step of the data analysis, the charging-transition voltages $V_\text{transition}$ of individual defects are converted to chemical potentials $\Delta$, like those illustrated in Figs.~\ref{fig:setup_experiment}\emph{D} and \emph{E}.
As detailed in \emph{SI Appendix}, section~\ref{sec:converting_Vtransion}, this is done by comparing the CPD reference potential (i.e., the globally average CPD, $\overline{V_\text{CPD}}$), representing the zero-bias condition between the backgate and the tip in the absence of charged defects, to the charge-transition voltages found in the previous paragraph.
Because the device is heterogeneous, with several different materials workfunctions, we relate the Fermi level of the tip to the chemical potential of the defect through the locally varying vacuum potential (purple lines in Figs.~\ref{fig:tip_shape_fitting}\emph{A} and \ref{fig:backgate_0V_1V_defectEnergy}).
Our calculations make use of two COMSOL simulations: a simulation of the vacuum potential at the globally averaged CPD condition (\emph{SI Appendix}, section~\ref{sec:converting_Vtransion}) and a simulation of the lever arm relating the voltage bias $V_b$ to the shifted electrostatic potential at the location of the defect $\Delta\phi({\mathbf r}_d)$ (\emph{SI Appendix}, section~\ref{sec:lever_arm}).
We finally express the defect chemical potential relative to the valence-band edge, as required for comparisons with DFT calculations, through knowledge of the electron affinity and the band gap of the Al$_2$O$_3$ oxide.
The final conversion is illustrated in the inset of Fig.~\ref{fig:tip_characterization}\emph{F}.

\vspace{.1in}\noindent
\textbf{DFT calculations.} 
To assist in the identification of defect species, we use DFT calculations to compute
the stabilities and defect levels for point-defect  candidates in model Al$_2$O$_3$ structures. 
The defects considered here include native defects and impurities such as carbon and hydrogen-containing complexes, with defect formation energies calculated using a standard supercell approach with 120-atom supercells for corundum Al$_2$O$_3$~\cite{RevModPhys.86.253}.
All simulations are performed using the HSE06 screened hybrid functional, a plane-wave energy cutoff of 400 eV, and a Brillouin zone sampling with a 2$\times$2$\times$2 Monkhorst-Pack $k$-point density mesh, using VASP version 5.4.~\cite{Heyd2003, KRESSE199615, Marsman_2008}.
A fraction of 32\% exact exchange is included in the hybrid functional, giving a band gap of $E_g$=8.82 eV and the optimized lattice constants $a$=4.74~\AA~ and $c$=12.93~\AA~for the corundum structure~\cite{Heyd2003}.
Defect formation energies for charged systems are corrected for finite-size effects using the methodology of Freysoldt, Neugebauer and Van de Walle, using a static dielectric constant of $\epsilon=9.35\epsilon_0$ and 11.4$\epsilon_0$ for values perpendicular and parallel to the $c$-axis, respectively~\cite{RevModPhys.86.253, PhysRevLett.102.016402}.

We have aligned the measured transition energies to the DFT-calculated defect transition energies by referencing both to the Al$_2$O$_3$ valence band maximum (VBM). This is a choice that requires no addition calculations for the DFT defect energies. To align the measured transition energies to the Al$_2$O$_3$ VBM we have computed the energy difference between the vacuum level and the Al$_2$O$_3$ VBM via two calculations.
First, we perform a bulk Al$_2$O$_3$ calculation to establish the value of difference between the Al$_2$O$_3$ VBM and the Al$_2$O$_3$ average electrostatic potential.
Second, we perform an Al$_2$O$_3$ slab calculation to obtain the difference between the Al$_2$O$_3$ average electrostatic potential and the vacuum potential. 
The slab calculation assumes four layers of Al$_2$O$_3$ terminated by OH and 12~\AA \ of vacuum.

\vspace{.1in}\noindent
\textbf{ACKNOWLEDGMENTS.} 
This research was supported by grant \#DE-SC0020313, funded by the U.S. Department of Energy, Office of Science.
The portion of this work performed at Lawrence Livermore National Laboratories (LLNL) was completed under Contract DE-AC52-07NA27344.
S.N.C.\ acknowledges useful conversations with D.~L.~Blank.

\bibliography{DefectIdentificationEFM}


\onecolumngrid
\clearpage

\beginsupplement

\section*{SI Appendices for ``A scanning probe microscopy approach for identifying defects in aluminum oxide''}

In these Appendices, we provide details for the methods and calculations described in the main text.

\section{Determining the absolute separation $h_\text{ts}$ between the oxide sample and the EFM tip}
\label{sec:tip-sample_separation}

\begin{figure}[b]
    \centering
\includegraphics[width=0.55\textwidth]{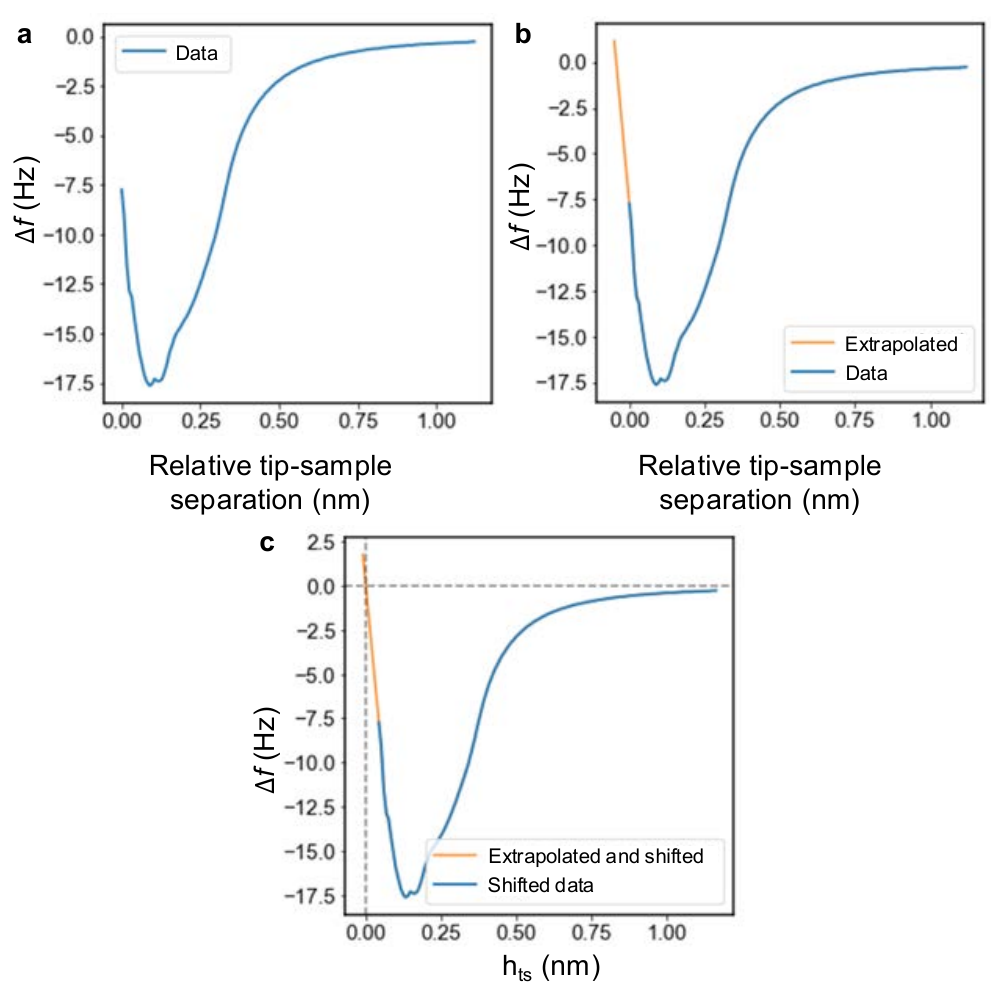}
    \caption{
    Calibrating the tip-sample interaction curve.
    (a) Shifts in the EFM resonant frequency $\Delta f$ are measured while maintaining the CPD condition, as a function of tip height.
    For small oscillation amplitudes, this curve provides a good estimate for the tip-sample interaction; however, the absolute value of the tip-sample separation $h_\text{ts}$ remains to be calibrated.
    (b) Due to the initial steepness of the interaction curve, its extrapolated crossing with the $\Delta f=0$ line should occur very near $h_\text{ts}=0$, providing the desired calibration.
    (c) Calibration is achieved by shifting the data horizontally, so the extrapolated curve passes through the origin.}
\label{fig:lennard_jones_curve_method1}
\end{figure}

The tip-sample separation $h_\text{ts}$ is a key parameter in all EFM experiments.
Unfortunately, without crashing the tip into the sample (an undesirable occurrence), it is not possible to directly determine this parameter.
Here, we describe an alternative method for determining $h_\text{ts}$.

We make use of the fact that tip-sample interactions are strongly repulsive in the absence of electrostatic effects~\cite{SaderJarvis}
We therefore perform EFM measurements as a function of tip height, while tuning the backgate to its Contact Potential Difference (CPD) bias, causing the tip-sample interactions to vanish.
The resonant frequency shift of the tip $\Delta f$ is then measured as a function of tip height, while maintaining the CPD condition, yielding the results shown in Fig.~\ref{fig:lennard_jones_curve_method1}(a).
At this point, we do not know the absolute tip-sample separation, only the relative separation.
In the limit of small-amplitude oscillations, as appropriate for this experiment, the frequency shift $\Delta f$ is proportional to the tip-sample force gradient $\partial f_\text{ts}/\partial z$.
Figure~\ref{fig:lennard_jones_curve_method1}(a) therefore provides a direct measurement of the tip-sample interaction, showing behavior consistent with a Lennard-Jones potential: a steep short-range repulsive interaction crossing over to a gradually decreasing attractive interaction at larger distances.
While the expected behavior of the Lennard-Jones potential diverges for small tip-sample separations, extrapolating the curve to $\Delta f=0$, as shown in Fig.~\ref{fig:lennard_jones_curve_method1}(b), gives a good approximation for $h_\text{ts}$~\cite{SaderJarvis}, with small errors on the order of $\Delta h_\text{ts}\approx 0.1$~nm.
A simple horizontal shift of the data, as shown in Fig.~\ref{fig:lennard_jones_curve_method1}(c), therefore provides the desired calibration of the interaction curve.

\begin{figure}[t]
    \centering
\includegraphics[width=0.3\textwidth]{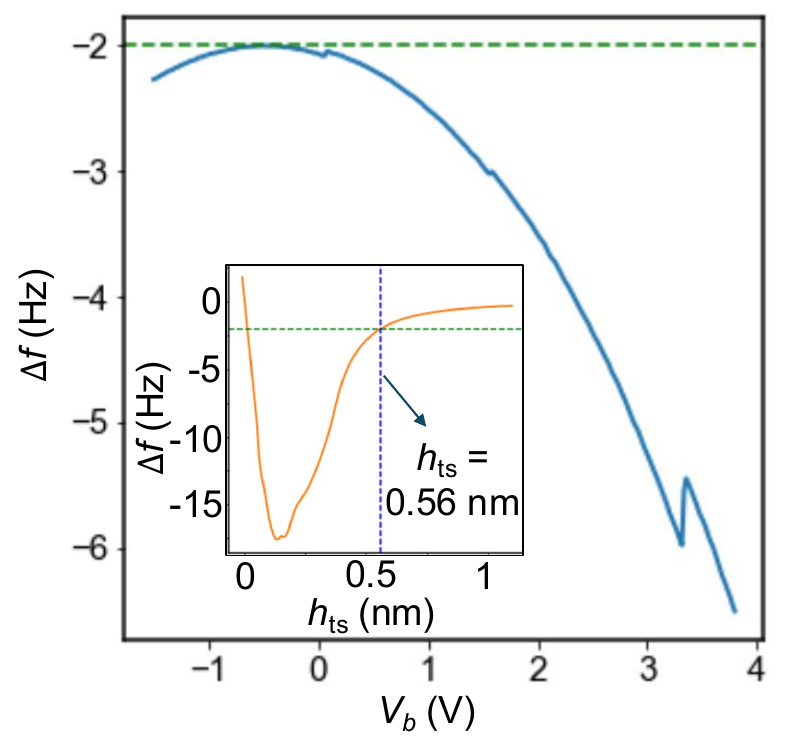}
    \caption{
    Determining $h_\text{ts}$ for an arbitrary tip height, using the calibrated $\Delta f$-$h_\text{ts}$ curve from Fig.\ref{fig:lennard_jones_curve_method1}(c) (also reproduced in the inset, here).
    The main panel shows an $f$-$V$ voltage sweep obtained at this tip setting.
    The $\Delta f$ value at the CPD condition is given by the dashed green line.
    The inset shows the same CPD condition (horizontal dashed green line), and its crossing point on the calibrated interaction curve (vertical dashed blue curve), indicating $h_\text{ts}$ for these experimental conditions.}
\label{fig:lennard_jones_curve_method2}
\end{figure}

Based on Fig.~\ref{fig:lennard_jones_curve_method1}(c), we may now estimate $h_\text{ts}$ for any experiment, as follows.
For a given EFM tuning, we identify the CPD condition in an $f$-$V$ curve, as indicated by the green dashed line in the main panel of Fig.~\ref{fig:lennard_jones_curve_method2}.
Since typical EFM measurements are performed at larger tip-sample separations $h_\text{ts}\gtrsim 0.5$~nm, we focus in the right-hand side of the calibrated interaction curve, which is reproduced in the inset of Fig.~\ref{fig:lennard_jones_curve_method2}.
We can then read $h_\text{ts}$ directly off of the curve: 
here, the horizontal dashed green line represents the measured CPD value and the vertical dashed blue line represents the estimated $h_\text{ts}$ value for this EFM setting.

\section{Characterizing the tip geometry}
\label{sec:Tip_Geometry}

\begin{figure}[!htb]
    \centering
    \includegraphics[width=0.9\textwidth]{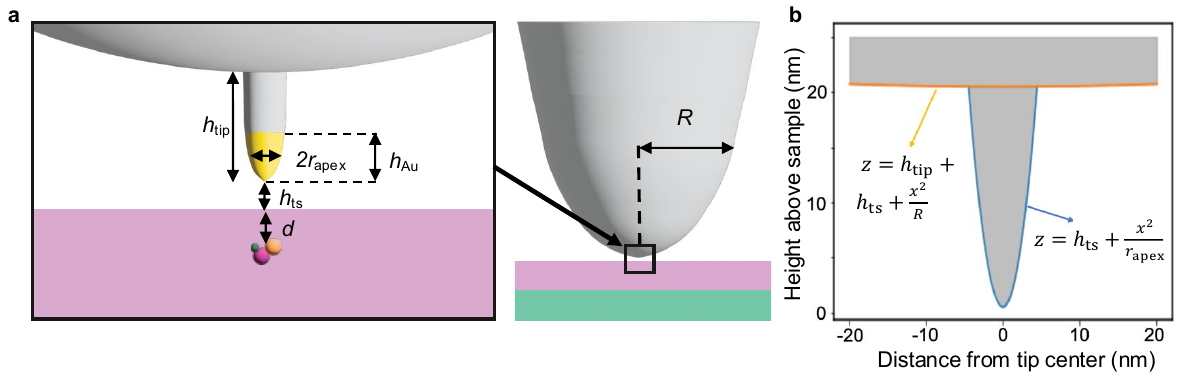}
    \caption{
    Two-section model for the EFM tip.
    (a) A wide view (right) and a blown-up view (left) of the tip, showing the various model parameters. 
    (Reproduced from Fig.~\ref{fig:tip_shape_fitting}\emph{B} of the main text.)
    (b) The parabolic geometry parameters used to model the tip shape in the simulations of Sec.~\ref{sec:Tip_Geometry} and the rest of this paper.
    Here, $R$ is the radius of the bulk or ``big'' tip and $r_\text{apex}$ is the radius of the fine tip.}
    \label{fig:parabolic_tippy_tip}
\end{figure}

As noted in Fig.~\ref{fig:tip_shape_fitting}\emph{B} of the main text, we model our EFM tip in two sections, with a fine tip attached to a bulk or ``big'' tip, as illustrated again in Fig.~\ref{fig:parabolic_tippy_tip}. 
The various parameters of the tip are labeled in Fig.~\ref{fig:parabolic_tippy_tip}(a) and the parabolic parameters used in the simulations are indicated in Fig.~\ref{fig:parabolic_tippy_tip}(b), similar to the model described in Ref.~\cite{Behn2021}. 
To perform accurate simulations of the full system, including the sample, we need to first characterize the tip parameters, including the big tip radius \textit{R}, the fine tip height $h_\text{tip}$, the fine tip radius $r_{\text{apex}}$, and the height of gold coating on the fine tip $h_\text{Au}$. 
Our first set of EFM measurements, described below, is performed on a gold sample.
This is practical and convenient, because gold's work function is well known, which simplifies our procedure.
Additional calibration steps are then performed on an oxide sample containing a defect, as explained below.
All simulations of the tip-sample system are performed using COMSOL software, assuming a 2D axisymmetric geometry for the tip.
The key physical quantity allowing us to make contact between the experiments and the simulations is the tip-sample capacitance $C$, and specifically, the second derivative of $C$ with respect to the tip-sample separation.

\begin{figure}[b]
    \centering
    \includegraphics[width=0.6\textwidth]{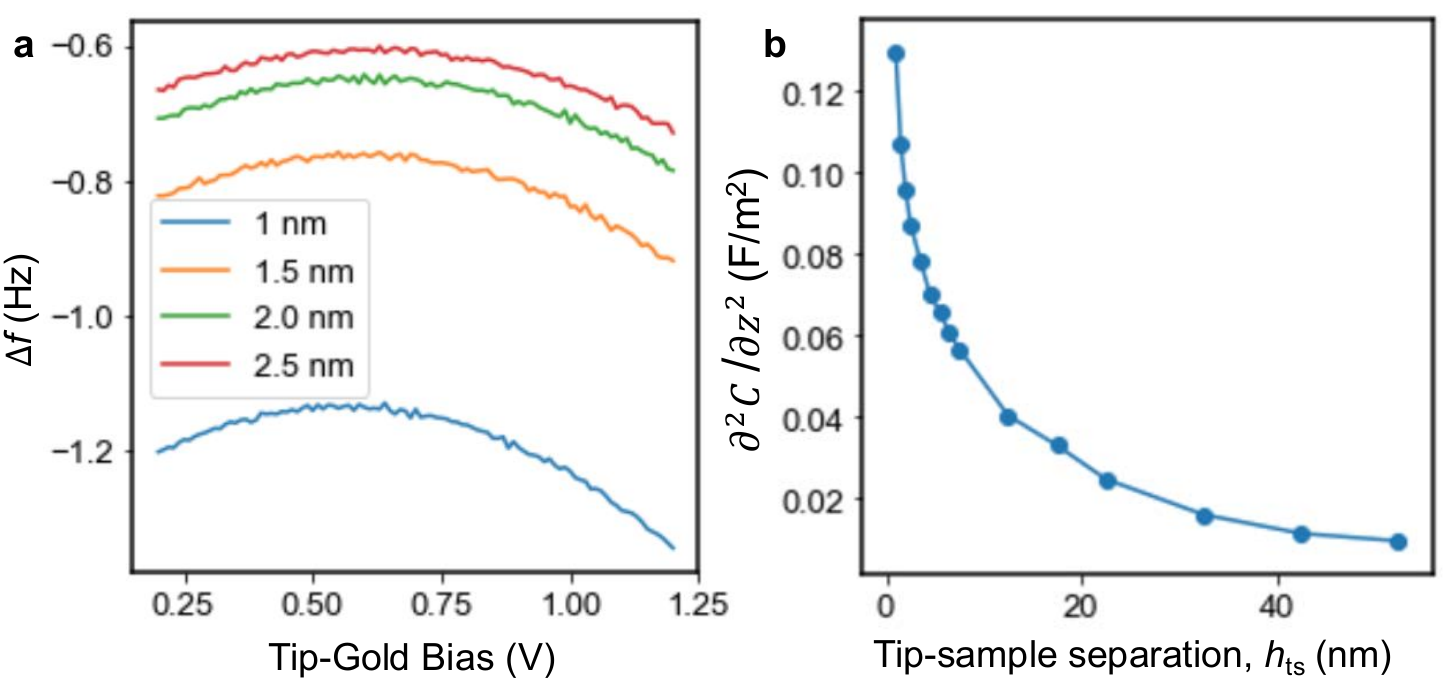}
    \caption{
    Determining $\partial^2C/\partial z^2$ from experiments.
    (a) Experimental measurements of variations of the tip resonance frequency $\Delta f$ as a function of the voltage bias $V_b$ between the tip and a gold sample for four different tip-sample separations $h_\text{ts}$, as indicated.
    These frequency variations are converted to capacitance curvature $\partial^2C/\partial z^2$ using Eq.~(\ref{gold_frequency_shift_capacitance}).
    (b) Results for $\partial^2C/\partial z^2$ are shown as a function of $h_\text{ts}$.}
    \label{fig:kpfm_curve_second_deriv_capacitances}
\end{figure}

\subsection{Determining $\partial^2C/\partial z^2$ from experiments and using it to determine work functions}
\label{AuMeasurements:simulations}

The second derivative of the tip-sample capacitance can be related to variations of the tip resonance frequency, $\Delta f$ through the equation~\cite{neff_insights_2015}
\begin{equation}
\Delta f = -\frac{f_{0}}{2 {k}_{0} }\frac{\partial F_{\text{el}} }{\partial z} = -\frac{f_{0}}{4{k}_{0}}\frac{\partial^2 C}{\partial z^{2} } V_{\text{eff}}^{2} .
\label{gold_frequency_shift_capacitance}
\end{equation}
Here, $F_{\text{el}}$ is the tip-sample electrostatic force, $f_{0}$ = 2865 Hz is the free resonant frequency of the oscillating tip (i.e., far away from the sample), $k_{0}$ = 1800 Nm$^{-1}$ is the spring constant of the oscillating tip, $C$ is the tip-sample capacitance in the absence of any charges inside the sample (motivating our use of a gold sample, since this does not contain any charged defects), and $z$ is the tip height.
For simple tip geometries, $V_{\text{eff}}=V_{b}-\Delta W/e$, where $V_{b}$ is the bias between the tip and the sample, and $\Delta W$ is their work function difference. 
We therefore expect the frequency variations of the tip to show a quadratic dependence on the tip-sample bias, $V_{b}$.
This is indeed observed in our experimental measurements, as seen in Fig.~\ref{fig:kpfm_curve_second_deriv_capacitances}(a), for several tip-sample separations $h_\text{ts}$.
Here, $\partial^2 C/\partial z^2$ characterizes the curvature of the $f$-$V$ curves, and has a strong dependence on tip-sample separation, as seen in Fig.~\ref{fig:kpfm_curve_second_deriv_capacitances}(b).

In our experiment, the work function of the tip is more complicated because the bulk portion of the tip is formed of tungsten, while the fine portion of the tip has a coating of gold at the end of the tip.
As noted above, the gold portion of the tip has a well-characterized work function with a value of 5.30~eV~\cite{sachtler1966work}.
To characterize the full tip, we consider the $V_b$ scans shown in Fig.~\ref{fig:kpfm_curve_second_deriv_capacitances}(a).
As the tip-sample separation increases, the effective (total) tip work function changes, approaching the value for the bulk tungsten portion of the tip for the largest tip-sample separations.
To characterize this changing behavior, we analyze here the CPD condition ($\partial \Delta f/\partial V_b=0$), which describes the tip-sample bias that exactly offsets the work function difference, causing tip-sample interactions to vanish.
In Fig.~\ref{fig:kpfm_curve_second_deriv_capacitances}(a), this condition is met at the maxima of the individual parabolas.
Importantly, we see that the CPD bias $V_b=V_\text{CPD}$ changes as a function of tip-sample separation.
To extract the CPD values for each curve, we fit the experimental data to an inverted parabola of the form $\Delta f=aV_b^2 +bV_b + c$. 
The CPD bias is then given by $V_b=V_\text{CPD}=-b/2a$. 

\begin{figure}[b]
    \centering
    \includegraphics[width=0.3\textwidth]{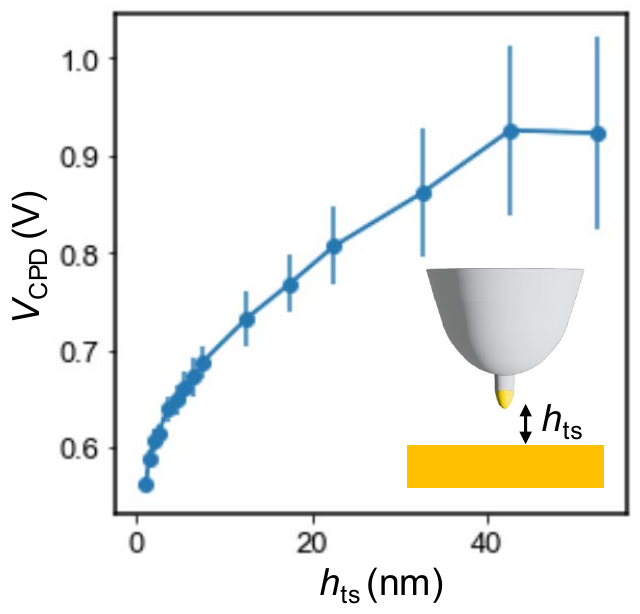}
    \caption{
    Experimental determination of the CPD voltage bias $V_\text{CPD}$ as a function of the tip-sample separation.
    Spatial variations arise in part from the fact that the tip is inhomogeneous, containing both gold and tungsten materials.
    Tip-sample interactions are dominated by the big (tungsten) portion of the tip for large $h_\text{ts}$ values, yielding the estimate for the tungsten work function used in this work: $W_\text{W}=4.38 \pm 0.10 $ eV.}
    \label{fig:cpd_variations_gold}
\end{figure}

In the limit where the tip is far from the sample, we expect that its work function will be dominated by the bulk tungsten rather than the small amount of gold at the end of the fine tip.
In our experiments, the largest tip-sample separation is given by 52.5~nm, as calibrated in Sec.~\ref{sec:tip-sample_separation} above.
At this separation we obtain $V_{\text{CPD}} = 0.92 \pm 0.10$~V, as shown in Fig.~\ref{fig:cpd_variations_gold}.
Taking into account the work function of  the gold sample, the relation $\Delta W=eV_\text{CPD}$ then predicts a work function for the tungsten tip of $4.38\pm 0.10$~eV.
This should be compared to the quoted value of 4.55~eV~\cite{crc_handbook}, which is not as well known as gold.
In our case, the small amount of gold on the EFM tip contributes some uncertainty to our estimate.
Additionally, at the largest tip-sample separations applied in our experiments, the  asymptotic behavior of the CPD values may not be fully realized, as apparent in Fig.~\ref{fig:cpd_variations_gold}.
However, this represents our best estimate for the tungsten work function, and we adopt the tungsten work function value of $W_\text{W}=4.38 \pm 0.10 $ eV in the remainder of this work.
Note that the uncertainty of the parabolic fits at different tip-sample separations were used to estimate the uncertainty of the CPD.
From the discussion of the parabolic fits, we therefore define the CPD error, arising from the fitting procedure, to be $\Delta V_\text{CPD} = \left| \frac{b}{2a}\right| \sqrt{ (\frac{\Delta a}{a})^{2} + (\frac{\Delta b}{b})^{2} }$.
Here, we have combined the separate error contributions in quadrature, and we have also included variations in the fitting results obtained from forward and backward voltage sweeps. 

For measurements performed on our oxide sample, we also need to account for the work function of the molybdenum backgate, especially since the reported values for molybdenum are somewhat variable or uncertain, ranging in the literature from 4.36 to 4.95~eV~\cite{crc_handbook}.
In our experiments, we also face uncertainties arising from the fact that the presence of charged defects in the oxide affects the accuracy of Eq.~(\ref{gold_frequency_shift_capacitance}).
To address this issue, we assume that the charge state of the defects at the CPD condition is neutral, on average, despite displaying local variations.
(Also, see the discussion in Sec.~\ref{sec:localcharge}, below.)
We therefore map out $V_\text{CPD}$ across a scan region of $25\times 25$~nm$^2$, using the parabolic fitting method described above, to obtain the average value of $\overline{V_\text{CPD}}=-0.25$~V.
Using this result, we then obtain a work function value of $W_\text{Mo}=4.63$~eV for our molybdenum backgate.

\subsection{Determining $\partial^2C/\partial z^2$ from simulations}
\label{simulation_second_derivative}

The tip parameters discussed in the following subsections are determined by comparing experimental and simulation results for $\partial^2C/\partial z^2$, where the latter are calculated for the model geometry of Fig.~\ref{fig:parabolic_tippy_tip}(b).
The experimental methods for finding $\partial^2C/\partial z^2$ are described in the previous subsection.
For the simulations, we employ a standard technique in which the tip-sample voltage bias is set to $V_b=1$~V, and Laplace's equation is solved numerically to determine the induced charge density on the gold sample surface.
Integrating this density to obtain the total charge $Q$ give the capacitance from the equation $C=Q/V_b$.
All simulations here include the full tip geometry and appropriate work functions.
The simulations are repeated as a function of tip height $z$, giving $C(z)$, and the results are fitted to a 5$^\text{th}$-degree B-spline, from which we compute the second derivative to finally give $\partial^2C/\partial z^2$.
The system size and the discrete step size $\Delta z$ are both checked for convergence.

\subsection{Determining the big tip radius $R$ and the fine tip height $h_\text{tip}$}
\label{sec:R_and_htip}

\begin{figure}[t]
    \centering
    \includegraphics[width=0.3\textwidth]{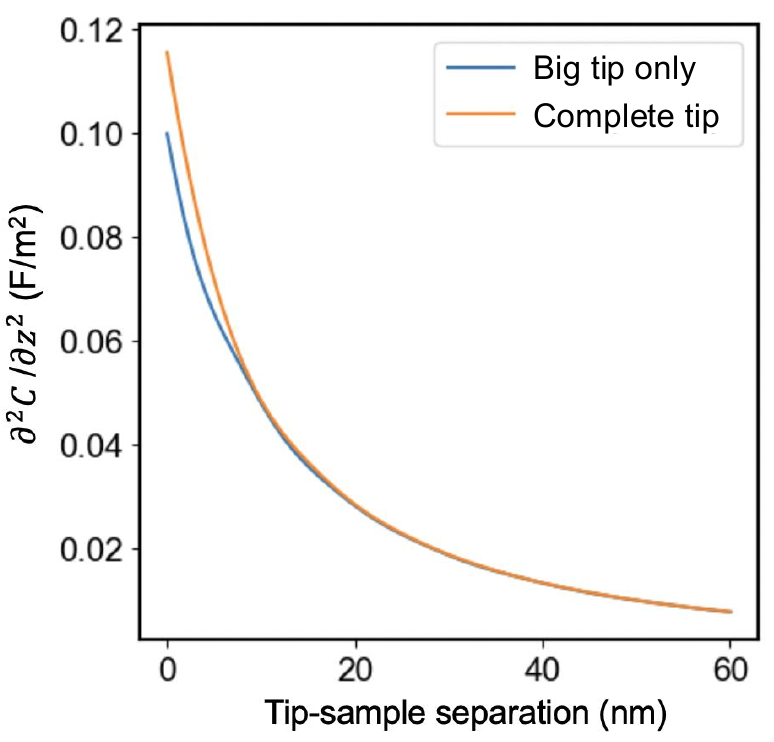}
    \caption{
    Simulation results for the capacitance curvature vs tip-sample separation for two different tip geometries.
    For the ``complete tip,'' we consider the two-section geometry illustrated in Fig.~\ref{fig:parabolic_tippy_tip}(b). 
    Here we use the parameters $R=1800$~nm, $h_\text{tip}=22.8$~nm, $h_\text{Au}=0.5$~nm, and $r_\text{apex}=1.5$~nm. 
    We define the tip-sample separation to be $h_\text{ts}=0$ when the tip touches the gold sample.
    The ``big tip'' simulation follows the same procedure, but with the fine-tip removed from the model geometry.
    The two simulations agree well for $h\gtrsim 10$~nm, with residual errors on the order of 1-2\% when the Complete tip is simply replaced by the Big tip.}
    \label{fig:SeparationBigTipWholeTip}
\end{figure}

\begin{figure}[b]
    \centering
    \includegraphics[width=0.6\textwidth]{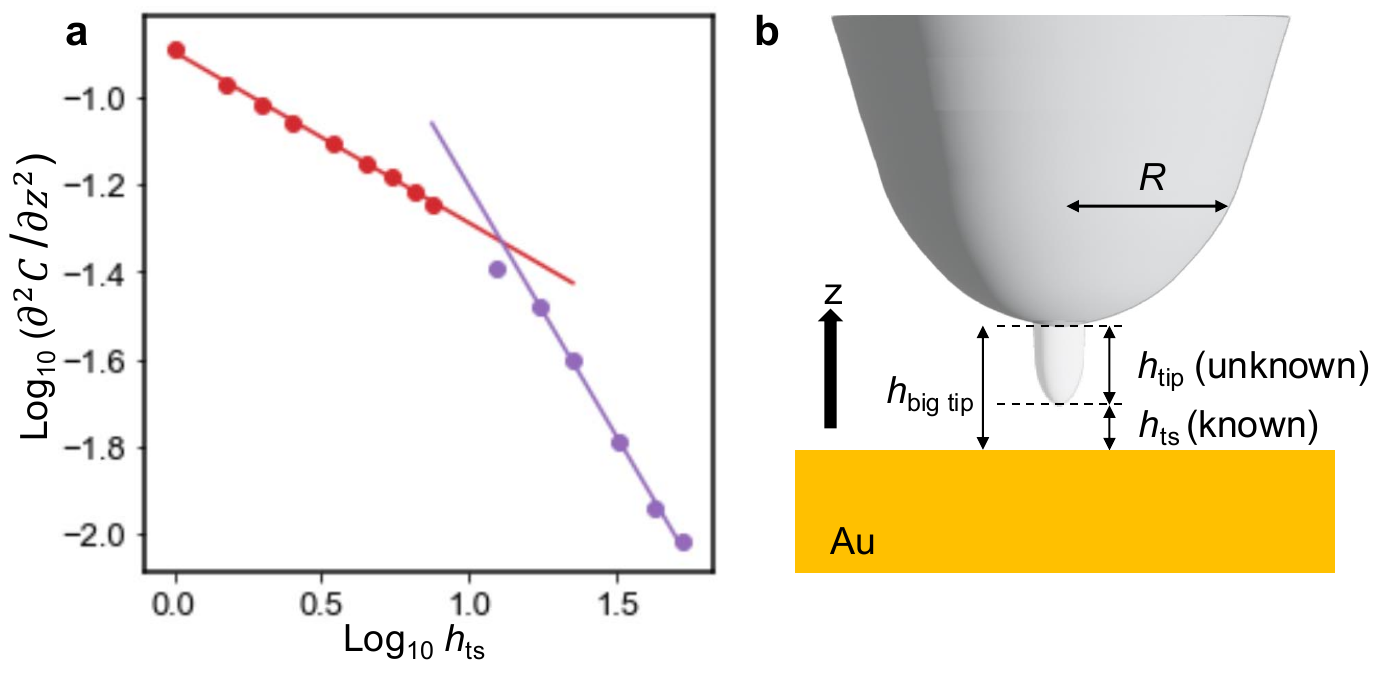}
    \caption{
    Log-log plot of the experimentally measured capacitance curvature vs tip-sample separation above a gold sample.
    (a) The data indicate two distinct types of behavior reflected in different power laws, where the red data are affected by interactions between the sample and the full tip, while the purple data are dominated by interactions between the sample and the big tip only.
    The crossover between these two regimes occurs at $h_\text{ts}\approx 12.5$~nm.
    (b) Illustration of the system parameters relevant for analyzing the purple data in (a).
    The fine tip length $h_\text{tip}$ is originally unknown and must be determined from the analysis.}
    \label{fig:loglog_capacitanceDeriv}
\end{figure}

\begin{figure}[t]
    \centering
    \includegraphics[width=0.6\textwidth]{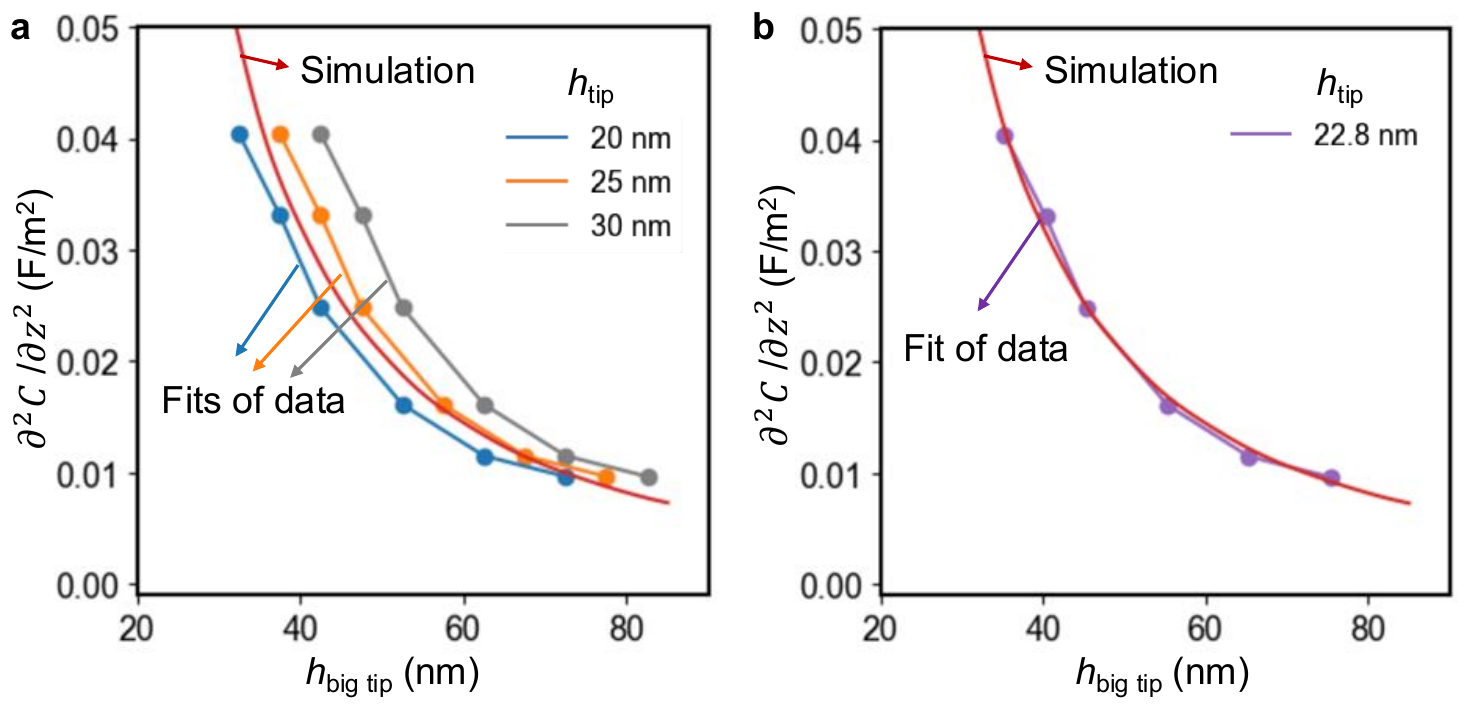}
    \caption{
    Fitting procedure to determine the best value of $h_\text{tip}$ for a given $R$.
    (a) Experimental measurements of the capacitance curvature as a function of the separation $h_\text{big tip}$ between the big tip and the sample, illustrated in Fig.~\ref{fig:loglog_capacitanceDeriv}(b), for several values of the unknown parameter $h_\text{tip}$ (blue, orange, gray curves).
    Simulation results are also shown, assuming a big-tip radius of $R=1800$~nm (red curve, ``Simulation'').
    (b) For this value of $R$, the best match between simulations and experiments is given by $h_\text{tip}=22.8$~nm.}
    \label{fig:optimizingSiloHeight}
\end{figure}

\begin{figure}[b]
    \centering
    \includegraphics[width=0.75\textwidth]{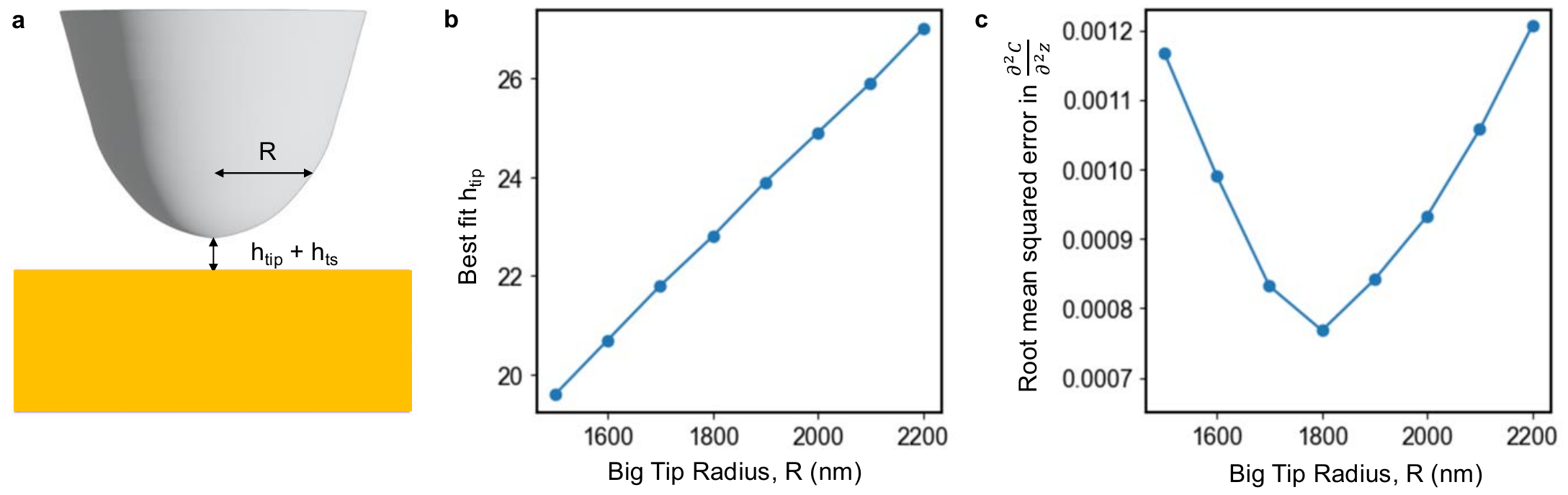}
    \caption{
    Simultaneous fitting procedure for the tip parameters $R$ and $h_\text{tip}$.
    (a) The simulation geometry.
    [See also Fig.~\ref{fig:loglog_capacitanceDeriv}(b).]
    Note that $h_\text{ts}$ was determined in Sec.~\ref{sec:tip-sample_separation}.
    (b) Best-fit results for $h_\text{tip}$, for a range of $R$ values, following the procedure described in Fig.~\ref{fig:optimizingSiloHeight}(b).
    (c) RMS deviations between the simulationaal and experimental data is shown for each of the points in (b).
    The optimal value, $R=1800$~nm, is obtained at the minimum RMS value. 
    We define the uncertainty of this calculation to be the RMS threshold occurring 25\% above the minimum value: $R=(1800\pm 200)$~nm.
    The corresponding error in $h_\text{tip}$ is determined from the slope in (b): $h_\text{tip}=(22.8\pm2.1)$~nm. }
    \label{fig:rms_err_bigTipRad}
\end{figure}

Figure~\ref{fig:SeparationBigTipWholeTip} shows typical simulation results for $\partial^2C/\partial z^2$ as a function of $h_\text{ts}$.
Here, the orange curve corresponds to the ``complete'' solution (big tip and fine tip), while the blue curve shows the solution where we include just the big tip, but we exclude the fine tip. 
We see that the two solutions converge for large enough $h_\text{ts}$ values, as expected, since the bulk tip dominates over the much smaller fine tip in this limit.
It is clear that the big tip and fine tip offer separate contributions to the total tip-sample capacitance.
We can observe similar behavior in the experimental data in Fig.~\ref{fig:loglog_capacitanceDeriv}(a), where the results are plotted on a logarithmic scale.
Here, two distinct power-law behaviors are observed, with the red data dominated by the fine tip and the purple data dominated by the big tip.
A well-demarcated crossover occurs at $h_\text{ts}\approx 12.5$~nm.

Since the purple data are dominated by the big tip, we can use this subset of data to characterize the big-tip radius $R$ and the separation $h_\text{big tip}$ between the sample and the big tip, as illustrated in Fig.~\ref{fig:loglog_capacitanceDeriv}(b).
For simplicity, we remove the fine tip and perform simulations of just the big tip.
Typical simulation results are plotted in \ref{fig:optimizingSiloHeight}(a) as a function of $h_\text{big tip}$ (red curve), for a fixed value of $R=1800$~nm.
We also replot the purple experimental data of Fig.~\ref{fig:loglog_capacitanceDeriv}(a) here, although we do not know the absolute value of $h_\text{big tip}$ yet.
(Recall that we calibrated the tip-sample separation $h_\text{ts}$ in Sec.~\ref{sec:tip-sample_separation}; however, we have not yet determined the length of the fine tip, $h_\text{tip}=h_\text{big tip}-h_\text{ts}$.)
Thus, determining the unknown parameter $h_\text{tip}$ would also allow us to calibrate the absolute value of $h_\text{big tip}$.
For example, the three values of $h_\text{tip}$ considered in Fig.~\ref{fig:optimizingSiloHeight}(a) cause horizontal shifts in the data as shown.
The optimal value value of $h_\text{tip}$ should be determined by matching the experimental and simulation results, as in Fig.~\ref{fig:optimizingSiloHeight}(b).

\begin{figure}[t]
    \centering
    \includegraphics[width=0.65\textwidth]{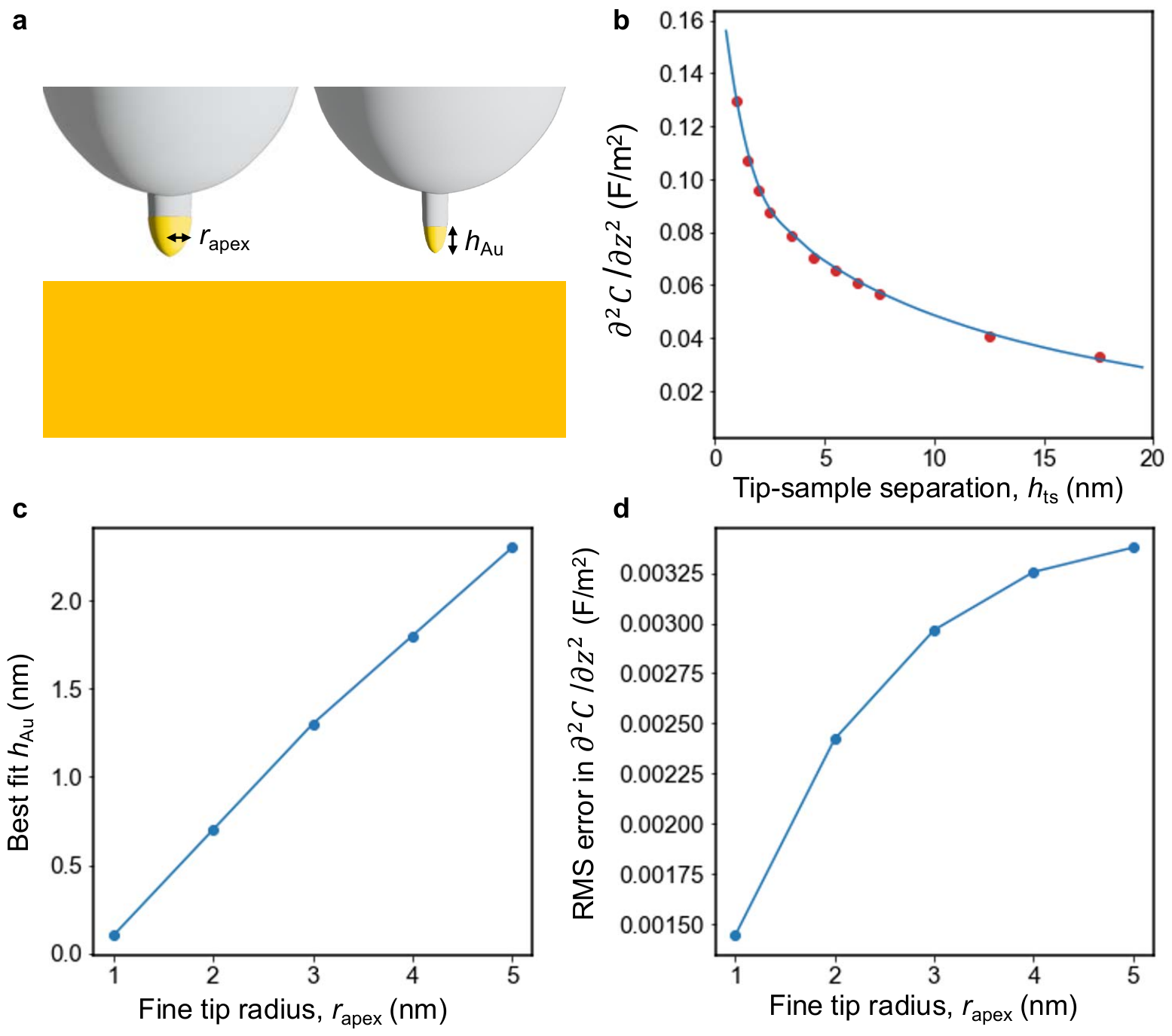}
    \caption{
    Simultaneous fitting procedure for the fine-tip parameters $r_\text{apex}$ and $h_\text{Au}$, illustrated in (a).
    (b) Setting $r_\text{apex}=1$~nm, $h_\text{Au}$ is taken as a fitting parameter in the simulations of the capacitance curvature as a function fo $h_\text{ts}$ (blue curve) to obtain the best match to the experimental data (red points).
    (c) This fitting procedure is repeated for different $r_\text{apex}$ values.
    (d) The RMS errors of the fits are shown for each point in (c).
    These results point towards an optimal value of $r_\text{apex}<1$~nm, which appears unphysical, suggesting that we develop an alternative fitting procedure.
    }
    \label{fig:bestFits_goldht_rapex}
\end{figure}

\begin{figure}[t]
    \centering
    \includegraphics[width=0.55\textwidth]{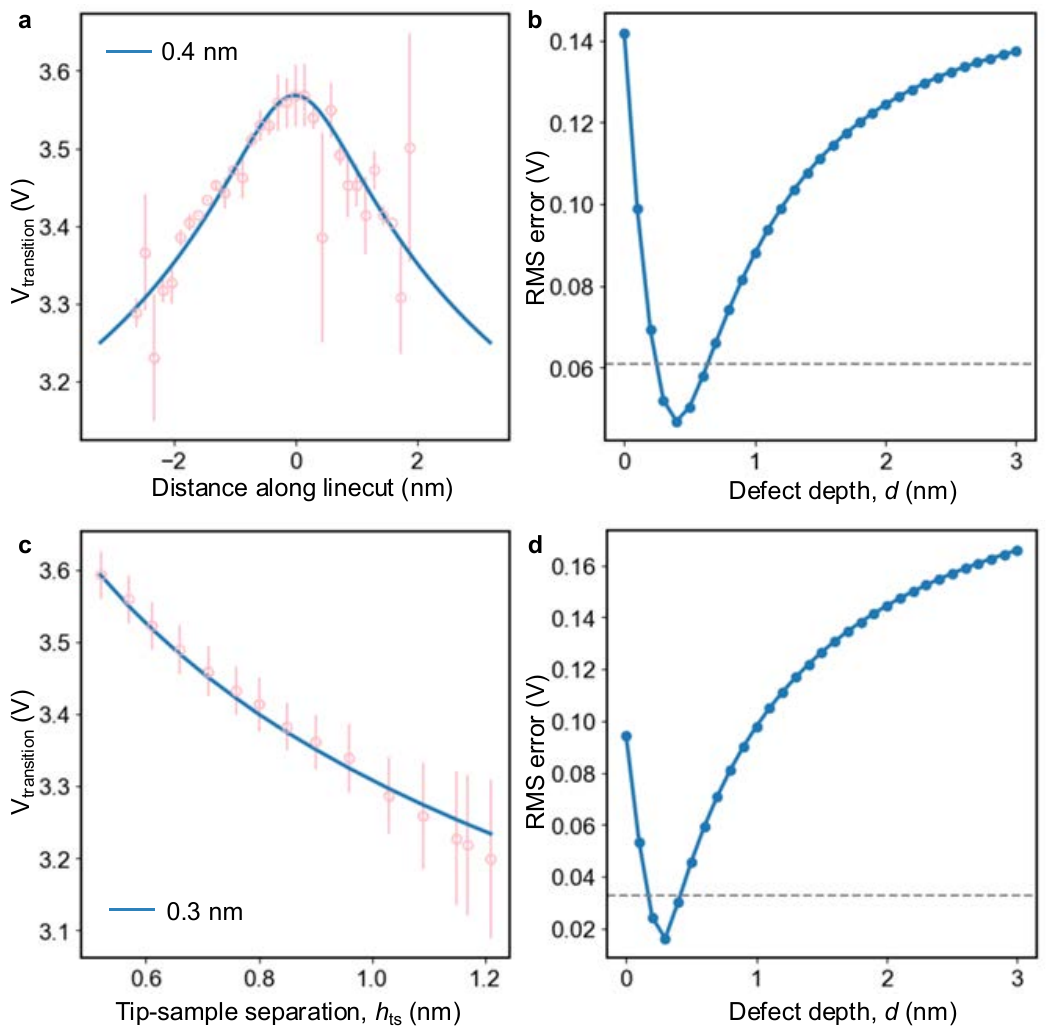}
    \caption{
    Fitting procedure for identifying optimized triplets of parameters $(d,h_\text{Au},r_\text{apex})$, given the pre-optimized pairs of parameters $(h_\text{Au},r_\text{apex})$ obtained in Fig.~\ref{fig:bestFits_goldht_rapex}(c).
    (a) Charging transition voltages obtained along a lateral EFM scan directly over a defect, as a function of position (red circles).
    Simulation-based estimates for $V_\text{transition}$ for the same tip locations, following the procedure described in the main text, for a given set of parameters $(d,h_\text{Au},r_\text{apex})$.
    RMS values between the experimental data and the simulations are minimized to determine the optimal value of $d$ corresponding to the values of $(h_\text{Au},r_\text{apex})$ used in the simulations.
    (b) This optimization procedure is repeated for different values of $d$, yielding the RMS values shown here.
    The dashed horizontal line represents about 10\% of the maximum RMS values, measured from the minimum RMS value.
    The results in (a) are obtained for $d=0.4$~nm, with a corresponding uncertainty of 0.2~nm.
    (c), (d) Here, we show the same type of results as (a) and (b) for the case of a vertical EFM scan, where the minimum tip-sample separation corresponds to the tip location at the center of the lateral scan in (a), which is directly above the defect.
    The results in (c) are obtained for $d=0.3$~nm, with a corresponding uncertainty of 0.1~nm. }
    \label{fig:defect5_lateralFit}
\end{figure}

\begin{figure}[t]
    \centering
    \includegraphics[width=0.3\textwidth]{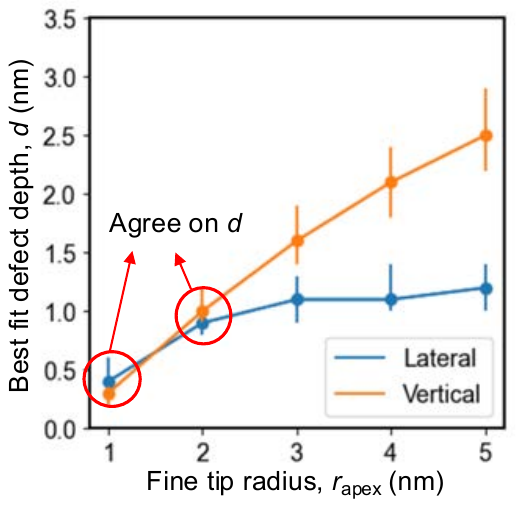}
    \caption{Joint fits for the defect depth $d$, obtained from the lateral and vertical scans described in the main text and in Fig.~\ref{fig:defect5_lateralFit}.
    The error bars shown here are determined in Figs.~\ref{fig:defect5_lateralFit}(b) and (d).
    Since the scans are performed above the same defect, self-consistency requires that they should give the same answer, yielding the optimal values indicated by red circles.
    We thus obtain best fit results given by $r_\text{apex}=(1.5\pm 0.5)$~nm and $d=(0.65\pm 0.30)$~nm.}
    \label{fig:defect5_jointFit}
\end{figure}

While the parameter $h_\text{tip}$ mainly affects the horizontal shift of the data in Fig.~\ref{fig:optimizingSiloHeight}(a), the shape of the curve is more strongly affected by the remaining big-tip parameter $R$.
We therefore perform a simultaneous fitting of the parameters $h_\text{tip}$ and $R$.
Specifically, we repeat the fitting procedure demonsstrated in Fig.~\ref{fig:optimizingSiloHeight}(b) for many different $R$ values, obtaining the results shown in Fig.~\ref{fig:rms_err_bigTipRad} for the optimal correspondence between $h_\text{tip}$ and $R$.
As noted above, the main point of comparison between simulations and experiment is through the capacitance curvature.
Hence, the quality of each point in Fig.~\ref{fig:rms_err_bigTipRad}(b) can be quantified by the root-mean-square (RMS) comparison of the experimental and simulational capacitance curvatures, giving the results shown in Fig.~\ref{fig:rms_err_bigTipRad}(c). 
Choosing the value with minimum error, we obtain the fully optimized results $h_\text{tip}=(22.8\pm 2.1)$~nm and $R=(1800\pm 200)$~nm.
Here, we have defined the uncertainty for $R$ as the threshold value in Fig.~\ref{fig:rms_err_bigTipRad}(c) for which the RMS value is 25\% above its minimum value.
We then convert this to an uncertainty in $h_\text{tip}$ via the slope shown in Fig.~\ref{fig:rms_err_bigTipRad}(b).

\subsection{Determining the fine tip radius $r_\text{apex}$ and the height of the gold coating on the tip $h_\text{Au}$}
\label{sec:rapex_and_hAu}

To determine $r_\text{apex}$ and $h_\text{Au}$, we consider the red data subset in Fig.~\ref{fig:loglog_capacitanceDeriv}, and the two-section parabolic tip model shown in 
Figs.~\ref{fig:bestFits_goldht_rapex}(a).
We first try to perform simultaneous fits for these two parameters by comparing simulations and experiments, similar to the approach used in the previous subsection.
For example, Fig.~\ref{fig:bestFits_goldht_rapex}(b) shows experimentally determined capacitance curvature (red points), compared to simulation results (blue curve) obtained for a fixed value of $r_\text{apex}$, with $h_\text{Au}$ taken as a fitting parameter. 
This procedure is repeated for different values of $r_\text{apex}$ in Fig.~\ref{fig:bestFits_goldht_rapex}(c), with corresponding RMS estimates shown in Fig.~\ref{fig:bestFits_goldht_rapex}(d).
Here we see that the best result is obtained at the boundary value $r_\text{apex}=1$~nm, suggesting that the actual optimum occurs at even smaller values of $r_\text{apex}$, with a corresponding value of $h_\text{Au}\rightarrow 0$.
Unfortunately, simulations with vanishing $r_\text{apex}$ are not physically reasonable, since we know the tip has some amount of gold, suggesting some type of ambiguity in our simulations, in this limit.
We therefore explore an alternative simulation approach, to supplement the results shown in Fig.~\ref{fig:bestFits_goldht_rapex}.

A partial explanation for the challenges encountered in Fig.~\ref{fig:bestFits_goldht_rapex} is that the parameters $r_\text{apex}$ and $h_\text{Au}$ have a similar effect on $\partial^2C/\partial z^2$ when they are varied.
The gold sample in this experiment has a smooth, uniform surface that exacerbates the problem.
We therefore consider a different sample -- the oxide-on-Si sample that forms the focus of this work -- because it exhibits lateral inhomogeneity, due to localized charged defects.
The most distinctive property of a defect, which we study in this work, is its charging transition --- specifically, the tip-backgate bias $V_\text{transition}$ that induces the transition.
However, adding defects to the analysis also means we need to include another new parameter in our simulations: the depth $d$ of the defect below the surface of the oxide.
We now describe the new fitting procedure.

From our experimental measurements, we choose a single defect with a well-defined charging transition that can be tracked as the EFM tip is scanned both vertically and laterally across the sample.
At each tip location, we record the transition voltage $V_\text{transition}$ of the lowest charging transition within the experimental voltage range; these are shown as red points in Figs.~\ref{fig:defect5_lateralFit}(a) (horizontal scan) and \ref{fig:defect5_lateralFit}(c) (vertical scan).
Here, each data point $V_\text{transition}$ describes the average value of the charge-transition voltages obtained in the forward and backward voltage sweeps, taken over several measurements.
The error bars describe the standard deviation of these measurements, with relatively large values reflecting the stochastic nature of the tunneling process.
The two scans (vertical and lateral) overlap at the point of closest separation between the tip and the defect.
As expected from the simulations described below, the charging-transition voltage takes its maximum value at this location, with $V_\text{transition}\approx 3.6$~V, and decreases for larger tip-defect separations.

Simulations are then used to fit the experimental data in these vertical and lateral scans.
We assume the two-section tip model defined in Fig.~\ref{fig:parabolic_tippy_tip}.
Optimized pairs of parameters $(h_\text{Au},r_\text{apex})$ were previously obtained in Fig.~\ref{fig:bestFits_goldht_rapex}(c), and we consider one of these pairs, and one value of the defect depth $d$.
The goal of the simulations is to choose the optimal values for these parameters.

We first consider the common point in the vertical and lateral scans -- the point of closest separation between the tip and defect.
We perform a simulation of the tip-sample system, using the experimentally determined value of $V_b=V_\text{transition}$ at that tip location.
The output of interest from the simulation is the electrostatic potential measured at the location of the defect $\phi({\mathbf r}_d)$, which we use as a proxy for the value of the chemical potential for the charging transition.
We then consider other positions of the tip along the vertical and lateral scans.
At each tip location, we use simulations to determine the value of $V_b$ that gives the same value of $\phi({\mathbf r}_d)$, since this corresponds to the modified value of the gate bias that gives the same chemical potential that causes a charging transition.
Repeating this procedure for every tip position gives a theoretical prediction for $V_\text{transition}$ as a function of position along the scan, which can be compared to the experimental data, as shown in Figs.~\ref{fig:defect5_lateralFit}(a) and (c) (blue line).
The RMS value describing the difference between the experimental data and the theoretical fit is then computed.
This whole procedure is repeated for the same value of $d$, while choosing new values of the correlated $(h_\text{Au},r_\text{apex})$ pair.
In this way, we identify the $(h_\text{Au},r_\text{apex})$ pair that minimizes the RMS for a given $d$, yielding optimized triplets of parameter values $(d,h_\text{Au},r_\text{apex})$.

We now repeat this whole procedure for different values of $d$, yielding  slightly different results for the lateral scan vs the vertical scan, as shown in Figs.~\ref{fig:defect5_lateralFit}(b) and (d).
This method provides estimates for the uncertainty in our optimization procedure, as defined by the horizontal dashed lines in the figure panels, corresponding to about 10\% of the largest reported RMS value, measured from the minimum RMS value.
However, the optimal values of $d$ do not necessarily match for the lateral and vertical scans, as observed in Fig.~\ref{fig:defect5_jointFit}.
We therefore insist that the values of $d$ should match, because they arise from the same defect.
The last step in the procedure is therefore to identify points of overlap in these two curves, as indicated here by red circles.
We accept the average of these data points as our overall optimized value for the $(d,h_\text{Au},r_\text{apex})$ triplet, giving the final results for our fitting procedure, which are given by $d= (0.65\pm 0.30)$~nm, $r_\text{apex}=(1.5\pm 0.5)$~nm, and $h_\text{Au}=(0.5\pm 0.4)$~nm.

As a final comment, we note that this technique also provides an estimate for the defect depth $d=(0.65\pm 0.30)$~nm.
We emphasize that this amounts to a very laborious scheme for finding $d$, which we do not repeat for every defect. 
However, the final result corroborates our intuition that defect charging transitions, induced by electrons tunneling to or from the EFM tip, can only occur when defects are situated very near the surface.

\subsection{Determining the tip-defect lever arm $\alpha_d$ and the charging transition voltage $V_\text{transition}$ as a function of tip position}
\label{sec:lever_arm}

The tip-defect lever arm $\alpha_d$ describes the capacitive interaction between the tip and a defect, and plays an important role in determining the defect's charge-transition voltages $V_\text{transition}$.
$\alpha_d$ clearly depends on the tip location, relative to the defect, and must therefore be re-computed for every new tip-defect configuration.
For example, in the preceding subsections, we considered lateral and vertical scans over a defect, for which the lever arm varies with the tip position.

To compute the lever arm for a given tip location, we apply a voltage bias $V_b$ between the tip and backgate in COMSOL simulations, by specifically setting the tungsten portion of the tip to 0~V and the backgate to $V_b$.
(For example, see the diagram in Fig.~\ref{fig:tip_shape_fitting}\emph{A} of the main text.)
To account for the work function difference $W_\text{Au}-W_\text{W}$ between the tungsten and gold portions of the tip, we set the voltage of the gold portion of the tip to -0.92~V, as a boundary condition.
(Also see Sec.~\ref{sec:converting_Vtransion}, below.)
We then perform two simulations at each tip location; in both cases, we solve for the electrostatic potential $\phi({\mathbf r}_d)$ at the location of the defect.
In the first simulation, we set $V_b=0$, while in the second simulation, we set $V_b=1$~V, yielding the results $\phi({\mathbf r}_d)=\phi_\text{0V}$ and $\phi({\mathbf r}_d)=\phi_\text{1V}$, respectively.
From such simulations, we can then define
\begin{equation}
\alpha_d=(\phi_\text{1V}-\phi_\text{0V})/\text{(1 V)},
\end{equation}
which describes a specific tip-defect geometry.

To compute transition voltages as a function of tip position, such as the solid blue curves shown in Figs.~\ref{fig:defect5_lateralFit}(a) and (c), we compare the reference values ($V_\text{transition},\alpha_d,\phi_\text{0V}$), obtained at a fixed tip location, to the same quantities obtained at other tip locations, ($V'_\text{transition},\alpha'_d,\phi'_\text{0V}$).
The relation between these sets of parameters is given by 
\begin{equation}
V'_\text{transition}=(\alpha_dV_\text{transition}+\phi_\text{0V}-\phi'_\text{0V})/\alpha'_d .
\end{equation}

\section{Algorithms for identifying charging transitions in $f$-$V$ curves}
\label{sec:locating_transitions}

\begin{figure}[!hbt]
    \centering
    \includegraphics[width=0.8\textwidth]{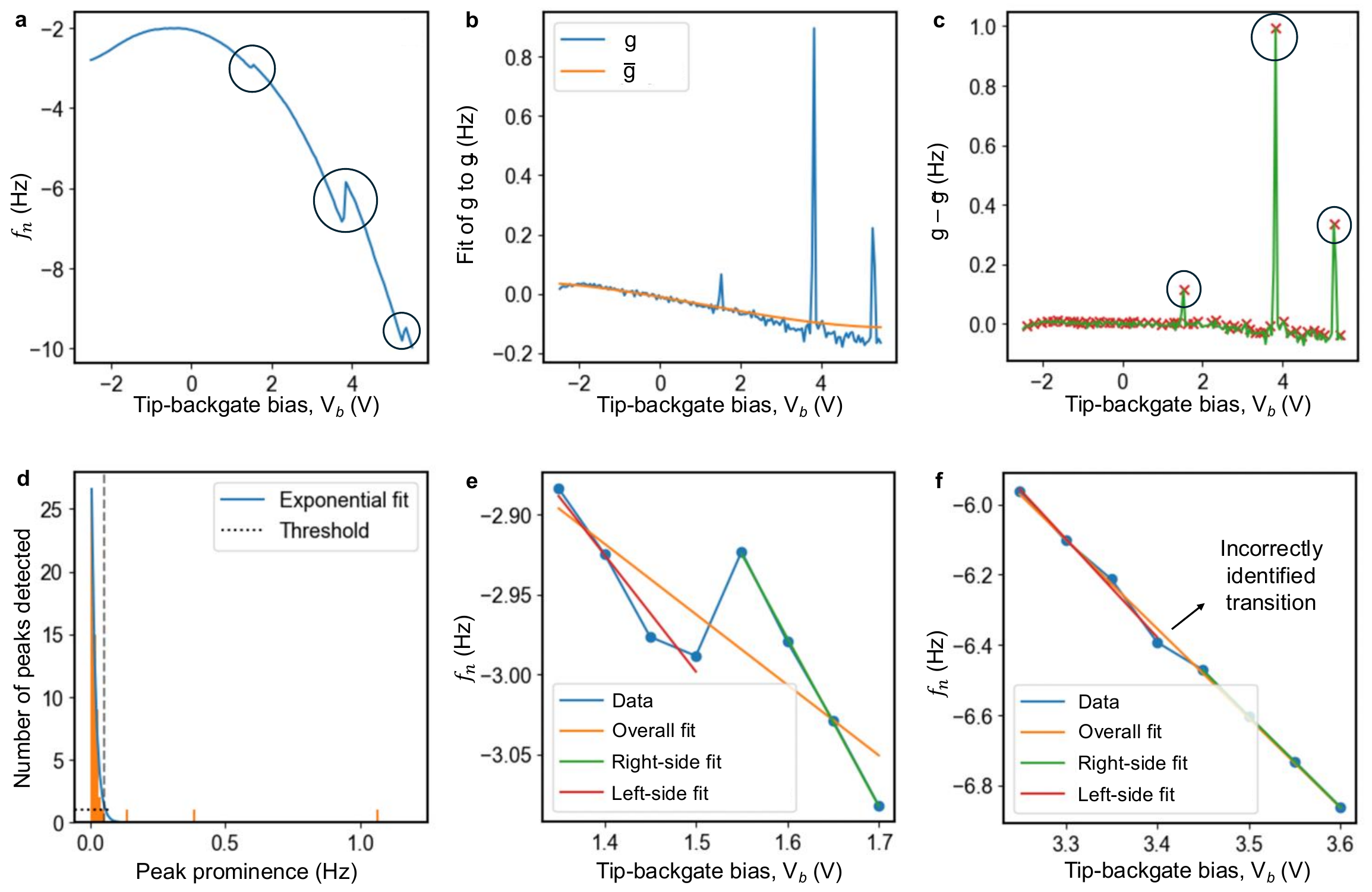}
    \caption{Algorithm 1 for detecting charging transitions and locating their position $V_\text{transition}$ in a voltage-bias sweep. 
    (a) A typical $f$-$V$ curve with three charging transitions (circled).
    The experimental data points are denoted as $f_n$.
    (b) The discrete numerical derivative $g_n$ of $f_n$, as defined in Eq.~(\ref{eq:gdef}) (blue curve).
    The orange curve shows a smooth approximation $\overline{g}$ for the background signal, obtained by fitting a polynomial of degree 3 to the $g$ data.
    (c) The difference $g-\overline{g}$ (green curve), approximating just the peak contribution to $g$.
    The Python routine \texttt{find\char`_peaks} identifies all possible peaks in this data (orange $\times$ markers), including legitimate transition peaks as well as noise peaks.
    (d) Histogram of the `peak prominence' values computed by \texttt{find\char`_peaks}, defined as the height of the peaks relative to their nearby surroundings (thin orange vertical bars).
    The corresponding distribution function (green curve) is determined by performing an exponential fit to the histogram.
    We define a threshold peak prominence value below which the distribution falls below 1 (dashed lines), as described in the main text.
    Peaks occurring on the right-hand side of this threshold are accepted as legitimate charging transitions.
    (e),(f) Two examples of a method used to identify false positives in (d):
    for eight $f_n$ data points, centered at a transition identified by Algorithm 1, straight lines are fit to the first four points, the last four points, and all eight points. 
    If the slope of the eight-point fit falls between the slopes of the other two fits, as it does in (f), then the transition is rejected.
    Otherwise, the transition is accepted, as in (e).}
    \label{fig:algo1_procedure}
\end{figure}

Charging events are observed as jumps in $f$-$V$ curves, with several examples shown in Fig.~\ref{fig:algo1_procedure}(a).
A very large number of transitions can be found in a large $f$-$V$ map.
In this section, we describe two algorithms used to automatically determine the voltage biases $V_\text{transition}$ where these transitions occur. 

Algorithm 1:
\begin{enumerate}
  \item Compute a pointwise derivative of the $f$-$V$ curve:
  if $\{f_n\}$ are consecutive data points in the $f(V)$ data set, then the numerical differences are given by
  \begin{equation}
      g_{n} = f_{n+1} - f_{n}.
      \label{eq:gdef}
  \end{equation}
  For example, the $f$ data in Fig.~\ref{fig:algo1_procedure}(a) yields the $g$ data in Fig.~\ref{fig:algo1_procedure}(b, shown in blue).
  \item Fit a polynomial of degree 3 to the $g$ data, yielding the smoothed orange curve $\overline{g}$ in Fig.~\ref{fig:algo1_procedure}(b). 
  This fit is intentionally designed to smooth over transitions in the $g$ data, yielding an approximation for the smooth background, which excludes the transitions. 
  The difference $g-\overline{g}$ then provides an approximation for just the transition, with the background excluded, as shown by the green curve in Fig.~\ref{fig:algo1_procedure}(c).
  \item Apply the routine \texttt{find\char`_peaks}, from the \texttt{scipy.signal} Python library code base, to identify all peaks in $g-\overline{g}$. 
  Based on the range of data shown in Fig.~\ref{fig:algo1_procedure}(c), we set the peak `prominence' parameter to span the range of $[0,5]$~Hz, so that it captures all possible peaks in the data set, even tiny peaks caused by noise.
  Here, the term `prominence' refers to the height of the peak compared to its nearby surroundings.
  In Fig.~\ref{fig:algo1_procedure}(c), all peaks identified by this routine are indicated with orange $\times$ markers.
  \item Next, consider the distribution of the peak prominence values, as shown by thin vertical bars in Fig.~\ref{fig:algo1_procedure}(d). 
  This distribution is fit to an exponentially decaying function, shown by the green curve.
  We define a crossover threshold for this distribution function at the point where it drops below 1, as indicated by the dashed lines. 
  All peaks occurring to the left of this threshold are considered to be noise, while peaks to the right are considered to be legitimate charging transitions.
  The specific threshold value was determined by applying the procedure to a set of 50 $f$-$V$ curves for which charging transitions were identified manually; the value of 1 was found to maximize the number of correct identifications, which we found to be 84.6\% of total transitions in the test data.
\end{enumerate}

Algorithm 1 is found to be effective, but also yields false negative and false positive identifications of charging transitions.
To address the problem of false positives, we apply the simple procedure shown in Figs.~\ref{fig:algo1_procedure}(e) and (f).
Here, we consider seven $f_n$ data points centered around a candidate transition.
We then fit straight lines to the first three data points, the last three data points, and the full set of data points.
If the slope of the fit to the full set lies between the slopes of the left and right sets, the curve is classified as `smooth,' meaning that it does not contain a transition; non-smooth curves are then accepted as legitimate transitions.
Using this procedure, the featue in Fig.~\ref{fig:algo1_procedure}(e) is accepted as a legitimate transition, while the feature in Fig.~\ref{fig:algo1_procedure}(f) is not.

\begin{figure}[!hbt]
    \centering
    \includegraphics[width=0.6\textwidth]{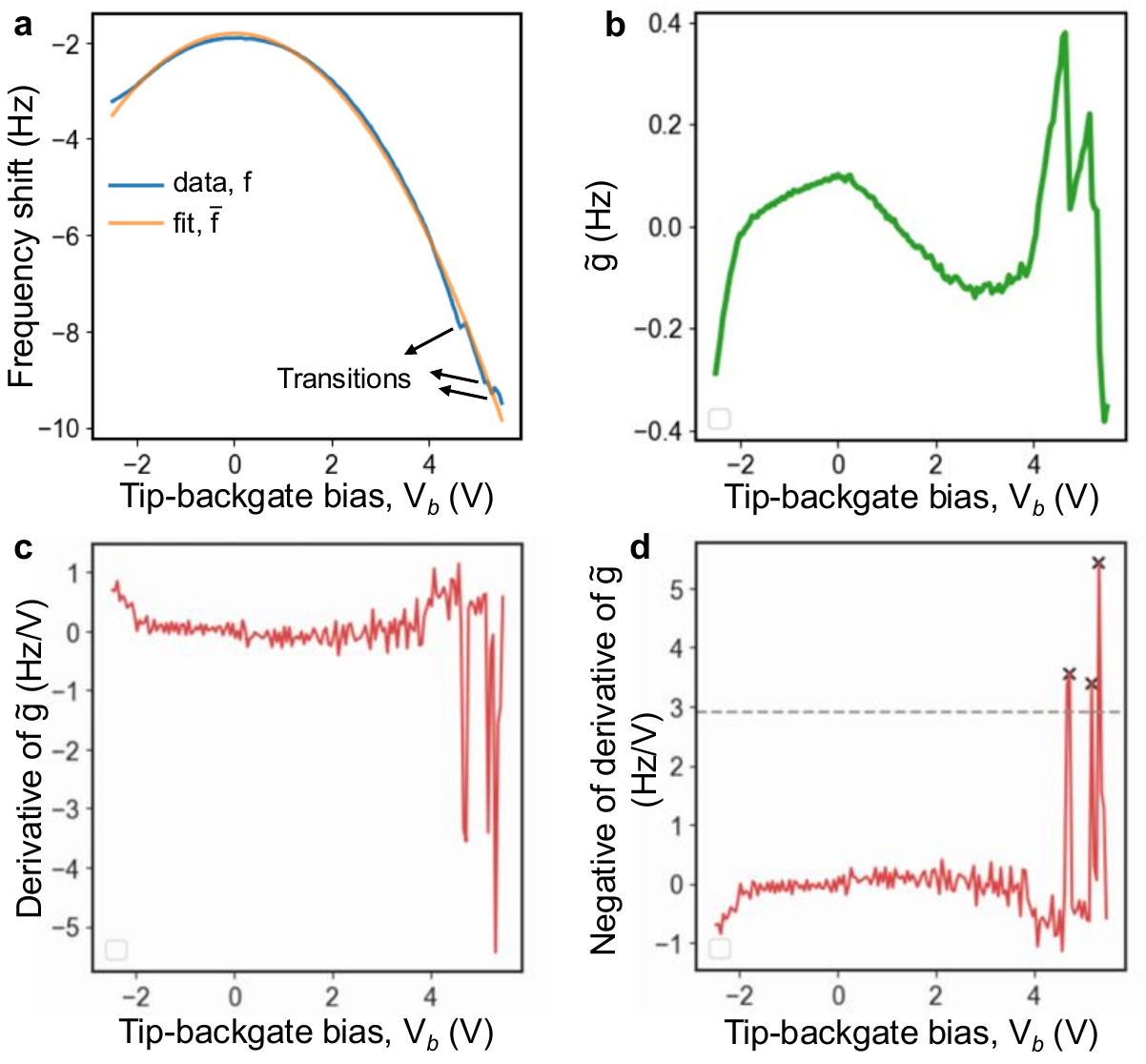}
    \caption{Algorithm 2 for detecting charging transitions and locating their position $V_\text{transition}$ in a voltage-bias sweep.  
    (a) A typical $f$-$V$ curve is shown (blue curve), with three charging transitions as indicated.
    A parabolic fit $\overline{f}$, shown in orange, describes the background signal for $f$.
    (b) The difference $\tilde{g}=\overline{f}-f$ highlights the charging transitions as singularities.
    (c),(d) The numerical derivative of $\tilde{g}$, and the same curve with its sign changed, which can both exhibit prominent upward peaks.
    A threshold value for identifying legitimate peaks is given by $4\sigma_\text{diff}$,  as indicated by the dashed line in (d).
    Here, $\sigma_\text{diff}$ is the standard deviation of the numerical derivative of the $\tilde{g}$ data shown in (c).
    (See main text for explanation.)
    Upward-pointing peaks larger than this threshold are identified by the \texttt{find\char`_peaks} numerical routine, as indicated by $\times$ markers.
    These peaks are added to the transitions identified by Algorithm 1, if not already present.
    In this example, no additional peaks were identified in (c).}
    \label{fig:algo2_procedure}
\end{figure}

To address the problem of false negatives when identifying charging transitions, we apply a second algorithm.

Algorithm 2:
\begin{enumerate}
    \item First fit the $\{f_n\}$ data set to a polynomial of degree 2, as shown by the orange curve in Fig.~\ref{fig:algo2_procedure}(a).
    We denote this fitted curve as $\overline{f}$, representing the smooth background of the $f$ data.
    \item Next, compute the difference $\tilde{g}=\overline{f}-f$ to highlight the charging transitions, as shown in Fig.~\ref{fig:algo2_procedure}(b).
    \item Compute the numerical derivative (difference) of $\tilde{g}$ between successive points, similar to Eq.~(\ref{eq:gdef}).
    A typical result is shown in Fig.~\ref{fig:algo2_procedure}(c).
    Note that if a transition occurs to the left of the CPD, it will appear as an upward peak in this derivative; here, the transitions occur to to the right of the CPD, and are therefore pointed downward.
    To treat these different cases using the same numerical subroutine, we also change the sign of the $\tilde{g}$ data, as shown in Fig.~\ref{fig:algo2_procedure}(d), which gives upward peaks.
    \item Again apply the routine \texttt{find\char`_peaks} to identify upward-pointing peaks in both the positive and negative derivative data.
    Here, we do not eliminate the small noise peaks analogously to Algorithm 1. 
    Instead, we choose a higher peak prominence threshold, as follows.
    First, we compute the standard deviation $\sigma_\text{diff}$ of the difference data in Figs.~\ref{fig:algo2_procedure}(c) or (d).
    We then choose a minimum peak-prominence threshold value of $4\sigma_\text{diff}$ [indicated by the dashed line in Fig.~\ref{fig:algo2_procedure}(d)], which we find to be optimal for distinguishing legitimate peaks  from noise, resulting in the peaks indicated by $\times$ markers in Fig.~\ref{fig:algo2_procedure}(d).
    Note that no (upward) peaks were identified by this method in Fig.~\ref{fig:algo2_procedure}(c), although we do find them in some cases.
    The $4\sigma_\text{diff}$ threshold value was determined after applying the algorithm to a set of 50 $f$-$V$ curves for which the charging transitions were identified manually; the factor of 4 was found to give the fewest false positive identifications among the transitions that were not previously identified by Algorithm 1. 
    \item The transitions determined by Algorithm 2 are then combined with those from Algorithm 1.
\end{enumerate}

\begin{figure}[hbt!]
    \centering
    \includegraphics[width=0.6\textwidth]{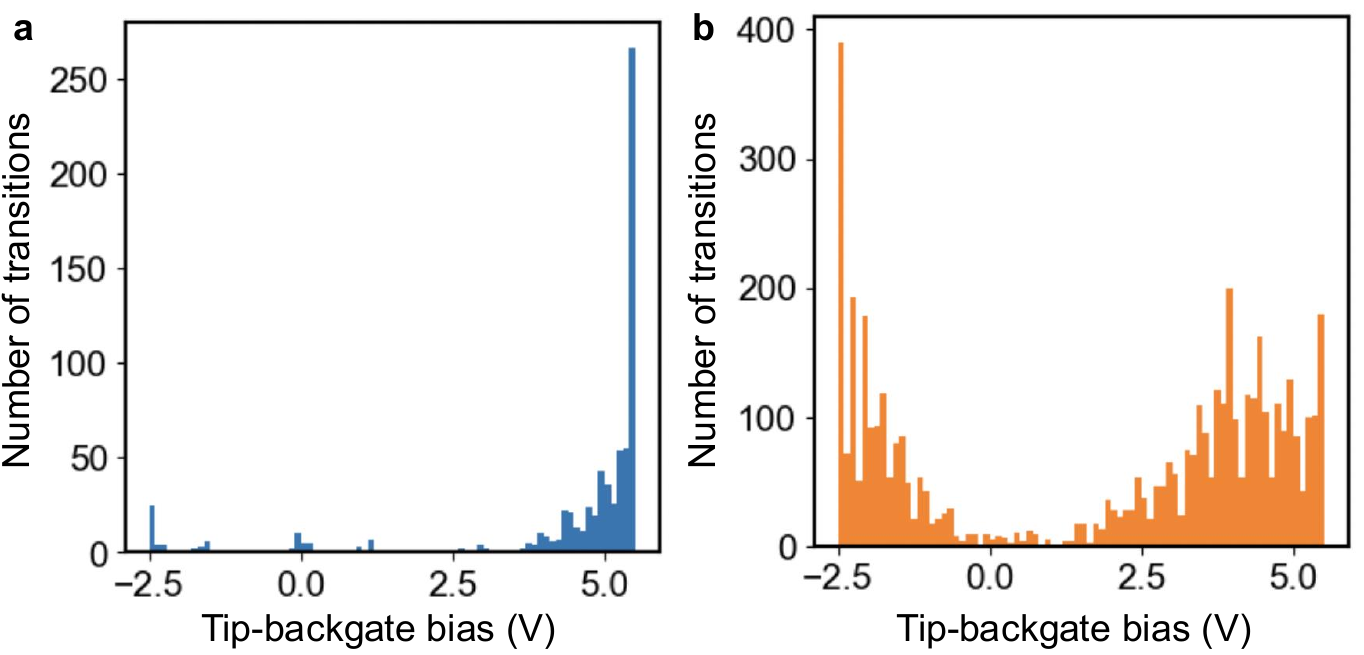}
    \caption{Histograms of charging transitions identified by Algorithm 2 that were not identified in Algorithm 1 for forward voltage sweeps (blue) and backward voltage sweeps (orange). }
    \label{fig:comparison_algo1_algo2}
\end{figure}

Figure~\ref{fig:comparison_algo1_algo2} shows histograms of transitions detected by Algorithm 2 that were not detected by Algorithm 1 for forward voltage sweeps (blue) and backward voltage sweeps (yellow).

\section{Identifying Individual Defects: Large-Map Method }
\label{sec:Identifying_Defects}

Chemical identification of defects in an oxide requires sorting through a large number of charging transitions to identify all the transitions for a single defect, while excluding transitions from nearby defects. 
This procedure is complicated by the fact that defects \emph{overlap} in our EFM maps, in the sense that the ``size'' of a defect (the spatial extent over which electrons tunnel between the tip and the defect) can be several nanometers, encompassing many pixels.
Hysteresis between charging transitions in forward and backward voltage bias sweeps further complicates the identification procedure. 
The most accurate identifications are obtained when this hysteresis is minimized, by centering the tip directly above the defect.

In this work, we consider two methods for identifying individual defects and locating their center pixel.
In this section, we describe a detailed approach for analyzing large transition maps, with the goal of isolating individual defects within a large, coarse data set.
In the next section, we describe a more-direct method, which is enabled by in-situ analysis of transition voltage hysteresis data.

\begin{figure}[t]
    \centering
    \includegraphics[width=0.7\textwidth]{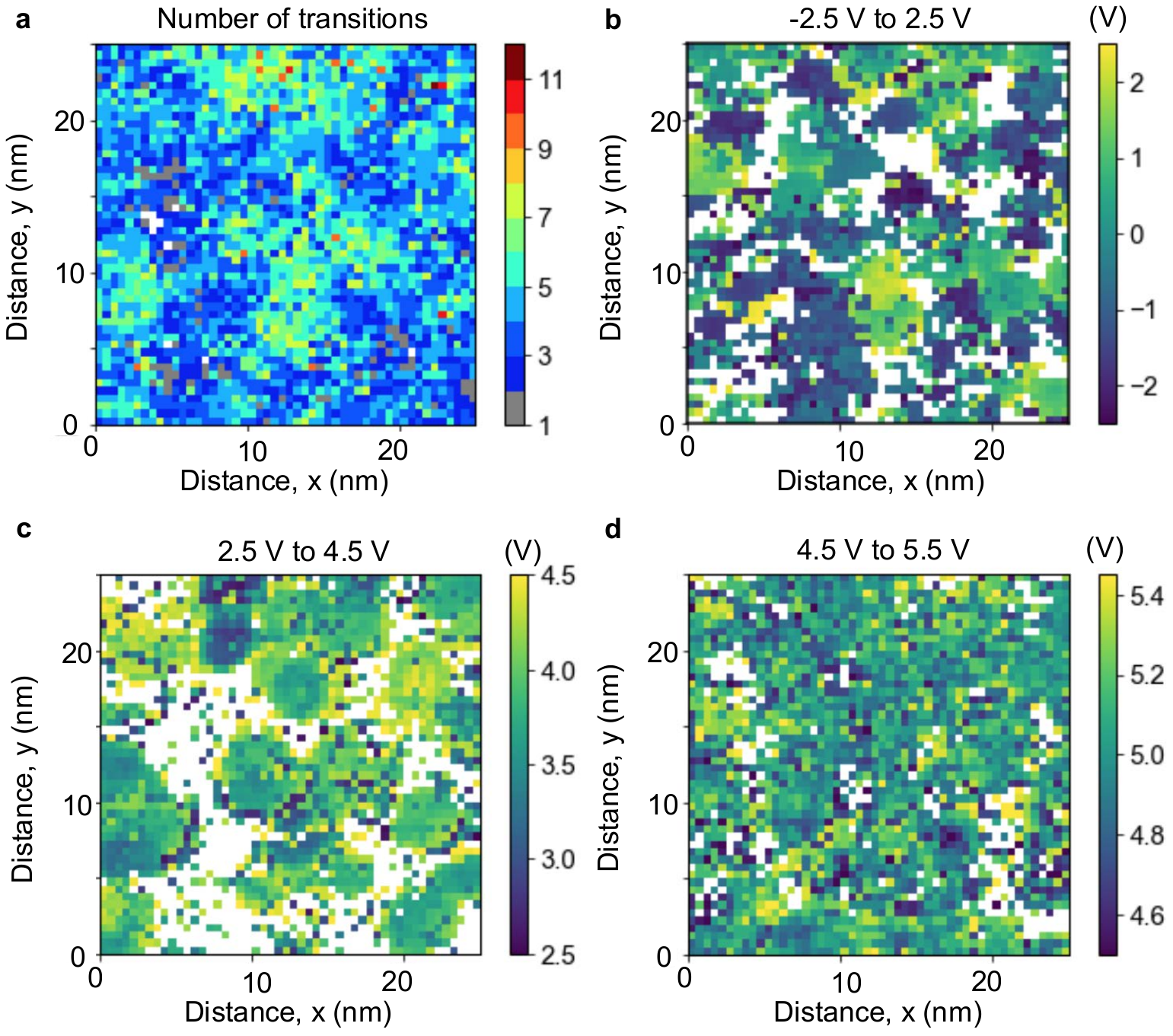}
    \caption
    {Maps of defect charging events and charging voltages.
    (a) Total number of charging transitions observed, per pixel, over the whole [-2.5, 5.5]~V sweep range of bias voltages.
    (b)-(d) Charging transition maps obtained in three different bias windows.
    Individual defects become more visible and isolated in narrower bias windows; however, transitions appear quite dense in the [4.5, 5.5]~V window, shown in (d), so that it becomes difficult to isolate individual defects.
    Note that when multiple transitions are present at a given pixel in (b)-(d), we only report the lowest voltage transition here.}
    \label{fig:transition_voltage_maps}
\end{figure}

\begin{figure}[t]
    \centering
    \includegraphics[width=0.6\textwidth]{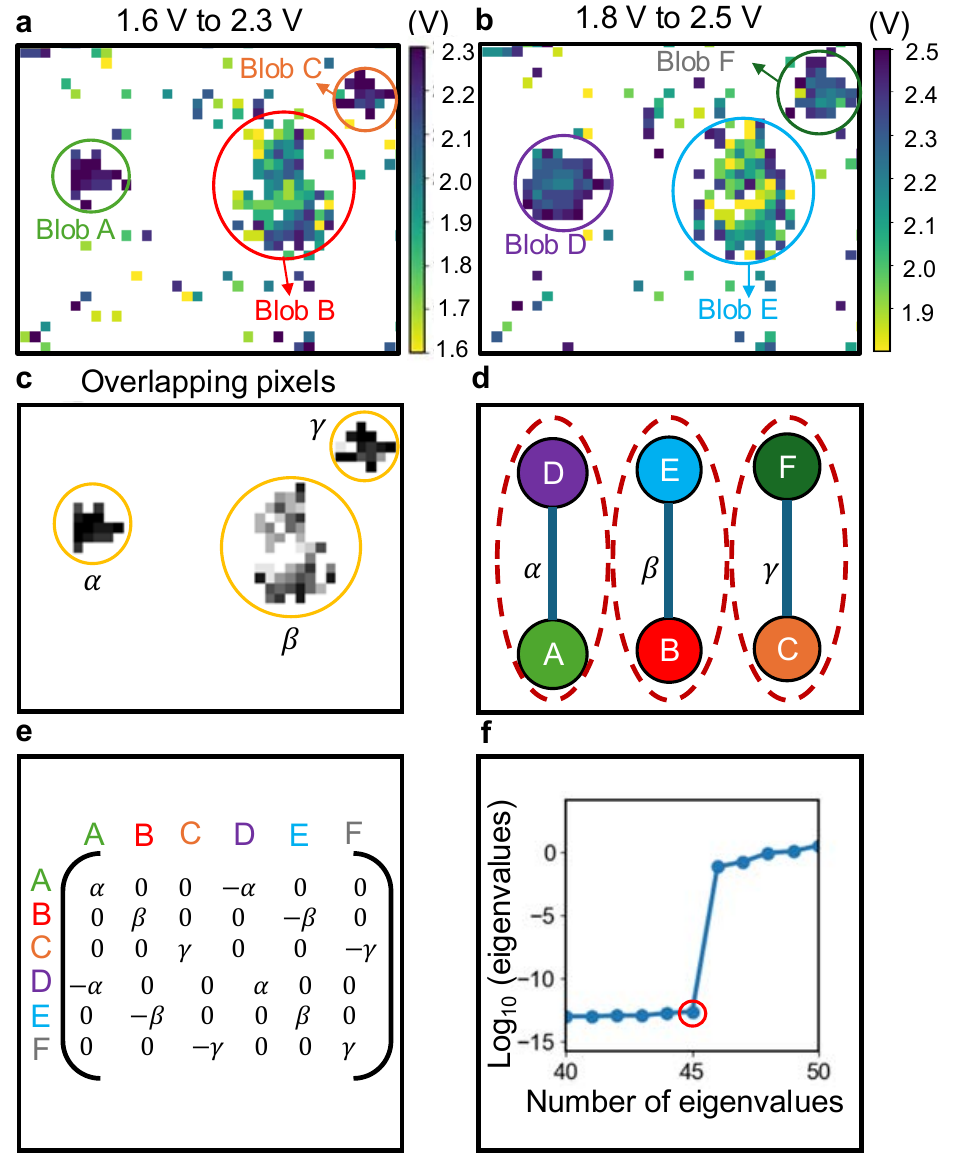}
    \caption{
    (a),(b) The CCA method, described in Sec.~\ref{sec:CCA}, identifies clusters of pixels, associated with a single defect, which are all associated with the same charging transition, using criteria based on pixel adjacency and cluster size.
    (If no charging transition is observed in the bias window,  for a given pixel, that pixel is shaded white.)
    In SCA parlance, these clusters are referred to as ``blobs.''
    In (a) and (b), blobs A and D most likely describe the same defect observed in different bias windows.
    Blobs B and E are similarly linked, as are blobs C and F.
    (c) The quantities $\alpha,\beta,\gamma$ describe the number of overlapping pixels in the blobs illustrated in (a) and (b).
    (d) In the SCA, described in Sec.~\ref{sec:SCA}, blobs are represented by nodes in an undirected graph, while the overlaps ($\alpha,\beta,\gamma$) describe the weights of the edges connecting those nodes.
    (e) The SCA Laplacian matrix corresponding to the graph shown in (d).
    (See text for discussion.)
    (f) Eigenvalues are shown for a typical Laplacian matrix.
    The number of eigenvalues near zero corresponds to the number of unique defects and transitions observed across multiple bias windows, while their corresponding eigenvectors describe the linkages between windows.
    In this example, 45 clusters are identified as unique defect transitions.
    The complete defect, observed across many windows, is then given by the union of pixels linked in this way, across those windows.
    }
    \label{fig:graph_nodes_clustering}
\end{figure}

\begin{figure}[t]
    \centering
    \includegraphics[width=0.3\textwidth]{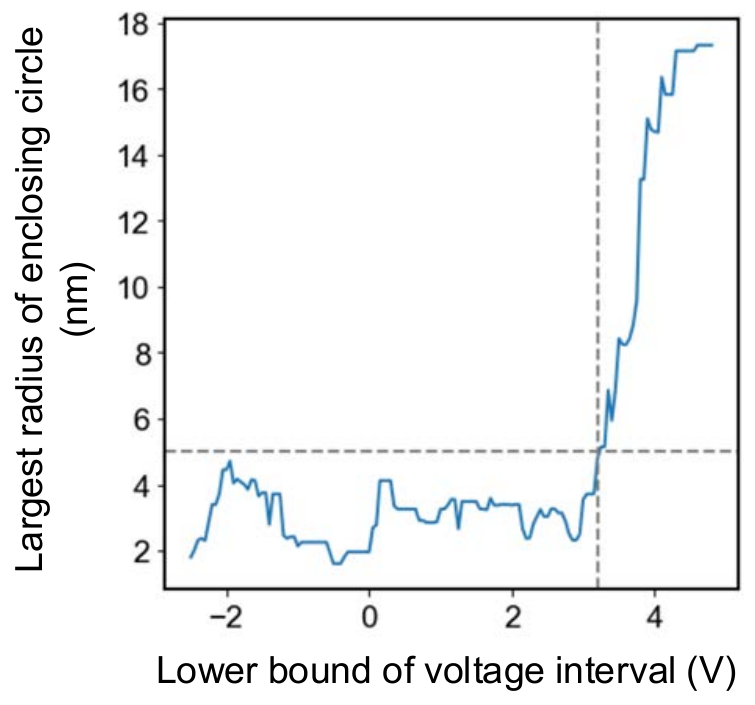}
    \caption{The radius of the minimum enclosing circle (see main text) is reported for each defect cluster identified by the CCA, as illustrated in Figs.~\ref{fig:graph_nodes_clustering}(a) and (b), for each bias window.
    Here, the bias windows are labeled by their lower-bound voltages.
    A sudden jump is observed at the lower-bound voltage of 3.2~V, indicating the point where defects are so dense that our CCA analysis fails, as it captures multiple defects rather than just one.
    Bias windows to the right of this dashed line therefore are excluded from further analysis.
    }
    \label{fig:max_radii_enclosing_circle}
\end{figure}

For the large-map analysis, we first note that most defect species exhibit multiple transitions, as confirmed in our DFT calculations.
This can be seen in Fig.~\ref{fig:transition_voltage_maps}(a), where transitions are enumerated at each pixel.
We can partially isolate individual defects by focusing on smaller bias-voltage windows, as shown in Figs.~\ref{fig:transition_voltage_maps}(b)-(d). 
Here, if multiple transitions occur at a given pixel, we only show the lowest transition.
In panels (b) and (c), we see individual defects begin to emerge from the data, although such behavior is less apparent for the voltage range shown in (d).
To identify all the pixels associated with a single charging transition and a single defect, we need to link clusters of pixels in different bias windows.
Fortunately, charging events associated with the same defect and transition occur at similar voltages on neighboring pixels, which makes our job easier.
Generally, we find that a bias range of 0.7~V represents a good ``Goldilocks'' value, which picks up most of the pixels belonging to a single defect and transition, while retaining a reduced density of transitions within the bias window.
We therefore adopt 0.7~V as the width of the bias windows in the following analysis.

To identify clusters of pixels belonging to the same defect and transition, we first apply a Connected Component Analysis (CCA)~\cite{opencv_library,LindaG_Shapiro1996} to every bias window.
We then link defects in different bias windows by applying a Spectral Clustering Algorithm (SCA)~\cite{towardsdatascienceSpectralClustering}.
If any defects overlap, particularly defects belonging to the same chemical species, they can be distinguished by the fact that their combined cluster forms an irregular, noncircular shape; a Gaussian Mixture Model (GMM) Algorithm~\cite{scikit_learn} is then used to split such clusters into smaller, regular shapes representing individual defects.
Finally, once all the pixels for a given defect and transition have been identified, we apply a Gaussian Blur procedure to approximately identify the centermost pixel of a cluster, where the charging transition hysteresis should be minimized.
We now describe these algorithms in more detail.

\subsection{Connected Component Analysis} \label{sec:CCA}
Figures~\ref{fig:transition_voltage_maps}(b) and (c) show charging transition maps, for bias windows smaller than the full range.
Here, the windows are still quite large, so that most of the pixels forming a single defect are displayed; however, the defects are seen to overlap significantly.
(Note that, if more than one transition is observed at a given pixel, only the lowest transition voltage is reported in the figure.)
As noted above, to help isolate the defects, we therefore consider smaller bias windows of width 0.7~V, yielding results like those shown in Figs.~\ref{fig:graph_nodes_clustering}(a) and (b), where the defects are largely distinct.
The same defect naturally appears in closely separated bias windows.
In this analysis, we divide the full bias scan of width [-2.5, 5.5]~V into a set of overlapping windows of width 0.7~V, offset by increments of 0.05~V. 
Thus, the full set of bias windows that could be considered is given by [-2.5, -1.8]~V, [-2.45, -1.75]~V, \dots, [4.5, 5.2]~V.
However, in practice, we reduce the upper range of this set, as described in the following paragraph. 

We now apply a Connected Component Analysis (CCA) to transition maps for each of the bias windows.
This procedure groups together all pixels that (i) show a transition in the given bias window and (ii) share an edge or a corner. 
(For a given pixel, there can be eight such nearest neighbors.)
Clusters containing ten or more pixels are then accepted as legitimate defects.
Figures~\ref{fig:graph_nodes_clustering}(a) and (b) illustrate some typical (circled) defects identified in this way.
To further analyze the defects, we apply the OpenCV python routine \texttt{minEnclosingCircle} to determine the circle of minimum radius that encloses a given cluster [yielding the circles shown in Figs.~\ref{fig:graph_nodes_clustering}(a) and (b)].
In Fig.~\ref{fig:max_radii_enclosing_circle}, we plot the enclosing radiuses as a function of the lower-bound voltage of each bias window. 
The results reveal a jump in radius size, beyond the (lower-bound) voltage bias of 3.2~V, indicating that beyond this point, the defects become so dense that it is impossible to identify individual defects by this method.
This is also consistent with our observation that typical defect radiuses fall in the range of 2-3~nm.
Thus we do not consider any (lower-bound) voltage biases beyond 3.2~V, i.e., we restrict our analysis to the voltage range [-2.5, 3.9]~V, and the highest bias interval we study is [3.2, 3.9]~V.
Using this CCA method, 70\% of the pixels showing transitions were grouped into defect clusters, while the remaining transitions (on isolated pixels) were regarded as ``noise'' and discarded.

\subsection{Spectral Clustering Algorithm} \label{sec:SCA}
\begin{figure}[t]
    \centering
    \includegraphics[width=0.65\textwidth]{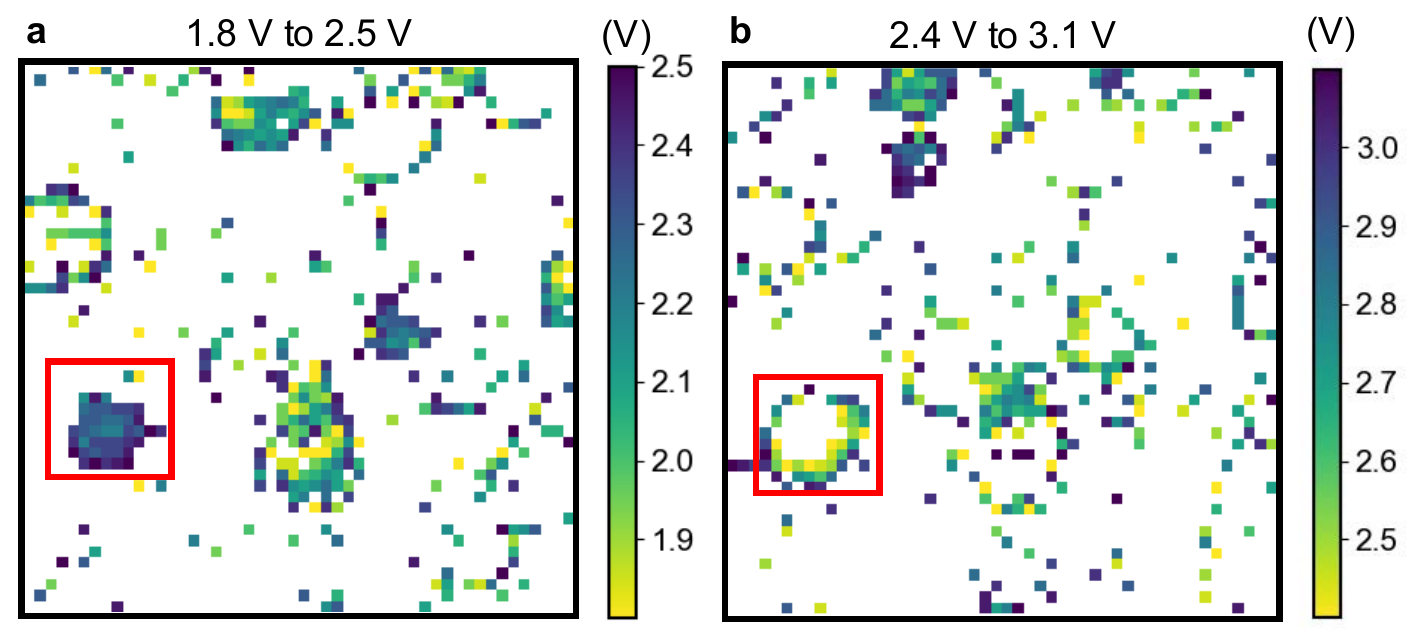}
    \caption{
    A single defect transition can appear in multiple bias windows, but it often takes a slightly different shape in those windows.
    Thus, while narrower bias ranges are useful for isolating individual defects, it is important to stitch together multiple windows, to obtain a full description of a defect.}
    \label{fig:defect_different_voltage_snapshots}
\end{figure}

The goal of this section is to develop an automated scheme to collect all the pixels associated with a defect, for each of its different charging transitions.
These defect maps can be observed across multiple bias windows, although the pixel arrangements may differ from window to window.
For example, in Figs.~\ref{fig:graph_nodes_clustering}(a) and (b), the pixels largely overlap, while in Fig.~\ref{fig:defect_different_voltage_snapshots}, they do not, although our analysis suggests they correspond to the same defect. 
(The bias windows in Fig.~\ref{fig:defect_different_voltage_snapshots} are well-separated in voltage space.)
To complete our task, and correctly link defects in different windows, we apply a Spectral Clustering Algorithm (SCA).

SCA is a graph theoretical technique, whose details are described in~\cite{towardsdatascienceSpectralClustering}. 
Here, we limit the discussion to our particular SCA implementation.
The first step in this procedure is to label the defect clusters identified by the CCA, for all bias windows.
[Two examples are shown in Figs.~\ref{fig:graph_nodes_clustering}(a) and (b).]
In SCA parlance, these are referred to as ``blobs.''
Next, we count the overlapping pixels in these blobs.
In the example illustrated in Fig.~\ref{fig:graph_nodes_clustering}(c), these overlaps are labeled $\alpha,\beta,\gamma$.
In the SCA, such overlaps ($\alpha,\beta,\gamma$) become weights assigned to the edges connecting the nodes (A-F) of an undirected graph, as illustrated in Fig.~\ref{fig:graph_nodes_clustering}(d).
In this particular example, the blobs are well-isolated and the graph is simple; however, the SCA robustly disentangles more complicated geometries containing multiply overlapping blobs.
We then define a Laplacian matrix, $L=D-A$, as illustrated in Fig.~\ref{fig:graph_nodes_clustering}(e), where the rows and columns represent the nodes in the graph.
The adjacency matrix $A$ describes the weights of the edges, with $A_{ij}$ representing the weight between nodes $i$ and $j$ and $A_{ii}=0$, while the diagonal matrix $D$ describes the total weight of each node: $D_{ii}=\sum_jA_{ij}$.
For the simple example shown in Fig.~\ref{fig:graph_nodes_clustering}, these diagonals are simply given by $\alpha,\beta,\gamma$, since the blobs are well-isolated.
Finally, by diagonalizing $L$, we obtain a set of nearly degenerate eigenvalues near zero, as illustrated in Fig.~\ref{fig:graph_nodes_clustering}(f), whose eigenvectors describe the properly linked blobs, according to SCA theory.

\subsection{Gaussian Mixture Model Algorithm}
The SCA is unable to distinguish overlapping defects appearing in the same bias window.
This is especially relevant for defects of the same chemical species, which naturally exhibit charging transitions in the same bias voltage range.
Only a few tools are available to potentially unravel such behavior, the most prominent being the irregular shape of a multi-defect cluster.
Since single-defect clusters are roughly circular, a Gaussian Mixture Model (GMM)~\cite{scikit_learn} algorithm can be effective for splitting up combined clusters.
Here, we use the routine \texttt{sklearn.mixture.GaussianMixture} from the scikit-learn python library to implement such a GMM algorithm. 

The number of component defects that a cluster should be divided into depends in part on its pixel count. 
Visual inspection suggests that the maximum size of single-defect clusters is about 50-60 pixels, as constrained by the maximum distance that electron can tunnel between the tip and a defect near the sample surface. 
For reference, many of the pixels considered in this work are of size 0.5$\times$0.5~nm$^2$.
Clusters larger than 60 pixels are found to be irregularly shaped, suggesting that they include multiple defects.
Clusters smaller than 50 pixels are typically circular, suggesting that they correspond to single defects.

\begin{figure}[t]
    \centering
    \includegraphics[width=0.5\textwidth]{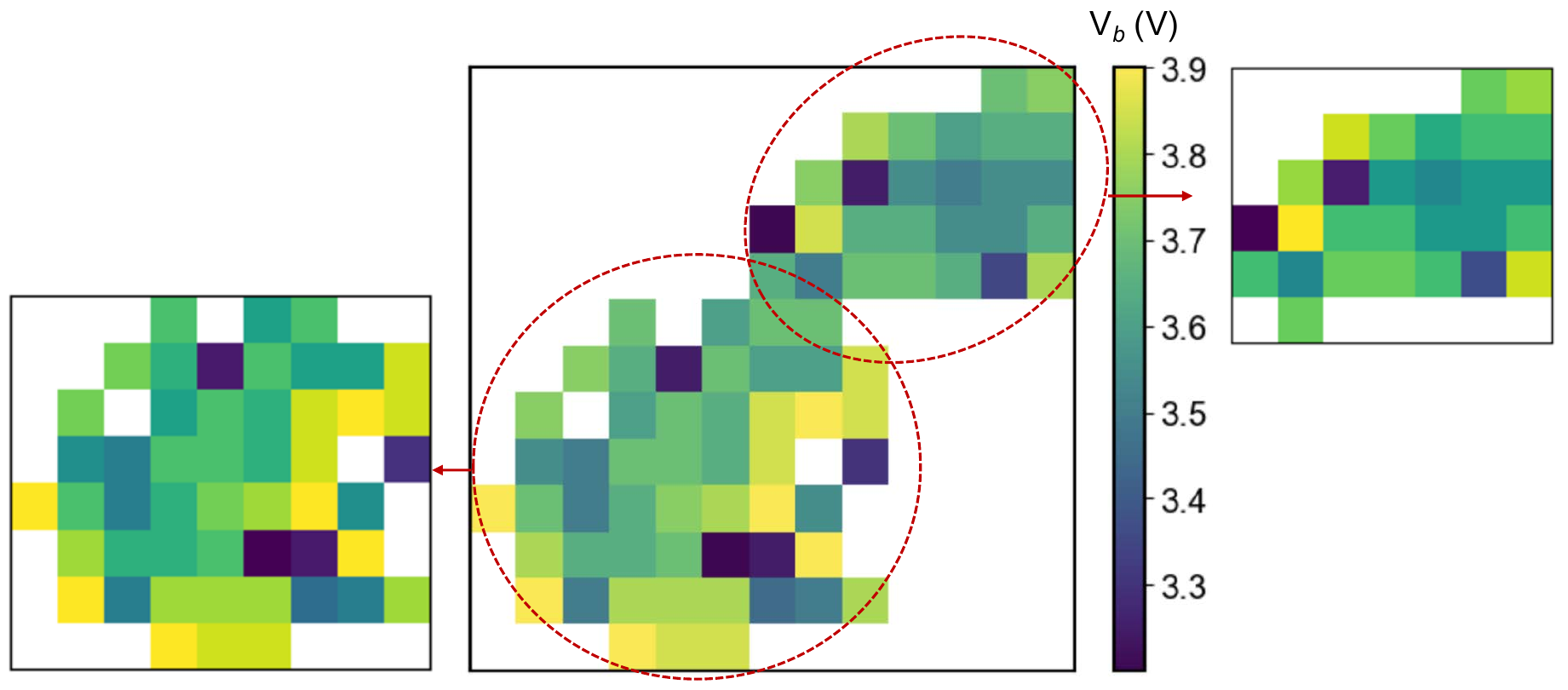}
    \caption{
    Gaussian Mixture Model.
    The SCA, described above, yields a potential defect cluster as shown in the center panel.
    The number $M$ of pixels in this cluster is $>$60, indicating that it probably contains more than one defect, as confirmed by its irregular shape.
    A GMM algorithm is applied to split the large cluster into $n=2$ smaller gaussians as shown on the left and right.
    Both of these smaller clusters are accepted as valid defects, based on the criteria described in the main text.}
    \label{fig:gaussian_mixture_subdefects}
\end{figure}

We apply the following procedure to split large clusters into smaller ones.
For clusters in the range of 50-60 pixels, we first apply the python routine \texttt{minEnclosingCircle} (see above) to determine the minimum circle enclosing the pixels constituting a defect. 
This is then used to determine the ``compactness'' of the defect, defined as the number of pixels in the cluster, divided by the the area of the minimum enclosing circle.
This compactness definition can also be thought of as ``circular-ness.'' 
If the compactness value is found to be lower than 0.6, indicating a non-circular shape, the GMM algorithm is applied. 
For clusters of size $>$60, the GMM is always applied.
The number $n$ of smaller clusters that a large cluster with $M$ pixels is split into depends on $M$: here, we choose $n=\lceil M/60\rceil$, where $\lceil \cdot\rceil$ is the integer ceiling function.
If one of the resulting small clusters has a pixel count $\geq$15 and a compactness value $\geq$0.4, it is accepted as a valid defect.
(We have found a compactness value of 0.6 to provide a good threshold for large defects with 50-60 pixels, while a compactness value of 0.4 is more appropriate for smaller defects.)
If its pixel count is $<$15, it is rejected and discarded. 
If the pixel count is $\geq$15 but its compactness value is $<$0.4, the GMM procedure is repeated once more, while setting $n=2$.
After this second iteration, if the defect still doesn't have a compactness value $>$0.4, we exclude from the compactness analysis any pixels that are only connected to the cluster by a corner, provided their total number doesn't constitute more than 25\% of the total. 
If the resulting compactness value is $>$0.4, the cluster is accepted.
If not, the GMM algorithm is repeated one more time.
Any clusters not meeting the appropriate criteria at this point are rejected.
A simple example of our GMM procedure is illustrated in Fig.~\ref{fig:gaussian_mixture_subdefects}

\begin{figure}[b]
    \centering
    \includegraphics[width=0.5\textwidth]{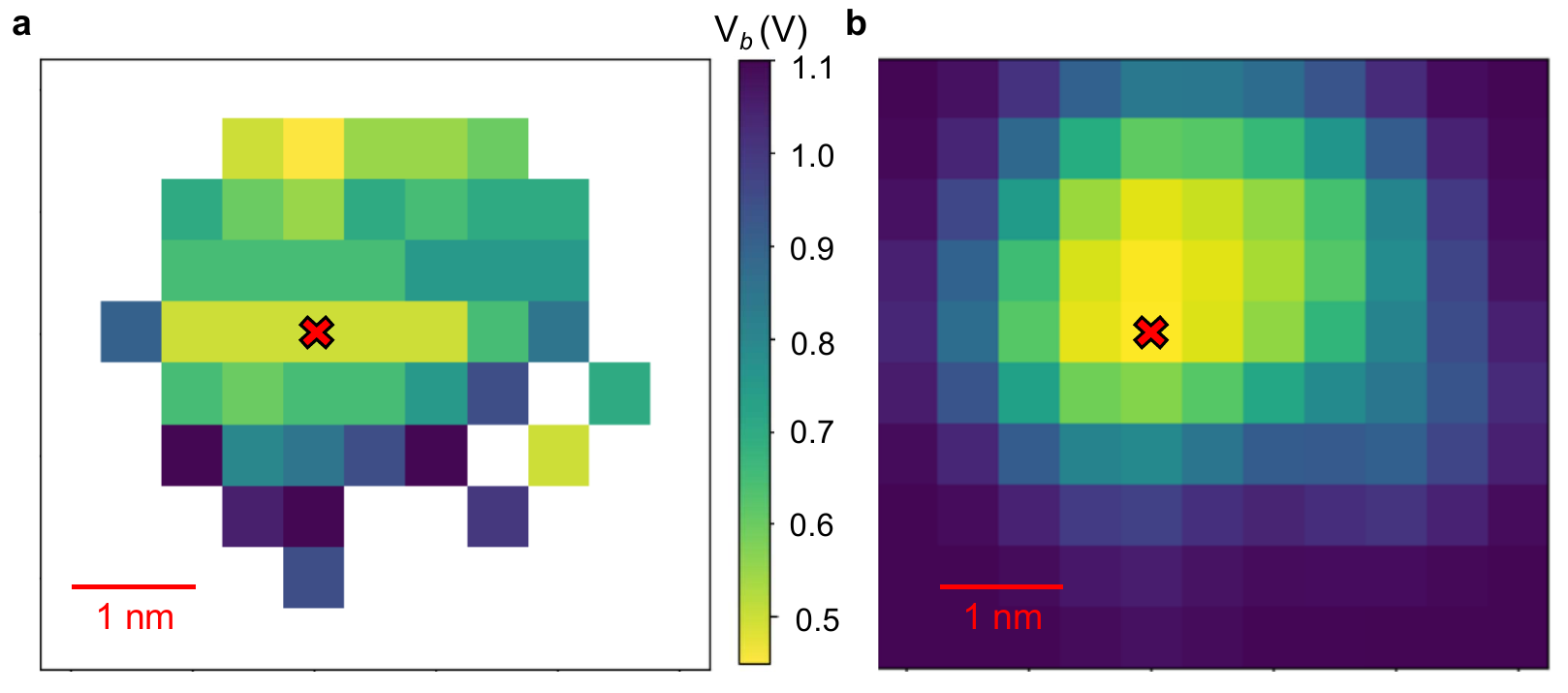}
    \caption{
    Identifying the center pixel of a defect.
    Following the CCA, SCA, and GMM procedures, the center pixel is determined for every defect-transition.
    (a) The resulting transition maps are noisy, complicating this task.
    (b) The same data is smoothed, using a gaussian blur technique, as described in the text.
    (The same color bar is used for both panels.)
    For a forward-bias sweep, the defect center is then chosen as the pixel with the least-positive or most-negative transition bias voltage.
    }
    \label{fig:gaussian_blur_determine_center}
\end{figure}

\subsection{Locating the defect center using a gaussian blur}
\label{sec:gaussian_blur}

It is important to identify the pixel at the center of a given defect, because the charging transition observed at this pixel should be minimally affected by hysteresis.
In a forward-bias sweep, this center point should occur at the lowest (i.e., the most negative or least positive) charging voltage of all the pixels comprising the defect.
However, such behavior can be obscured by hysteresis, stochastic tunneling, or experimental noise.
The ambiguity of the center point is apparent in Fig.~\ref{fig:gaussian_blur_determine_center}(a), where the map is even missing some pixels around the edges.

To overcome this problem, we apply a gaussian blur procedure to smooth out the data.
Specifically, we use the SciPy routine \texttt{gaussian\textunderscore filter}, adopting a standard deviation of $\sigma=1$~pixels for the gaussian, and a truncation width of 5~pixels.
A typical smoothed result is shown in Fig.~\ref{fig:gaussian_blur_determine_center}(b) for the same defect as Fig.~\ref{fig:gaussian_blur_determine_center}(a). 
After applying the blur, the defect center is identified as the pixel with the lowest voltage.

Using the combined techniques described above, our large-map method is able to confirm the presence of 15 defects across the $25\times 25$~nm$^2$ scan area and locate their center pixels.
In the analyses described below and in the main text, these 15 defects are labeled 1, 2, 4-13, 17, 19, and 20.

\begin{figure}[b]
    \centering
    \includegraphics[width=0.65\textwidth]{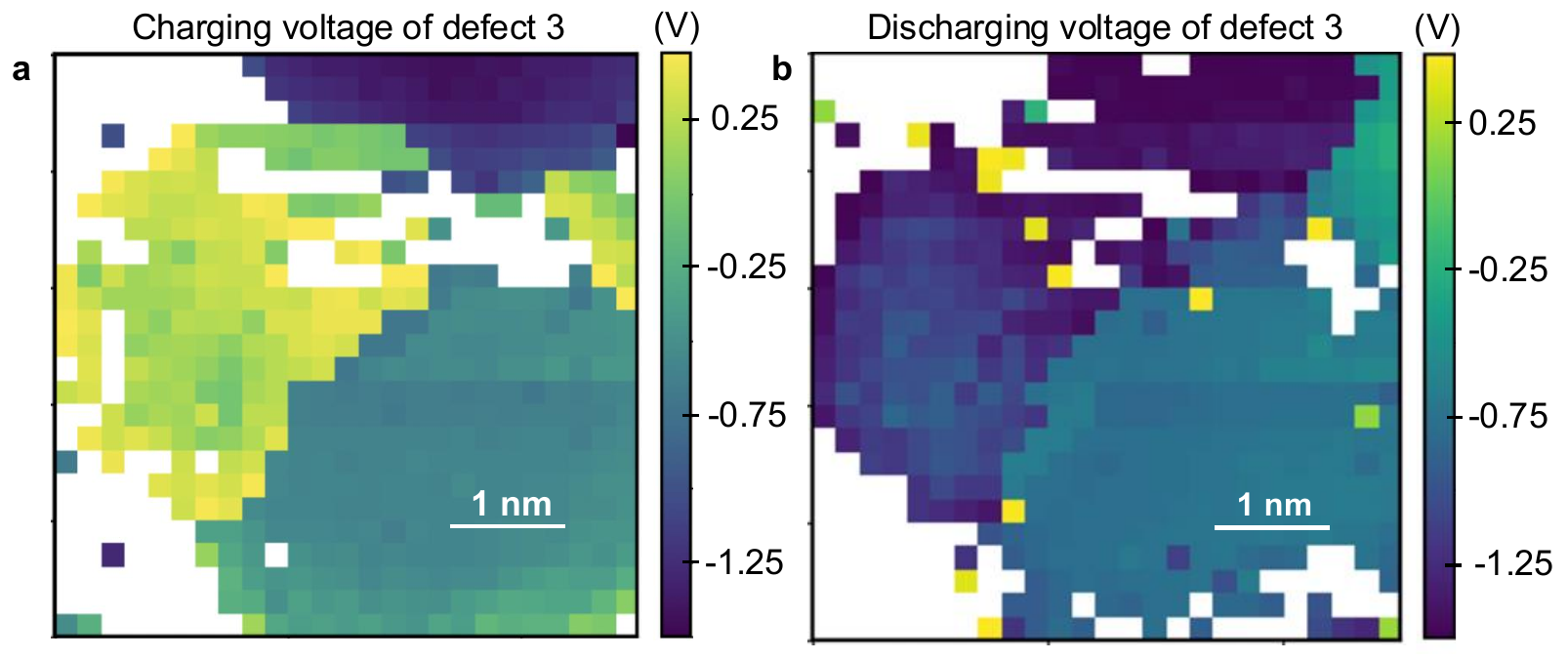}
    \caption{Charging-transition scans of defect 3 (at lower right), obtained at a pixel resolution of $0.2\times 0.2$~nm$^2$.
    (a) A charging scan, from a forward voltage sweep.
    (b) A discharging scan, from a backward voltage sweep.
    Other nearby defects can also be observed within the scan range.}
    \label{fig:highResData_defect1}
\end{figure}

\section{Identifying Individual Defects: Hysteresis Method }
\label{sec:Identifying_Defects2}

Following upon our new understanding of defect centers from the previous section, we now also apply a second method to identify defects and locate their centers.
This method involves acquiring finer scan date centered around the candidate defects.
Through repeated measurements, we identify the defect center as the location where the average transition-voltage hysteresis is minimized.
We then perform repeated voltage sweeps and vertical position scans at these center pixels, to obtain averaged results with  higher accuracy, and determine the corresponding standard deviations.
A total of five additional defects were identified this way, corresponding to the defects labeled  3, 14, 15, 16, and 18.
We now provide some details related to this method.

\subsection{High-resolution scans}
\label{sec:highres}

\begin{figure}[t]
    \centering
    \includegraphics[width=0.65\textwidth]{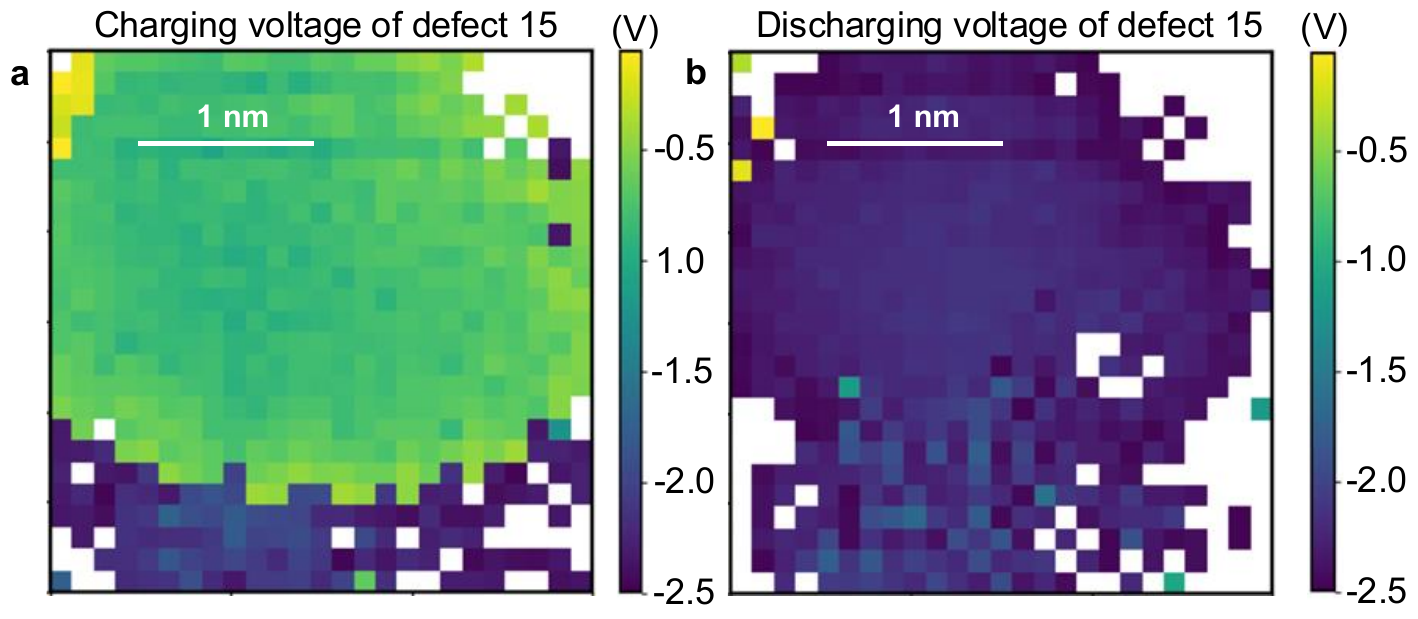}
    \caption{Charging-transition scans of defect 15 (at upper middle), obtained at a pixel resolution of $0.12\times 0.12$~nm$^2$.
    (a) A charging scan, from a forward voltage sweep.
    (b) A discharging scan, from a backward voltage sweep.
    Other nearby defects can also be observed within the scan range.}
    \label{fig:highResData_defect3}
\end{figure}

\begin{figure}[t]
    \centering
    \includegraphics[width=0.65\textwidth]{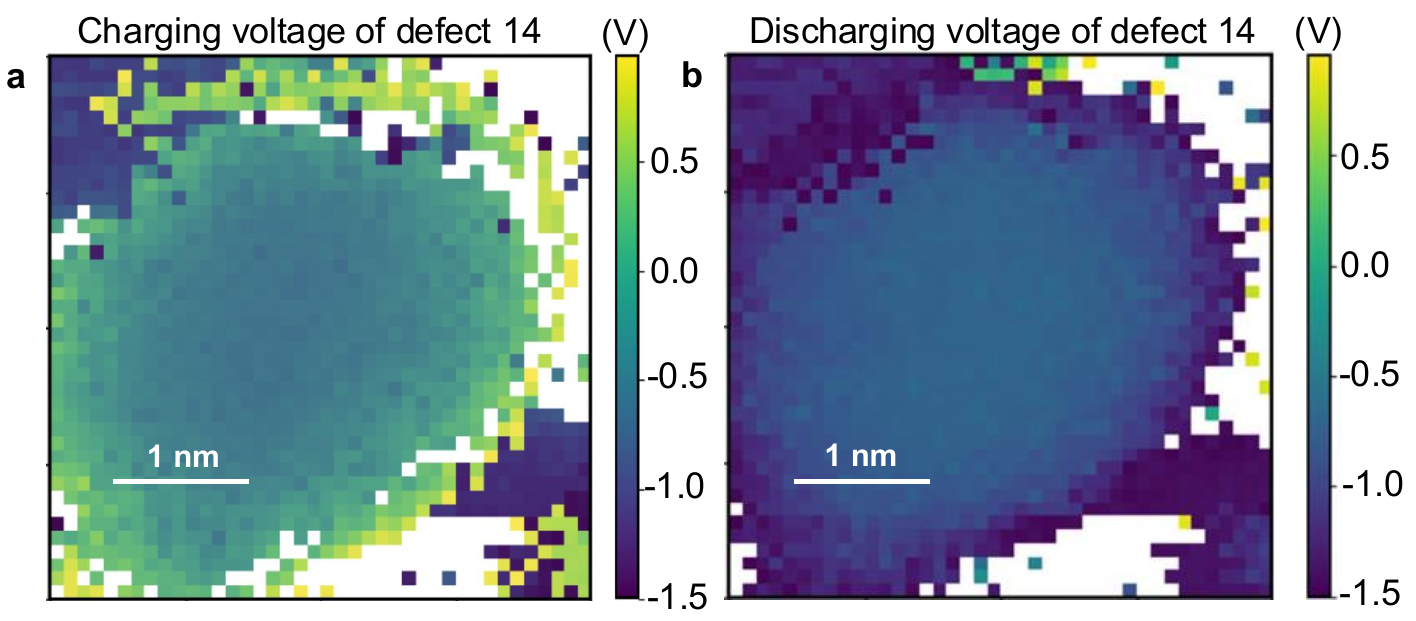}
    \caption{Charging-transition scans of defect 14 (at lower left), obtained at a pixel resolution of $0.1\times 0.1$~nm$^2$.
    (a) A charging scan, from a forward voltage sweep.
    (b) A discharging scan, from a backward voltage sweep.
    Other nearby defects can also be observed within the scan range.}
    \label{fig:highResData_defect4}
\end{figure}

Most of the scans in this work were performed at a pixel resolution of $0.5\times 0.5$~nm$^2$.
However, additional information is obtained from higher-resolution images, such as the $0.2\times 0.2$~nm$^2$, $0.12\times 0.12$~nm$^2$, and $0.1\times 0.1$~nm$^2$ pixel resolutions shown in Figs.~\ref{fig:highResData_defect1}-\ref{fig:highResData_defect4}, corresponding to Defects 3, 15, and 14, respectively.
As usual here, when multiple transitions are observed at a single pixel, only the lowest voltage transition is shown on the map. 
Additionally, we note these particular defects were chosen for further high-resolution study because they were already well-resolved. 
This helps to identify individual transitions, because the transition-identifying algorithms described in Sec.~\ref{sec:locating_transitions} can then be tuned by hand to provide very detailed images with great clarity, as seen in Figs.~\ref{fig:highResData_defect1}-\ref{fig:highResData_defect4}.
Such images provide insight into the spatial profile of the defects, including the transition voltage variations and the changes in hysteresis observed when moving away from the defect centers.
They also clearly illustrate the differences in the charging behaviors between forward (charging) scans and backward (discharging) scans.
Unlike the low-resolution results, the centermost pixel of the high-resolution images is relatively easy to to pick out, without applying a multistep procedure. 
These images are therefore also used to identify defects, together with the low-resolution images.

\begin{figure}[t]
    \centering
    \includegraphics[width=0.7\textwidth]{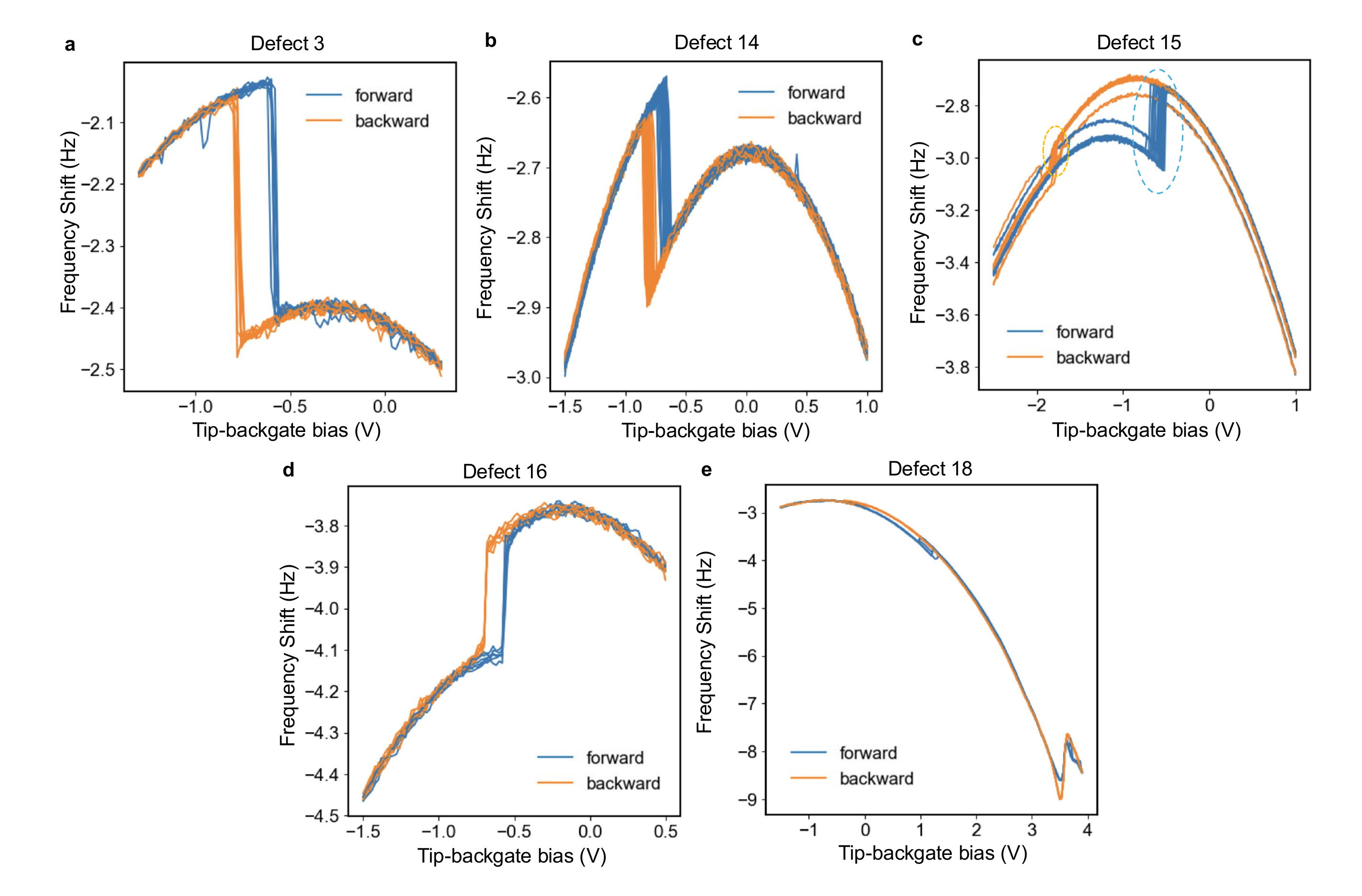}
    \caption{Forward (blue) and backward (orange) $f$-$V$ sweeps, with matched charging transitions, are obtained at five different tip-sample separations.
    (See Fig.~\ref{fig:highResData_chargingVoltages} for additional details.)
    Hysteresis is reflected by the fact that transitions occur at different voltages in the forward and backward sweeps.
    The hysteresis is found to increase with increasing tip-sample separations. 
    Here, if multiple transitions are present, we circle the main transition of interest. }
    \label{fig:highResData_curves}
\end{figure}

\begin{figure}[t]
    \centering
    \includegraphics[width=0.7\textwidth]{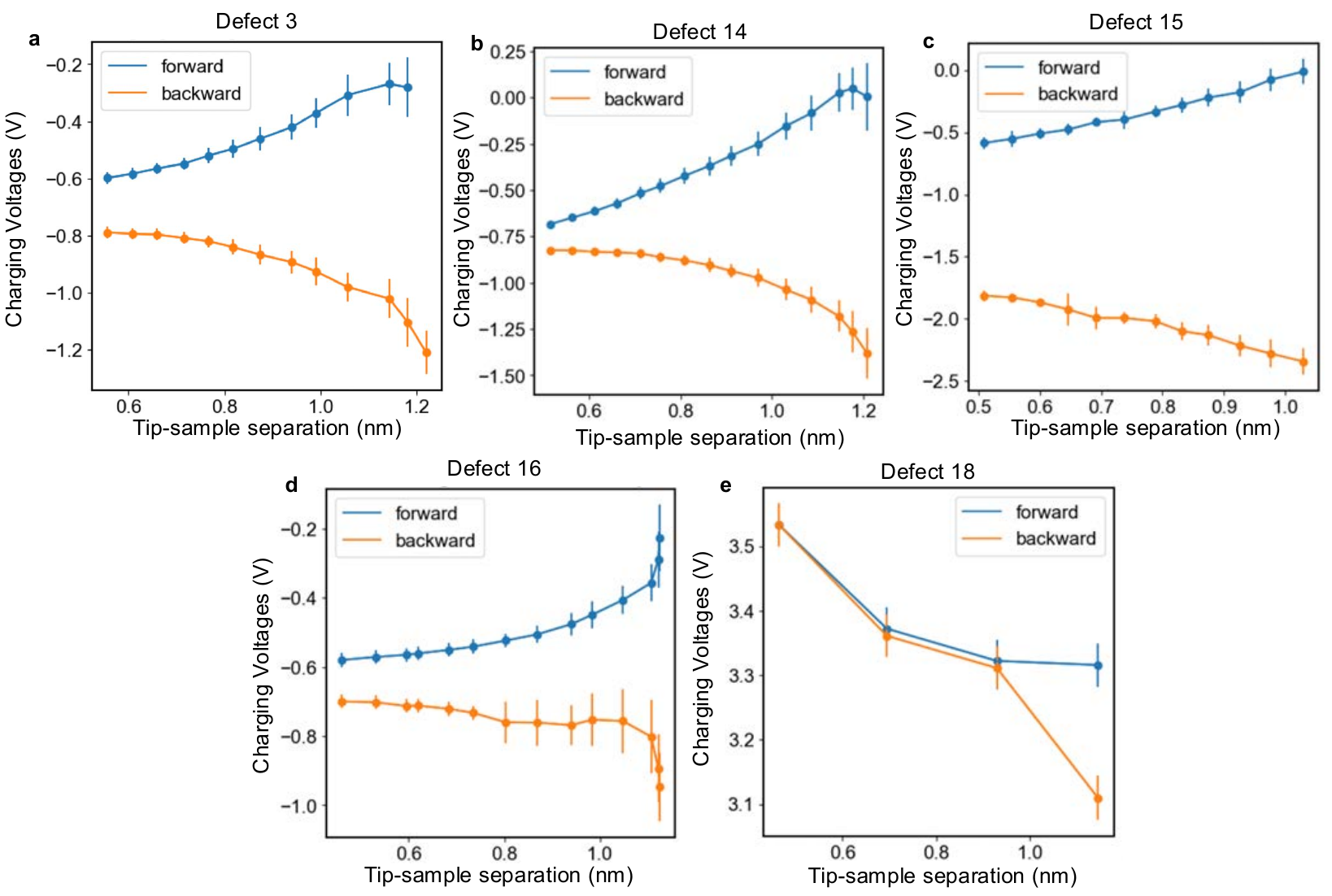}
    \caption{Charging transition voltages are reported for forward (blue) and backward (orange) $f$-$V$ sweeps, like those shown in Fig.~\ref{fig:highResData_curves}, for the same five defects studied there.
    Here, multiple sweeps are performed at each tip-sample separation, to obtain more-accurate averaged results, with error bars representing the standard deviation.
    The results reveal several interesting types of hysteretic behavior.  }
    \label{fig:highResData_chargingVoltages}
\end{figure}

\subsection{Dependence of hysteresis on tip height}

Figures~ \ref{fig:highResData_defect1}-\ref{fig:highResData_defect4} reveal charging-transition hysteresis as a function of tip-defect separation, as the tip is scanned laterally.
In this section, we also study hysteretic behavior as a function of tip-sample separation, while the tip is positioned directly above the defect.

In Fig.~\ref{fig:highResData_curves}, we plot $f$-$V$ curves for Defects 3, 14, 15, 16, and 18.
In each case, we perform forward and backward bias sweeps at the same tip-sample separation, observing matching charging transitions. 
We then repeat such voltage sweeps for five different tip-sample separations, observing changes in the transition voltages that can be ascribed to hysteresis.

In Fig.~\ref{fig:highResData_chargingVoltages}, we plot these forward and backward transition voltages for a wider range of tip-sample separations.
Interestingly, we observe a variety of hysteretic behaviors that could potentially yield useful information about the transitions and the chemical species of the defects, when combined with DFT calculations.
Such behavior will be explored in a future publication.

\section{Determining the sign of a charged defect}
\label{sec:sign}
\subsection{Evidence for local charge fluctuations}
\label{sec:localcharge}

\begin{figure*}[t]
    \centering
    \includegraphics[width=0.6\textwidth]{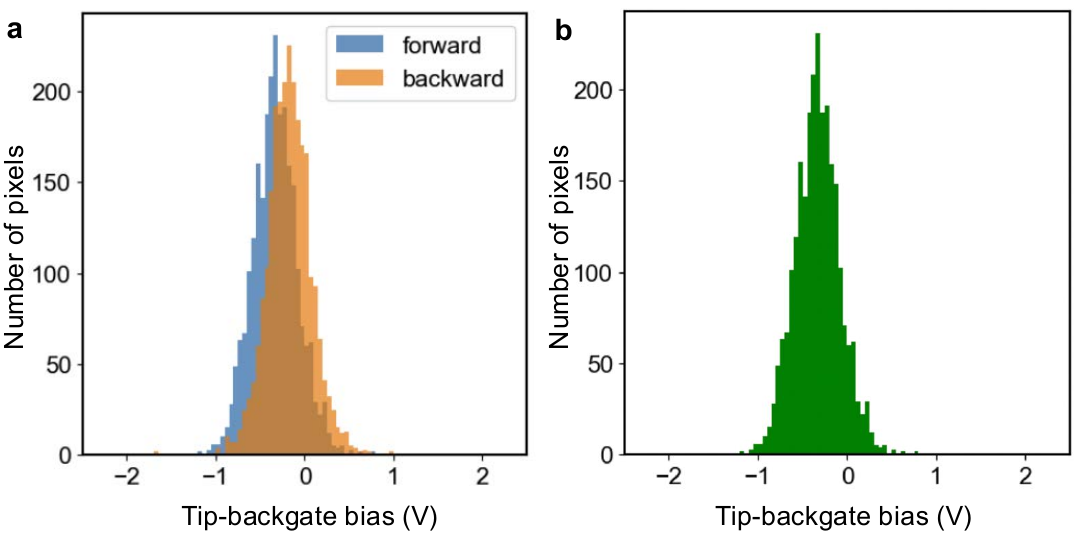}
    \caption{
    Histograms of CPD measurements obtained for every pixel in the $25\times 25$~nm${^2}$ scan range.
    (For this data, only one forward and one backward voltage sweep was performed at each pixel and no averages were performed.)
    (a) CPD values obtained from the forward (blue) and backward (orange) $f$-$V$ voltage sweeps.
    Note the presence of (i) hysteresis between the forward and backward sweeps, evidenced by the shifted distributions, and (ii) local fluctuations in the electrostatic landscape, caused by charged defects, as reflected in the widths of the distributions.
    The mean values of the blue and orange distributions are given by $\overline{V_\text{CPD}^\text{fwd}}=-0.33$~V and $\overline{V_\text{CPD}^\text{bwd}}=-0.17$~V, respectively.
    (b) The same histogram data as (a), where the CPDs from the forward and backward $f$-$V$ sweeps have been averaged.
    The mean value of this green distribution is given by $V_{\Delta\Phi}=-0.25$~V.}
    \label{fig:hist_cpds_fwd_bwd}
\end{figure*}

The CPD condition, $V_b=V_\text{CPD}$, corresponds to the backgate voltage at which an $f$-$V$ curve is maximized and $\partial f/\partial V=0$; it also represents the point where the tip-sample interactions vanish.
$V_\text{CPD}$ is found to vary as a function of the lateral position in an EFM scan across a sample, reflecting the presence of locally charged defects.
$V_\text{CPD}$ measurements also reveal hysteresis between forward and backward EFM bias sweeps, similar to the hysteresis observed in the charging transitions.
The local variations and hysteresis effects are evident in histograms of the CPD values shown in Fig.~\ref{fig:hist_cpds_fwd_bwd}(a).
Despite this variability, we argue here that, in the absence of tip-sample interactions, we should expect the total integrated charge across a sample to vanish.
It is therefore meaningful to compare locally varying CPD values to this globally averaged CPD value, defined as $V_{\Delta\Phi}=\overline{V_\text{CPD}}$, because the latter provides a reference value for quantifying the local fluctuations.
At the center of a defect, we assume that the local electrostatic fluctuations are dominated by the charge of the defect.
Thus, while the precise value of $V_\text{CPD}$ is affected by many nearby defects, the sign of $V_\text{CPD}-V_{\Delta\Phi}$ should reflect the sign of the charge state of the dominant defect, as measured by a nearby EFM tip. 

Determining the sign of the local charge state is complicated by the presence of hysteresis in the forward and backward sweeps, which introduces errors or uncertainty into the analysis.
Here, we separately consider the average CPD values for the forward and backward sweeps, $\overline{V_\text{CPD}^\text{fwd}}=-0.33$~V and $\overline{V_\text{CPD}^\text{bwd}}=-0.17$~V, such that $V_{\Delta\Phi}=(\overline{V_\text{CPD}^\text{fwd}}+\overline{V_\text{CPD}^\text{bwd}})/2=-0.25$~V.
For reference, we show in Fig.~\ref{fig:hist_cpds_fwd_bwd}(b) the histogram of CPD values obtained by averaging the forward and backward CPDs for a large number of sweeps.
We then declare the local charge state to be negative when $V_\text{CPD}>\overline{V_\text{CPD}^\text{bwd}}$, for either the forward or backward sweep, and we declare the local charge state to be positive when $V_\text{CPD}<\overline{V_\text{CPD}^\text{fwd}}$, for either the forward or backward sweep.

This leaves us with an exclusion zone defined by $\overline{V_\text{CPD}^\text{fwd}}<V_\text{CPD}<\overline{V_\text{CPD}^\text{bwd}}$, where we declare the local charge state to be approximately neutral, due to the uncertainty caused by hysteresis.
Note again, according to the definition above, that ``neutral'' refers to the case where the forward and backward CPDs both fall into the exclusion zone.
Here, we reject such neutral defects from further analysis, similar to other rejection criteria described in Secs.~\ref{sec:locating_transitions} and \ref{sec:Identifying_Defects}. 
We justify such an approach by again noting the uncertainty in our determination of a defect's charge state, due to hysteresis.
Additionally, we note that this rejection criterion helps to eliminate cases where two defects of opposite charge sign are observed in close proximity at a single pixel, since these would tend to approximately cancel.
It is preferable to ignore these cases because we find them difficult to classify via the procedures described here.
Finally, in addition to neutral defects, we also reject defects found to be positively charged in one voltage sweep direction and negatively charged in the opposite sweep direction, due to their ambiguous classification.

\subsection{Locating the CPD}

The identification of the CPD is complicated by the presence of charging transitions in the $f$-$V$ curves, as well as noisy data. 
Below, we describe our automated procedure for locating the CPD.

\begin{enumerate}
  \item The $f$-$V$ scan is first sectioned into regions separated by charging transitions, where the charging transitions were identified in Sec.~\ref{sec:locating_transitions}. 
  Each section of curve is fit separately to a polynomial of degree 10.
  The zeros of the first derivatives of this curve give a set of local extrema.
  From this set, we choose CPD values for which the second derivatives, evaluated at the extremum values, are negative.
  This gives a set of CPD candidates.
    
  \item If there is just one CPD candidate from this procedure, then it is accepted.

  \item If there are no candidates, we choose the voltage that maximizes the $f$-$V$ curve.  
  
  \item If there are multiple candidates, we reject any candidate that falls within 0.1~V of a charging transition, as the transitions are sometimes found to be broadened over several experimental voltage steps, which could lead to spurious CPD identifications. If there are still multiple candidates after this, we accept the candidate with a voltage closest to the maximum value of the $f$-$V$ curve.

\end{enumerate}

\section{Matching transitions in the forward and backward voltage sweeps}
\label{sec:matching_transitions}

For the defect analysis described in the main text, we only consider charging transitions that are observed in both the forward and backward $f$-$V$ scans.
This is important because of the hysteresis observed in many transitions. 
We therefore need to match corresponding transitions in the forward and backward scans, including for cases when multiple transitions are present.
As noted above, the ``fundamental'' charging transition voltage $V_\text{transition}$ is taken to be the average value of the forward and backward transitions, as determined for the closest separation between the defect and the EFM tip.
Although automated or semi-automated procedures were developed for many other aspects of our analysis, this matching procedure was performed by hand.
Fortunately, the matching only needs to be performed once per defect, so the human burden was not excessive.

\begin{figure}[htb!]
    \centering
    \includegraphics[width=0.5\textwidth]{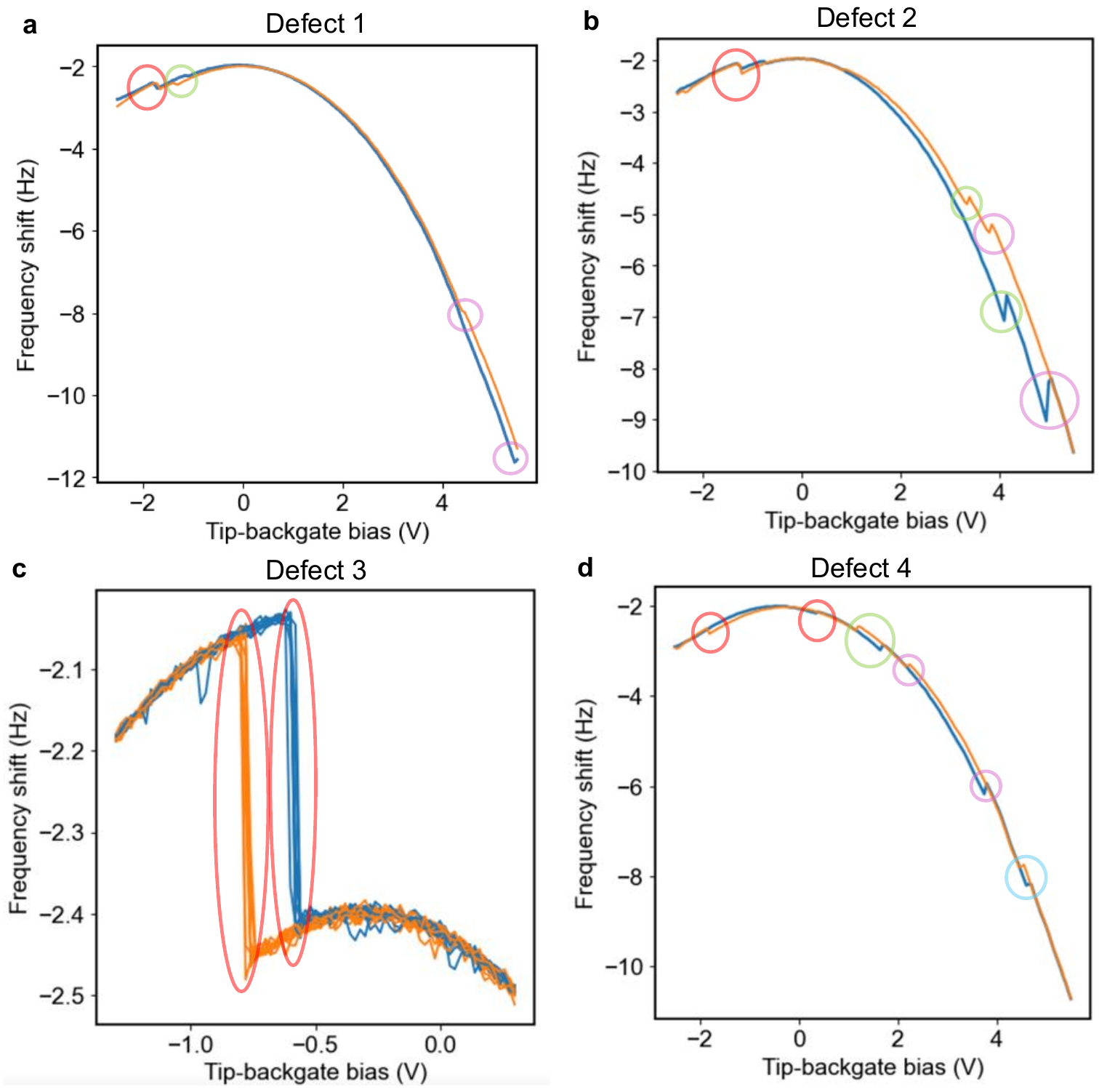}
    \caption{Matching charging transitions in forward (blue) and backward (orange) $f$-$V$ scans, for Defects 1-4.
    Circle colors indicate matched transitions; if transitions are closely separated, only one circle is shown.
    If multiple scans are shown, the transitions are assigned their average voltage values, for both forward or backward scans.
    Note that different vertical scales are used in the plots.}
    \label{fig:transitions_bwd_fwd_matched_part1}
\end{figure}

\begin{figure}[htb!]
    \centering
    \includegraphics[width=0.5\textwidth]{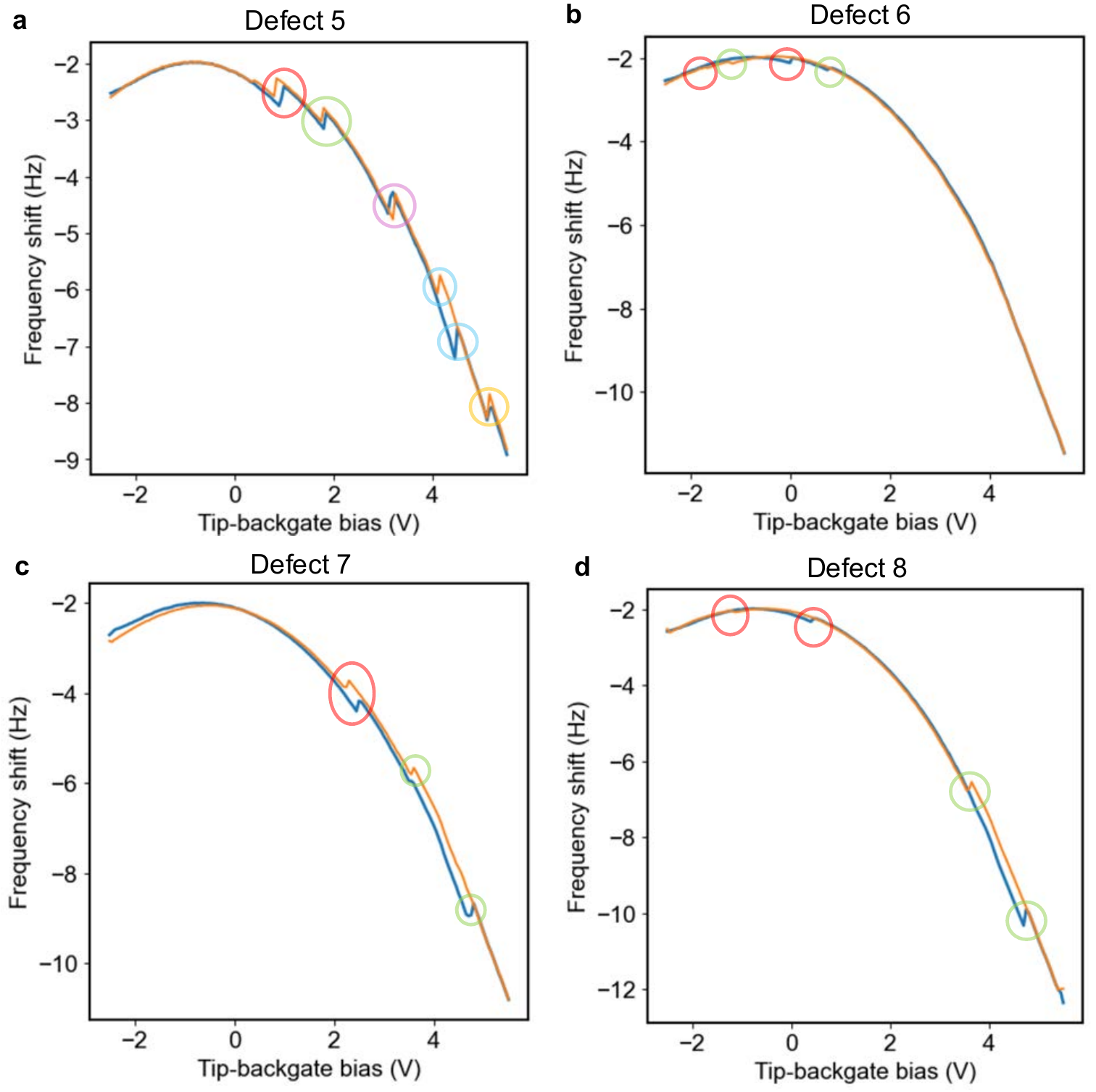}
    \caption{Matching charging transitions in forward (blue) and backward (orange) $f$-$V$ scans, for Defects 5-8.
    Circle colors indicate matched transitions; if transitions are closely separated, only one circle is shown.
    If multiple scans are shown, the transitions are assigned their average voltage values, for both forward or backward scans.
    Note that different vertical scales are used in the plots.}
    \label{fig:transitions_bwd_fwd_matched_part2}
\end{figure}

\begin{figure}[htb!]
    \centering
    \includegraphics[width=0.5\textwidth]{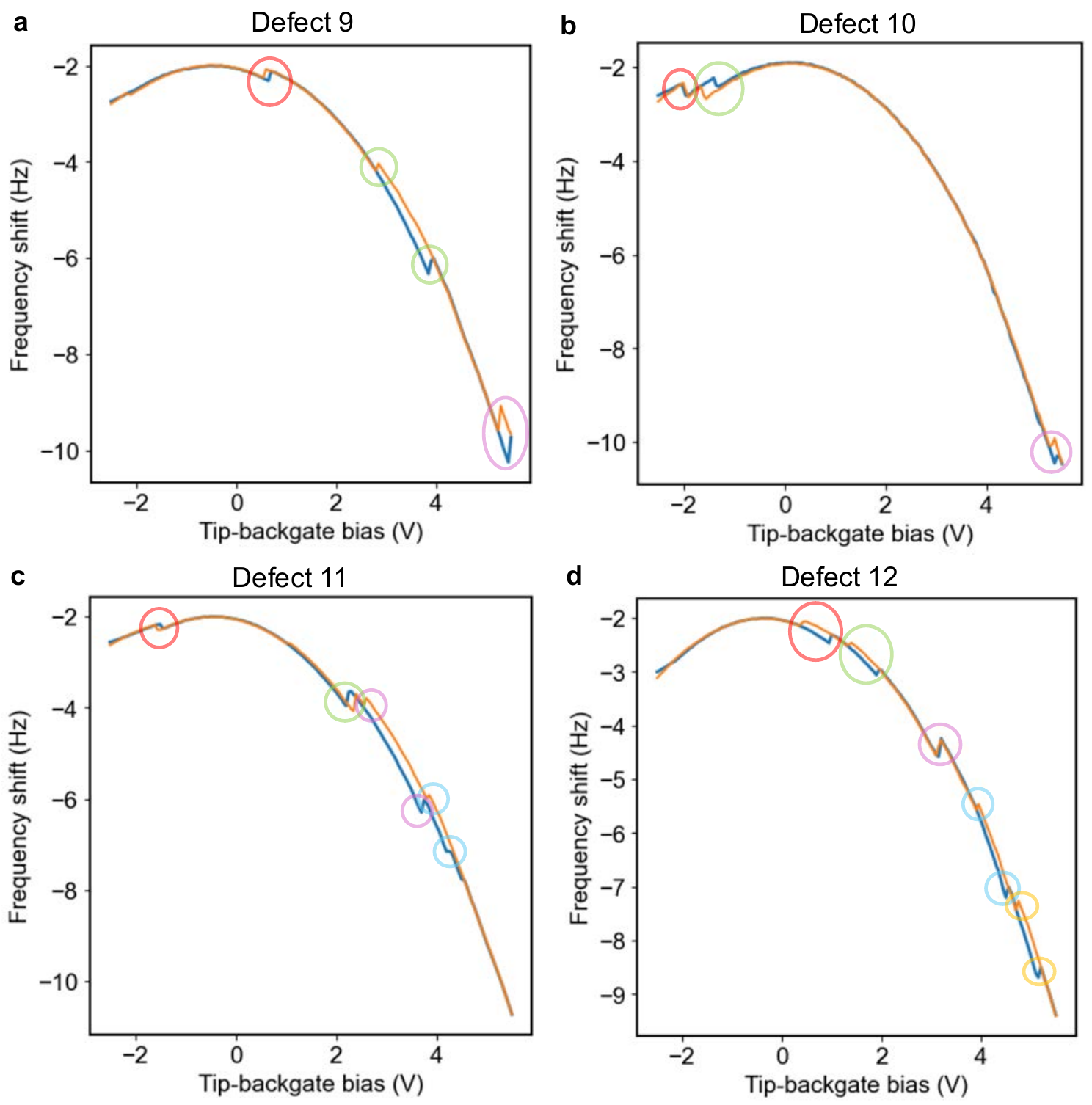}
    \caption{Matching charging transitions in forward (blue) and backward (orange) $f$-$V$ scans, for Defects 9-12.
    Circle colors indicate matched transitions; if transitions are closely separated, only one circle is shown.
    If multiple scans are shown, the transitions are assigned their average voltage values, for both forward or backward scans.
    Note that different vertical scales are used in the plots.}
    \label{fig:transitions_bwd_fwd_matched_part3}
\end{figure}

\begin{figure}[htb!]
    \centering
    \includegraphics[width=0.5\textwidth]{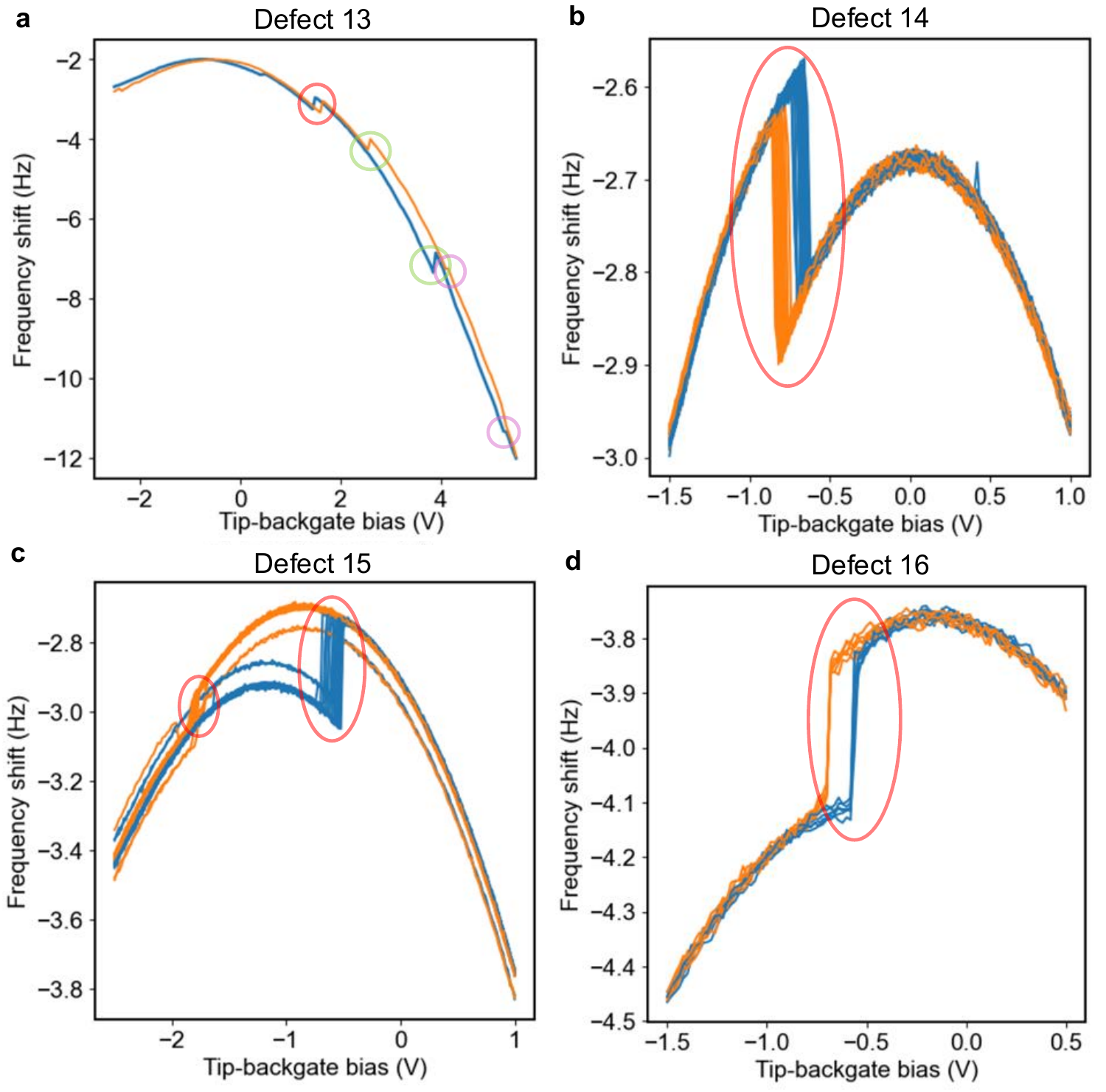}
    \caption{Matching charging transitions in forward (blue) and backward (orange) $f$-$V$ scans, for Defects 13-16.
    Circle colors indicate matched transitions; if transitions are closely separated, only one circle is shown.
    If multiple scans are shown, the transitions are assigned their average voltage values, for both forward or backward scans.
    Note that different vertical scales are used in the plots.}
    \label{fig:transitions_bwd_fwd_matched_part4}
\end{figure}

\begin{figure}[htb!]
    \centering
    \includegraphics[width=0.5\textwidth]{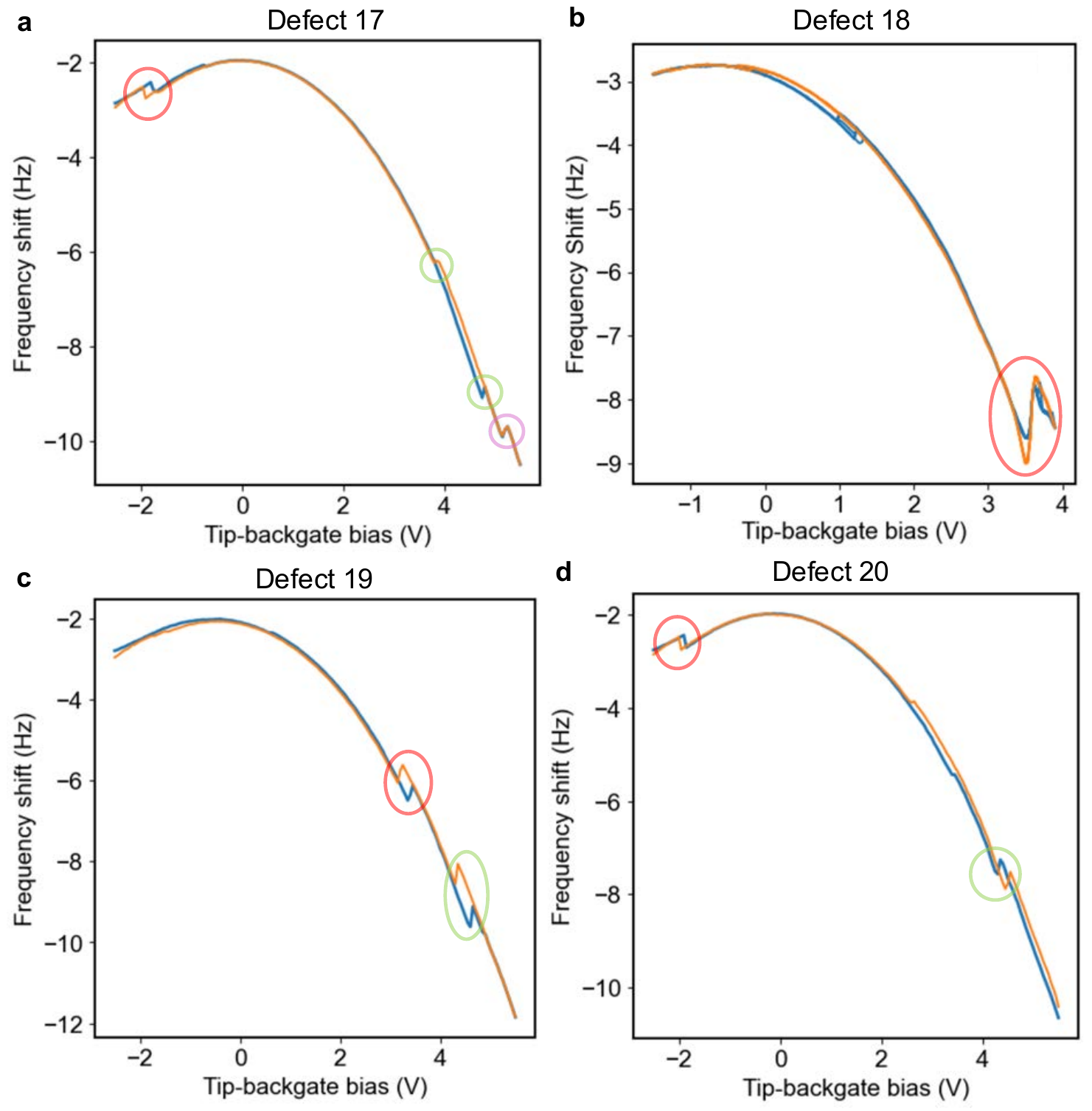}
    \caption{Matching charging transitions in forward (blue) and backward (orange) $f$-$V$ scans, for Defects 17-20.
    Circle colors indicate matched transitions; if transitions are closely separated, only one circle is shown.
    If multiple scans are shown, the transitions are assigned their average voltage values, for both forward or backward scans.
    Note that different vertical scales are used in the plots.}
    \label{fig:transitions_bwd_fwd_matched_part5}
\end{figure}

The procedure we use is as follows.
First, the center pixel of the defect is determined using methods described in previous sections.
At this center-pixel location, the charging transition signal is largest, and the hysteresis is smallest, due to close proximity of the tip to the defect.
Following the procedure described in Sec.~\ref{sec:locating_transitions}, valid transitions are identified and spurious transitions are rejected.
Matching then proceeds by considering both the size of the $f$-$V$ jumps and the proximity of the transitions along the voltage axis.
The procedure is somewhat self-explanatory, as seen in Figs.~\ref{fig:transitions_bwd_fwd_matched_part1}-\ref{fig:transitions_bwd_fwd_matched_part5}, where we show the $f$-$V$ curves for all 20 defects identified in this work.
Here, in all figures, the forward sweep is shown in blue, while the backward sweep is shown in orange.
For defects identified by the large-map method in Sec.~\ref{sec:Identifying_Defects}, only one forward and one backward voltage scan is obtained.
For defects identified by the hysteresis method in Sec.~\ref{sec:Identifying_Defects2}, multiple sweeps are obtained and plotted in the figures.
These repeated sweeps are used to obtain statistics for the transition voltages in forward or backward sweeps, including average values and standard deviations.  
In many cases, the matching is straight-forward (e.g., Defect 1).
In some cases, the size of the jumps in the forward and backward sweeps is not perfectly matched (e.g., Defect 7); however, the $f$-$V$ curves overlap well, outside the hysteretic region, giving more confidence in the matching assignment.
In cases where multiple transitions are present, we look for overlapping $f$-$V$ curves outside the hysteretic region.
In Figs.~\ref{fig:transitions_bwd_fwd_matched_part1}-\ref{fig:transitions_bwd_fwd_matched_part5}, matched transitions are indicated by the same color circles; if the transitions are very close together along the voltage axis, we only use a single circle.
We reject any transitions whose step size is less than 20\% of the largest transition, in either the forward or backward sweep; in such cases, it is assumed that the transition arises from a different, nearby defect (e.g., see Defect 10).

\section{Converting experimental charging-transition voltages $V_\text{transition}$ to defect energies $\Delta$}
\label{sec:converting_Vtransion}

\begin{figure}[h!]
    \centering
    \includegraphics[width=0.55\textwidth]{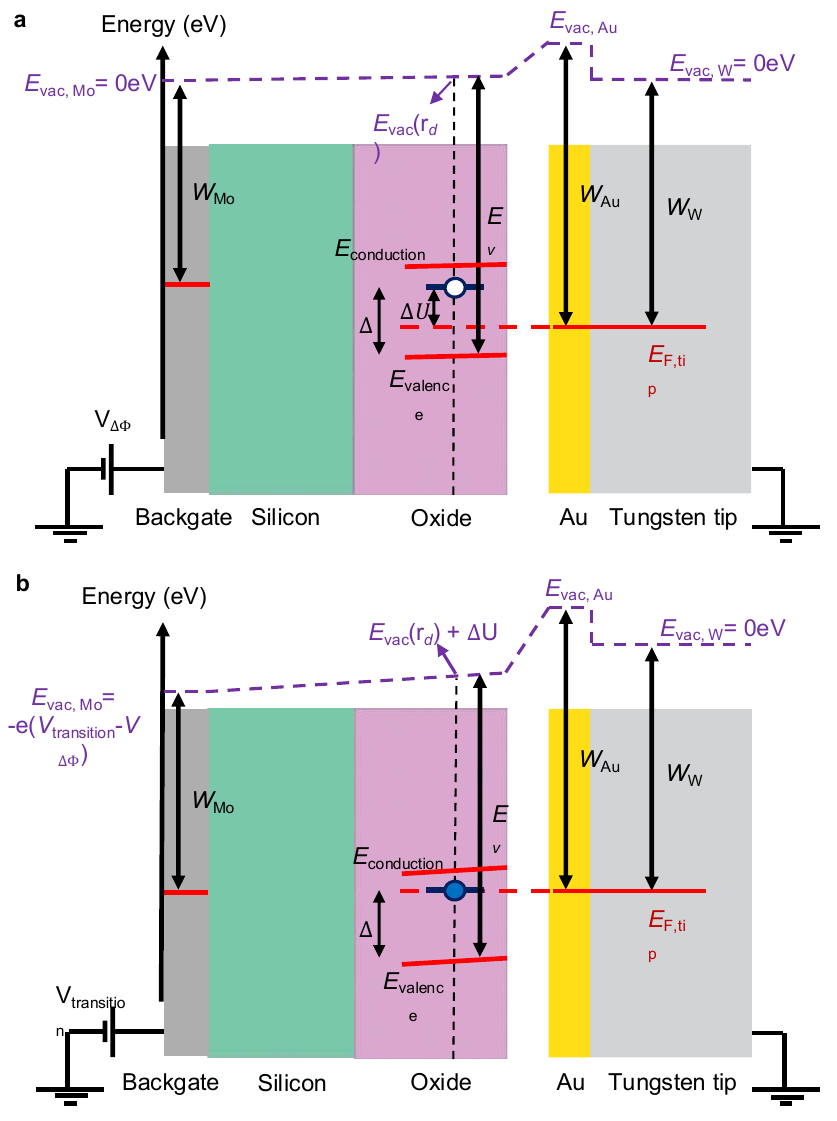}
    \caption{ 
    Defining two special values of the backgate bias: $V_{\Delta\Phi}$ and $V_\text{transition}$.
    (a) $V_{\Delta\Phi}$ corresponds to the special value that causes the vacuum levels of the tungsten portion of the tip and the backgate to align.
    To account for local fluctuations of the electrostatic potential, we take $V_{\Delta\Phi}$ to be the sample-averaged value of the CPD bias, since this should correspond to the sample-averaged charge-neutral state.
    When the voltage bias is given by $V_{\Delta\Phi}$, the vacuum energy $E_\text{vac}$ has the same value at the very top and very bottom of the device, representing a condition of vanishing tip-sample interaction.
    (b) $V_\text{transition}$ is the backgate bias that induces a charging event on the defect when its chemical potential aligns with the Fermi level of the tip, as described by Eq.~(\ref{defect_energy_eqn}).}
    \label{fig:backgate_0V_1V_defectEnergy}
\end{figure}

A charging transition occurs when the chemical potential of a defect crosses the Fermi level of the tip $E_{F,\text{tip}}$, and if the tip and defect are close enough for tunneling to occur.
The process is controlled by the tip-backgate voltage bias, as illustrated in Fig.~\ref{fig:backgate_0V_1V_defectEnergy}.
In this section, we explain the procedure for converting the experimentally determined transition bias voltage $V_\text{transition}$ to the chemical potential of the charging transition $\Delta$, where $\Delta$ is defined relative to the valence band edge, to allow for direct comparison with DFT predictions.

The conversion procedure involves relating $\Delta$ and $E_{F,\text{tip}}$ to the spatially varying vacuum energy $E_\text{vac}({\mathbf r})$, indicated by dashed purple lines in the figure.
Here, the sudden jump in $E_\text{vac}$ reflects the difference in work functions between the gold and tungsten portions of the tip while the slower changes in $E_\text{vac}$ reflect the spatially varying electrostatic potential energy.
The gold and tungsten work functions are given by $W_\text{Au}=5.30$~eV and $W_\text{W}=4.38$~eV, where $W_\text{Au}$ is reported in  \cite{sachtler1966work} and $W_\text{W}$ is measured in Sec.~\ref{AuMeasurements:simulations}, yielding the difference $ -(W_\text{Au}-W_\text{W})/e=-0.92$~V used in our simulations (for example, see Sec.~\ref{sec:Tip_Geometry}, and below).
The work function of the molybdenum backgate is taken to be $W_\text{Mo}= 4.63$~eV (see Sec.~\ref{AuMeasurements:simulations}). 
The energy difference between the vacuum level and the valence band edge (sometimes referred to as the ionization potential) is given by $E_v$=7.169~eV, as obtained by DFT and described in Materials and Methods.

Figure~\ref{fig:backgate_0V_1V_defectEnergy} illustrates two important bias conditions.
Figure~\ref{fig:backgate_0V_1V_defectEnergy}(a) shows the CPD condition, with the voltage bias $V_b=V_\text{CPD}$ corresponding to the point where the tip-sample interaction vanishes.
Due to the presence of charged defects, $V_\text{CPD}$ varies locally.
However, since our device is undoped, we expect the total integrated charge in the sample, including positively and negatively charged defects, to be zero at the CPD condition, since the sample as a whole is unaffected by the tip.
The globally averaged CPD voltage should therefore represent the bias for which the vacuum energies of the backgate and the tungsten portion of the tip line up.
In our experiments, this globally averaged CPD value is found to be $V_{\Delta\Phi}=\overline{V_\text{CPD}}\approx -0.25$~V, as also discussed in Sec.~\ref{sec:localcharge}.

With these definitions, we obtain the following relation between $V_\text{transition}$ and $\Delta$:
\begin{equation}
  \Delta = -e \alpha_{d}(V_\text{transition} - V_{\Delta\Phi}) + E_{v} - W_\text{W} - E_\text{vac}({\mathbf r}_d) .
 \label{defect_energy_eqn}
\end{equation}
Here, the lever arm $\alpha_d$ between the tip and the defect is computed as in Sec.~\ref{sec:lever_arm}, for the case when the tip is situated directly above the defect at a tip-sample separation of $h_\text{ts}=0.56$~nm, and $E_\text{vac}({\mathbf r}_d)$ is the vacuum energy at the location of the defect, when the backgate voltage bias is set to $V_b=V_{\Delta\Phi}$.
As usual in these simulations, we set the boundary condition on the tungsten portion of the tip to 0~V, while we set the gold portion of the tip to -0.92~V, to account for work function difference between the two metals, $W_\text{Au}-W_\text{W}$.
For simplicity here, since tunneling occurs over very short distances, we assume that the defect lies at the sample surface.
For these settings, we obtain $E_\text{vac}({\mathbf r}_d)$=0.120~eV and $\alpha_{d}$=0.243~eV/V.

Figure~\ref{fig:backgate_0V_1V_defectEnergy}(b) shows the bias condition corresponding to a charging transition where the chemical potential of the defect aligns with the Fermi level of the tip.
To compute the chemical potential $\Delta$ at this transition, we simply plug into Eq.~(\ref{defect_energy_eqn}) the experimentally determined value of $V_\text{transition}$, keeping all the other quantities in the equation fixed.

\section{Likelihood methods for assigning defect species}
\label{sec:likelihood}

\subsection{Likelihood methods}
\label{sec:likelihood_method}
We use the likelihood functional in Eq.~(\ref{eq:likelihood}) of the main text to provide our best guess for identifying the chemical species of a given defect.
The method is described in the main text, with additional details given below.

\emph{Single-defect identification.}
In the first set of columns in Table~1 of the main text, we list the probabilities for matching a subset of experimental and theoretical charging transitions, observed at the center pixel of a defect, to the predictions for a single defect species.
This identification is complicated by the fact that many $f$-$V$ scans exhibit transitions from multiple defects. 
Our identification method must therefore allow for experimental transitions that match the theoretical predictions, but are interspersed with unmatched transitions.
In principle, this is not a problem, since all the transitions in a given scan are analyzed, and low probability matches are simply discarded.
On the other hand, for each defect species considered, we insist that all theoretically predicted transitions falling within the experimental measurement window must be observed in the data. 

In all cases, only transitions with the correct charge sign are considered, as defined in Sec.~\ref{sec:sign}.
Thus, if the defect is determined to be positively charged, it can only be matched to the chemical species C$_\text{Al}$-H, C$_\text{O}$, C$_\text{O}$-H, or V$_\text{O}$, while a negatively charged defect can be matched to V$_\text{Al}$, V$_\text{Al}$-H, or V$_\text{Al}$-2H.
The likelihood of a given matching assignment is computed using the functional defined in Eq.~(\ref{eq:likelihood}) of the main text, where $i=1$ to $k$ enumerates the charging transitions of a given defect and $k$ is determined by the defect species.
We then convert this likelihood to a probability by computing the ratio of the maximum likelihood, corresponding to the case where each of the experimental measurements for the transition energy ($\Delta_i$) is equal to its theoretical estimate from DFT calculations ($E_i$). 
If $n$ is the number of charging transitions expected in the experimental range, for a given defect, then we must have $n\leq k$. 
We then compute the likelihood of all 
$\genfrac(){0pt}{2}{k}{n}$
transition matches and choose the result with the highest probability. 
As noted above, we allow for the possibility that a charging transition may not be observed if it falls close to the edge of the experimental sweep range.
Here, ``close'' refers to the experimental uncertainty $\rho_i$ [defined in Eq.~(\ref{defect_energy_uncertainty_eqn}), below] of the chemical potential assignment [Eq.~(\ref{defect_energy_eqn})], for a given transition. 
To account for this variable fitting range we apply the following strategy.
In the absence of experimental uncertainties, we have the default energy fitting window, given by $[E_\text{lower},E_\text{upper}]$, where $E_\text{lower}$ corresponds to the lower voltage bound of -2.5~V, and $E_\text{upper}$ corresponds to the upper voltage bound of 5.5~V.
Now including uncertainties, we consider an extended \emph{set} of fitting boundaries;
for example, the lower bound becomes $E_\text{lower}\rightarrow (E_\text{lower}-\rho_\text{lower},E_\text{lower},\text{ or }E_\text{lower}+\rho_\text{lower})$, and similarly for the upper bound, where $\rho_\text{lower}$ is the value of $\rho_i$ near $E_\text{lower}$ and $\rho_\text{upper}$ is the value of $\rho_i$ near $E_\text{upper}$.
Our procedure is then to fit defect transitions within each of these ranges, to obtain assignment probabilities, as described above.
Looking at all of these different results, we then choose the fit that gives the highest probability, which may or may not include edge-case transitions.
We note that for negatively charged defects, no transitions occur near $E_\text{upper}$, while for positively charged defects, no transitions occur near $E_\text{lower}$.
Tables S1 and S2 thus describe the defect transitions that fall within these different fitting ranges, for positively charged defects (Table~S1) or for negatively charge defects (Table~S2).
The resulting identifications based on this analysis are listed from best to worst in Table~1 of the main text, according to their probabilities.
We note that the ability to ignore some transitions could allow too much flexibility in assigning defects. 
This suggests applying a multiple-defect analysis where all charging transitions must be accounted for. 
On the other hand, the multi-defect scheme also has its own deficiencies, since it assumes that all possible defects have been characterized by DFT; hence, if one of the contributing defects has not been included in the DFT analysis, then none of the contributing defects can be properly identified.

\emph{Multiple-defect identification.}
In the second set of columns in Table~1, we list the probabilities for the simultaneous matching of multiple defects.
The matching procedure is similar to our previous approach for single defects; however in this case, we insist that all charging transitions observed in an $f$-$V$ sweep must be accounted for.
We do this by matching the experimental transitions to groups of theoretically predicted defects whose charging transitions have the same sign.
For the multiple-defect identification procedure, we insist that all theoretically predicted transitions falling within the experimental measurement window must be observed in the data, unless the theoretical prediction falls very close to the edge of the experimental range.
We define such edge cases as falling within a border of width $\rho_i$, as discussed above.
For other cases, where an expected transition is not observed, the match is considered to be unsuccessful. 
The defect groupings are also required to have the correct number of transitions in total.
Here again, we allow for the possibility of edge cases, such that transitions occurring within an experimental uncertainty of $\rho_i$, near the experimental measurement boundaries, may or may not be included in the analysis.
As for the single-defect analysis, the likelihoods of the individual defect matchings are computed and converted to probabilities. 
The probability of the defect \emph{grouping} is then obtained as the product of single-defect probabilities.

\begin{table}[]
\centering
\begin{tabular}{|c|c|} 
 \hline
 Upper bound (eV) & Chemical potentials at which \\ 
 & candidate defects   \\ 
 &  charge (eV) \\  
 \hline
  $E_\text{upper}-\rho_\text{upper}$ & $\text{C}_\text{Al}$-H : 3.35    \\
  & $\text{C}_\text{O}$ : 2.82 \\
  & $\text{C}_\text{O}$-H : 2.57 \\
  & V$_\text{O}$ : 3.2 \\
 \hline
 
 $E_\text{upper}$, $E_\text{upper}+\rho_\text{upper}$  & $\text{C}_\text{Al}$-H : 3.35   \\
  & $\text{C}_\text{O}$ : 2.82, 4.07 \\
  & $\text{C}_\text{O}$-H : 2.57, 4.06 \\
  & V$_\text{O}$ : 3.2, 4.1 \\
 \hline

\end{tabular}
\caption{ 
Chemical potentials computed by DFT, for the positively charged defects studied in this work.
As described in Sec.~\ref{sec:likelihood_method}, we consider three different upper bounds for the experimental measurement range, to account for experimental uncertainties.
For the upper bounds shown in the left column, we consider the DFT-computed transitions for different defect species shown in the right column, in both our single-defect and multi-defect analyses.
Here, $E_\text{upper}$ is the chemical potential corresponding to the upper bound of the experimental measurements, and $\rho_\text{upper}$ is the experimental uncertainty near this boundary. }
\label{table:table_positive_defects_bounds}
\end{table}

\begin{table}[]
\centering
\begin{tabular}{|c|c|} 
 \hline
 Lower bound (eV) & Chemical potentials at which \\ 
 & candidate defects   \\ 
 &  discharge (eV) \\  
 \hline
  $E_\text{lower}-\rho_\text{lower}$ & V$_\text{Al}$ : 1.84, 2.32, 2.81    \\
  & V$_\text{Al}$-H : 2.42 \\
  & V$_\text{Al}$-2H : 1.91 \\
 \hline
 
 $E_\text{lower}$ & V$_\text{Al}$ : 2.32, 2.81   \\
 & V$_\text{Al}$-H: 2.42 \\
 & V$_\text{Al}$-2H: 1.91 \\
 \hline
 $E_\text{lower} + \rho_\text{lower}$ & V$_\text{Al}$ :  2.32, 2.81   \\
 & V$_\text{Al}$-H: 2.42 \\
 \hline

\end{tabular}
\caption{ Chemical potentials computed by DFT, for the negatively charged defects studied in this work.
As described in Sec.~\ref{sec:likelihood_method}, we consider three different lower bounds for the experimental measurement range, to account for experimental uncertainties.
For the lower bounds shown in the left column, we consider the DFT-computed transitions for different defect species shown in the right column, in both our single-defect and multi-defect analyses.
Here, $E_\text{lower}$ is the chemical potential corresponding to the lower bound of the experimental measurements, and $\rho_\text{lower}$ is the experimental uncertainty near this boundary.   }
\label{table:table_negative_defects_bounds}
\end{table}

\subsection{Sources of experimental and theoretical uncertainty}
\label{sec:errors}
The likelihood functional makes use of estimates for both the experimental ($\rho_i$) and theoretical ($\sigma_i$) uncertainties, where $i=1$ to $k$ enumerates the charging transitions of a given defect and $k$ is determined by the defect species.
In Table~1 of the main text, we list the probabilities for assigning each of the identified defects to a particular chemical species.
As noted above, this is computed as the ratio of the likelihood for a given assignment, divided by its maximum value. 

The main theoretical uncertainties in the DFT calculations include the uncertainty in the band alignment between tungsten and the aluminum oxide and the uncertainty of the bandgap energy for the aluminum oxide.
The latter arises from our approximate treatment of the oxide as crystalline, although it is actually amorphous. 
In this work, we adopt a constant value of $\sigma_i=0.25$~meV for all charging transitions.

\begin{table}[b]
\centering
\begin{tabular} { |p{1.5cm}|p{3.8cm}|p{3.2cm}|p{3.0cm}|  } 
 \hline
 Quantity & Sources of   & Uncertainty in  & Total uncertainty \\ 
          &   uncertainty    &  constituent quantities  & of quantity    \\ 
 \hline
  $\alpha_{d} $ & Large tip radius and & 0.001 eV/V  & 0.05 eV/V \\
                & fine tip height  &        &       \\
  \cline{2-3}
                & Fine tip radius and  &  0.02  eV/V  &       \\
                & gold height      &        &        \\
    \cline{2-3}
                & Tip-sample separation  &    0.01 eV/V   &       \\ 
  \cline{2-3}
                & Defect depth  &    0.05  eV/V  &       \\
 \hline
   $E_\text{vac}({\mathbf r}_d)$ & Large tip radius and & 1e-6 eV & 0.1 eV \\
                & fine tip height  &        &       \\
  \cline{2-3}
                & Fine tip radius and  &   0.1 eV    &       \\
                & gold height      &        &        \\
    \cline{2-3}
                & Tip-sample  &    0.01 eV   &       \\
                & separation      &        &        \\ 
  \cline{2-3}
                & Defect depth  &   0.09  eV   &       \\
 \hline
  $V_\text{transition}$ & Transition Voltage in forward sweep, $V_\text{fwd}$ & Variable & For a single sweep: $ |V_{\text{fwd}} - V_{\text{bwd}}|/2; $  \\
  \cline{2-3}
                   & Transition Voltage in backward sweep, $V_{\text{bwd}}$ & Variable & For multiple sweeps: $ \sqrt{\Delta V_{\text{fwd}}^2 + \Delta V_{\text{bwd}}^2  }/2 $ \\ 
 \hline
 $V_{\Delta\Phi}$  & Variation in CPD measured across sample surface & 0.24 V & 0.24 V \\
 \hline
$W_\text{W}$  & Variation in CPD measured against gold & 0.1 eV&  0.1 eV \\
 \hline
\end{tabular}
\caption{
Error estimates for quantities appearing in Eq.~(\ref{defect_energy_eqn}).
Column 1: the quantities with uncertainties.
Column 2: the sources of uncertainty considered here.
Column 3: the computed uncertainty values.
As an example, it was found in Sec.~\ref{sec:R_and_htip} that the big-tip radius is given by $R=1800\pm200$~nm.
Furthermore, we can correlate uncertainties in $R$ with errors in $h_\text{tip}$, as seen in Fig.~\ref{fig:rms_err_bigTipRad}(b).
We therefore compute $\alpha_d$ as in Sec.~\ref{sec:R_and_htip} by considering pairs of $(R,h_\text{tip})$ values, with $R$ in the range of 1600-2000~nm, yielding uncertainties in the lever arm of $\delta\alpha_d=0.001$~eV/V, as shown in the first row of the table.
We also compute uncertainties in $E_\text{vac}$ in a similar fashion.
The uncertainties in $V_\text{transition}$ are determined by the differences in the values obtained between forward vs backward voltage sweeps, and they differ from pixel to pixel. 
They also depend on whether the uncertainties are calculated for pixels where only a single voltage sweep is performed, or for pixels where multiple sweeps are performed; the formulas used for each of these cases are shown in the next column.
The uncertainty in $V_{\Delta \Phi}$ is defined as the standard deviation of the globally averaged CPD value, as discussed in Sec.~\ref{sec:converting_Vtransion}.
The uncertainty in $W_\text{W}$ is determined as described in Fig.~\ref{fig:cpd_variations_gold}.
Column 4: the total uncertainty of a given quantity is computed from the uncertainties in column 3, combined in quadrature. 
Here, $\Delta V_{\text{fwd}}$ and $\Delta V_{\text{bwd}}$ refer to the uncertainties in the forward and backward transition voltages observed in the high-resolution sweeps shown in Fig.~\ref{fig:highResData_curves}. }
\label{table:table_uncertainty_contributions}
\end{table}

The experimental uncertainty has several contributions, including uncertainties in our estimates for the fine tip radius, the fine tip height, the height of the gold portion of the tip, and the big tip radius. 
As described in Sec.~\ref{sec:R_and_htip}, the quantities $R$ and $h_\text{tip}$ are interrelated, so their errors are also interrelated.
(We give an example of this in the caption of Table~\ref{table:table_uncertainty_contributions}, discussing the uncertainty calculations for $\alpha_d$.)
Similar relations apply to the quantities $r_\text{apex}$ and $h_\text{Au}$, as described in Sec.~\ref{sec:rapex_and_hAu}.
Additional experimental uncertainty sources are the tip-sample separation, the unknown defect depth, and the hysteresis of the charging transition voltages in the forward and backward bias sweeps.
Each of these sources contributes to the uncertainty of quantities appearing in Eq.~(\ref{defect_energy_eqn}), as described in Table~\ref{table:table_uncertainty_contributions}.
All of these effects are finally combined in quadrature to give the total experimental uncertainty:
\begin{equation}
\rho_i^2= (\Delta \alpha_{d})^2 (V_\text{transition} - V_{\Delta\Phi} )^2 + \alpha_{d}^2 \Delta V_\text{transition}^2 + \alpha_{d}^2\Delta V_{\Delta\Phi}^2  + (\Delta W_\text{W})^2 + [\Delta E_\text{vac}({\mathbf r}_d)]^2 ,
 \label{defect_energy_uncertainty_eqn}
\end{equation}
which differs for each charging transition.
Here, the $\Delta$ symbols indicate the error values for the various quantities.
Overall, we find that the first and fourth terms in Eq.~(\ref{defect_energy_uncertainty_eqn}) tend to dominate, although all the terms play a role.
For the large-map scans (Sec.~\ref{sec:Identifying_Defects}), we perform only one forward bias sweep and one backward bias sweep. 
In this case, the experimental uncertainty is defined as the observed difference in the charging and discharging transition voltages (i.e. the transition voltage hysteresis).
For the high-resolution scans (Sec.~\ref{sec:highres}), we perform multiple bias sweeps in each direction for higher accuracy.
We then compute the average charging and discharging voltages, the average transition voltage $V_\text{transition}$, and the corresponding standard deviation values $\Delta V_\text{fwd}$ and $\Delta V_\text{bwd}$ for each transition.

\begin{figure*}[t]
    \centering
    \includegraphics[width=0.6\textwidth]{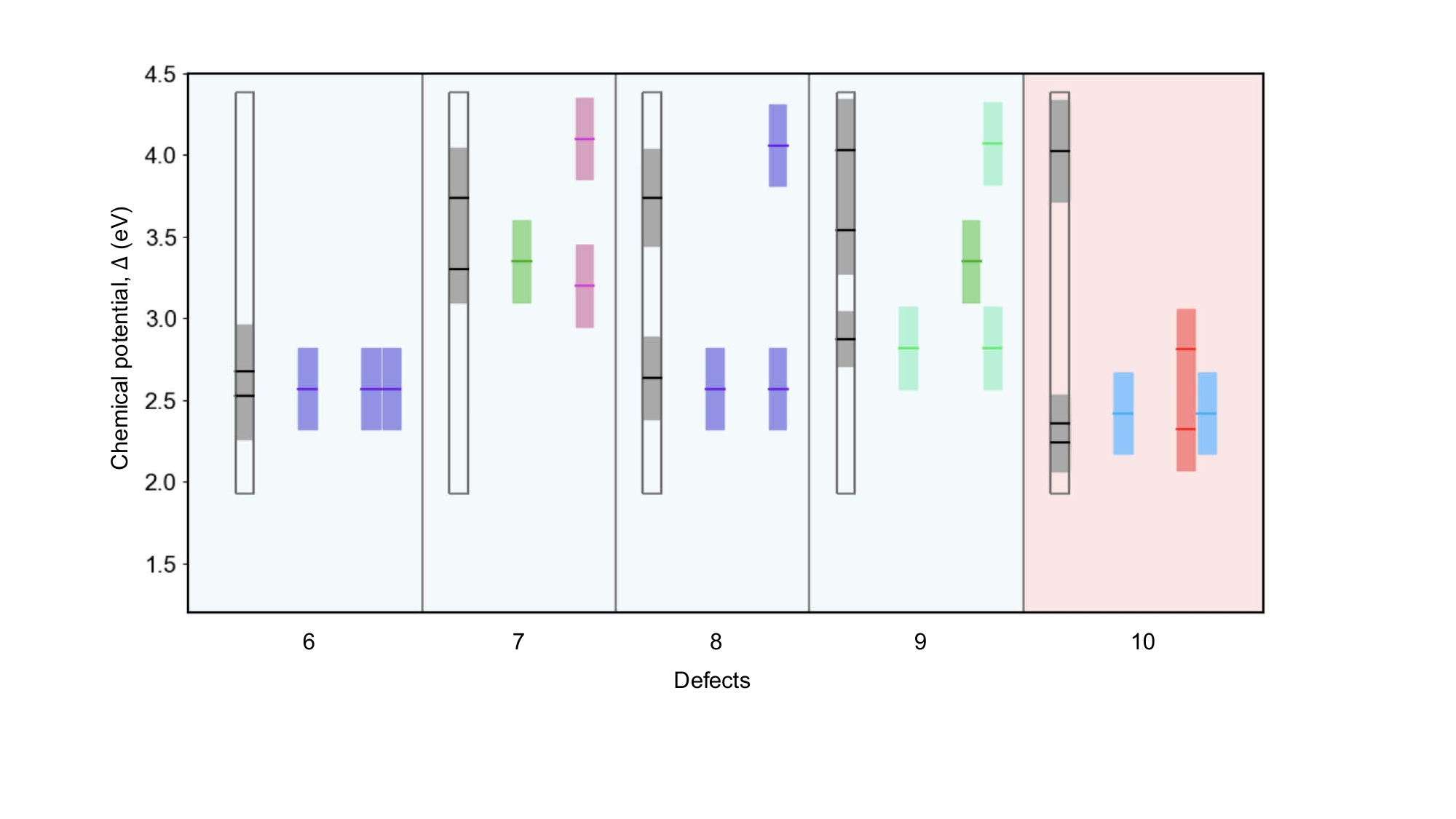}
    \caption{
    Chemical species identifications for Defects 6-10, using the same color codings as Fig.~\ref{fig:dft_energy_levels}\emph{A} and the same reporting scheme as Fig.~\ref{fig:dft_energy_levels}\emph{C}.
    For each defect, the first column shows the experimentally determined charging energy $\Delta$ (black line), with the corresponding experimental uncertainty level (gray box), and experimental measurement range (open black box).
    The second column corresponds to our best theoretical match to the experimental data based on the single-defect analysis (colored lines), with theoretical uncertainty levels shown (colored boxes).
    The third column corresponds to our best match based on the multi-defect analysis.
    Note that some of the colored boxes, describing theoretical results, fall outside the experimental measurement range, but are still shown here for completeness. }
    \label{fig:defects8to13}
\end{figure*}

\begin{figure*}[b]
    \centering
    \includegraphics[width=0.6\textwidth]{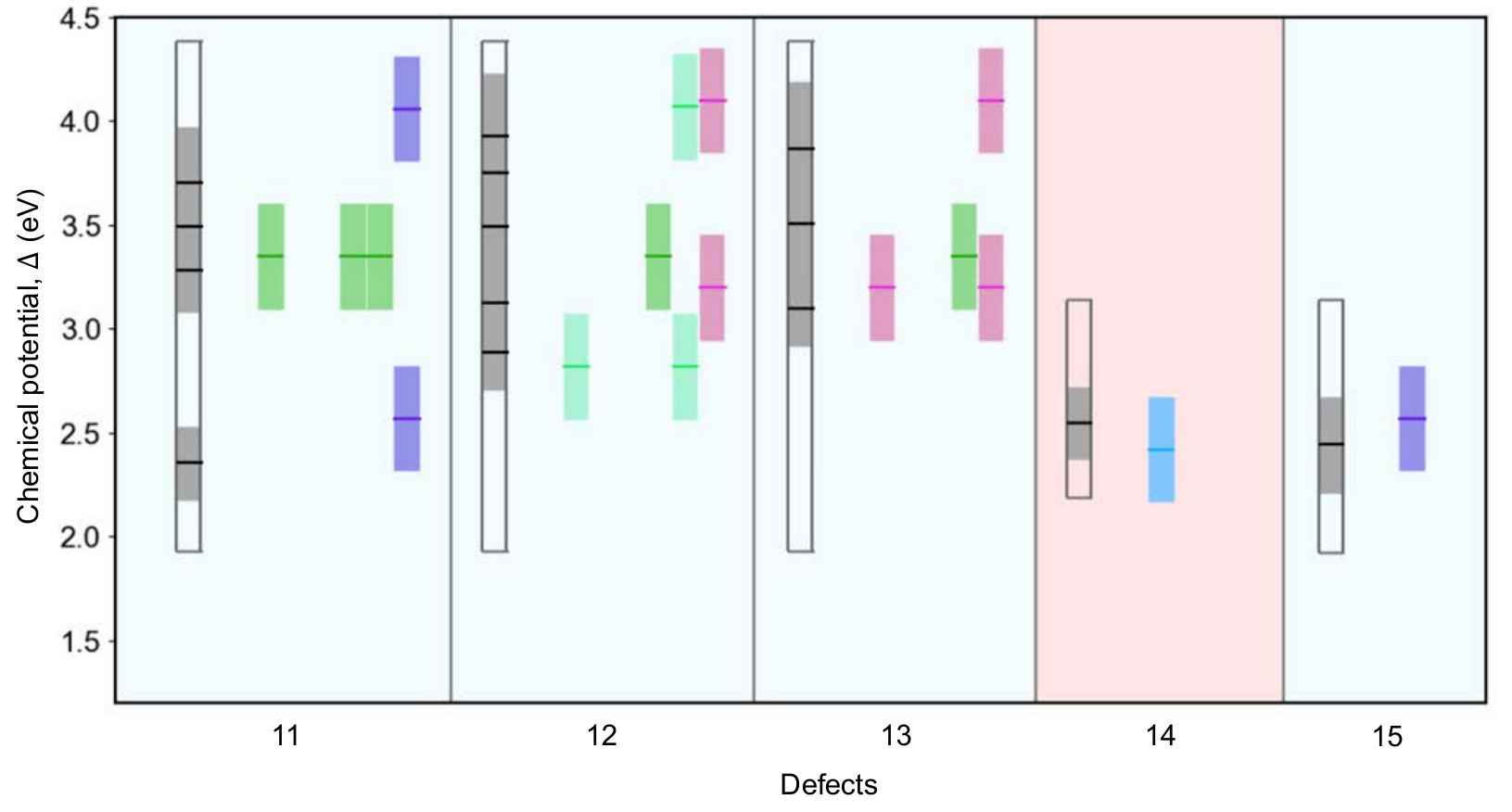}
    \caption{
    Chemical species identifications for Defects 11-15, using the same color codings as Fig.~\ref{fig:dft_energy_levels}\emph{A} and the same reporting scheme as Fig.~\ref{fig:dft_energy_levels}\emph{C}.
    For each defect, the first column shows the experimentally determined charging energy $\Delta$ (black line), with the corresponding experimental uncertainty level (gray box), and experimental measurement range (open black box).
    The second column corresponds to our best theoretical match to the experimental data based on the single-defect analysis (colored lines), with theoretical uncertainty levels shown (colored boxes).
    The third column corresponds to our best match based on the multi-defect analysis.
    Note that some of the colored boxes, describing theoretical results, fall outside the experimental measurement range, but are still shown here for completeness. }
    \label{fig:defects14to18}
\end{figure*}

\begin{figure*}[t]
    \centering
    \includegraphics[width=0.6\textwidth]{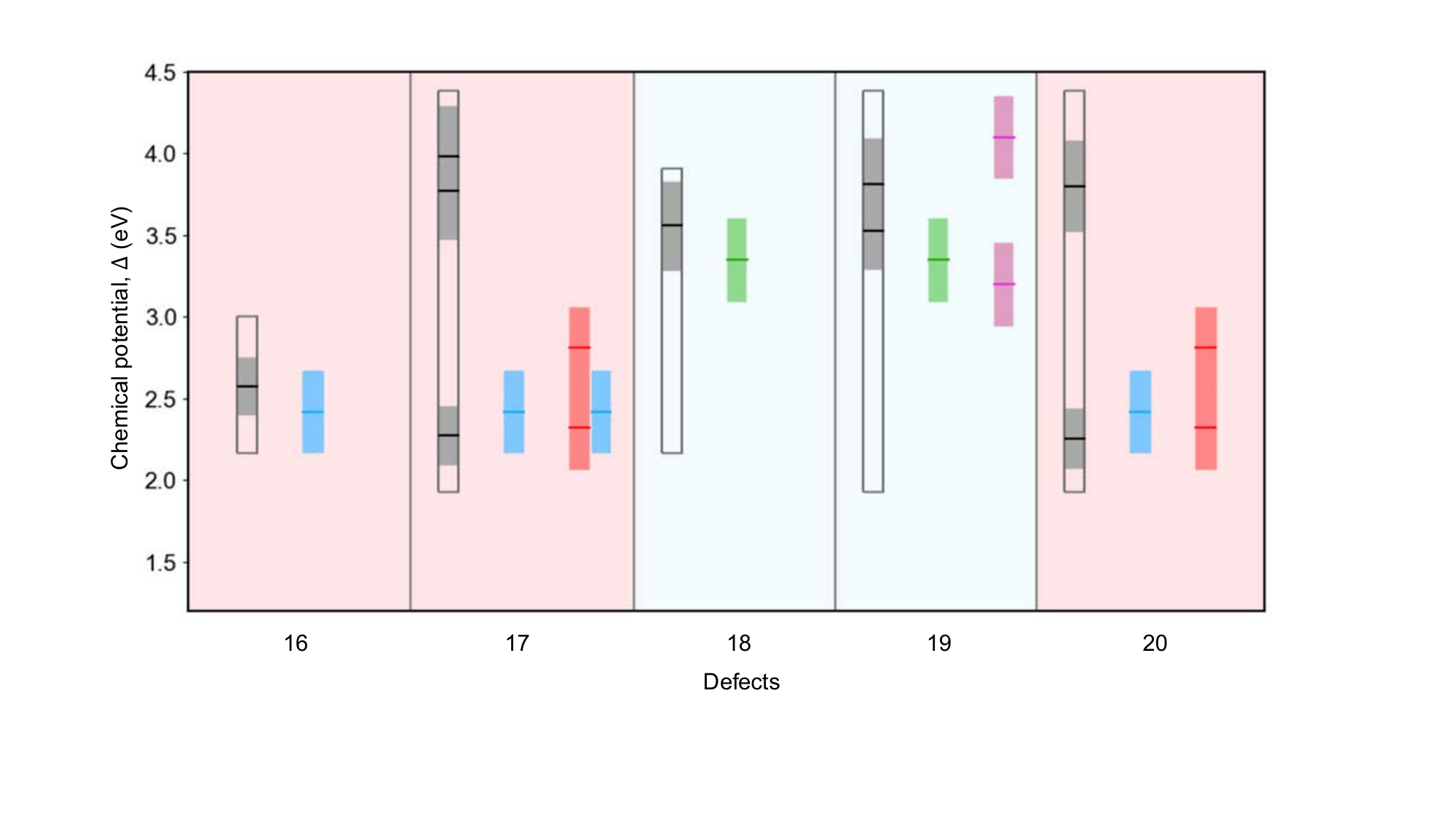}
    \caption{
    Chemical species identifications for Defects 16-20, using the same color codings as Fig.~\ref{fig:dft_energy_levels}\emph{A} and the same reporting scheme as Fig.~\ref{fig:dft_energy_levels}\emph{C}.
    For each defect, the first column shows the experimentally determined charging energy $\Delta$ (black line), with the corresponding experimental uncertainty level (gray box), and experimental measurement range (open black box).
    The second column corresponds to our best theoretical match to the experimental data based on the single-defect analysis (colored lines), with theoretical uncertainty levels shown (colored boxes).
    The third column corresponds to our best match based on the multi-defect analysis.
    Note that some of the colored boxes, describing theoretical results, fall outside the experimental measurement range, but are still shown here for completeness. }
    \label{fig:defects19to23}
\end{figure*}

\subsection{Chemical species assignments}

Our best guesses for the chemical species of Defects 1-5 are reported in Fig.~\ref{fig:dft_energy_levels} of the main text.
In Fig.~\ref{fig:dft_energy_levels}\emph{A}, the predicted transition energies for seven chemical species are shown with thick bars; color coding shows the corresponding charge states.
The same transition energies are shown as colored lines in the narrow bars on the right; here, the surrounding colored boxes show the theoretical uncertainties, and the colors of these narrow bars establish the color coding for the chemical species, which is used again in Fig.~\ref{fig:dft_energy_levels}\emph{C}.
In Fig.~\ref{fig:dft_energy_levels}\emph{A}, we note that the narrow boxes are only shown for transitions falling within the experimental measurement range.

\begin{figure*}[b]
    \centering
    \includegraphics[width=0.7\textwidth]{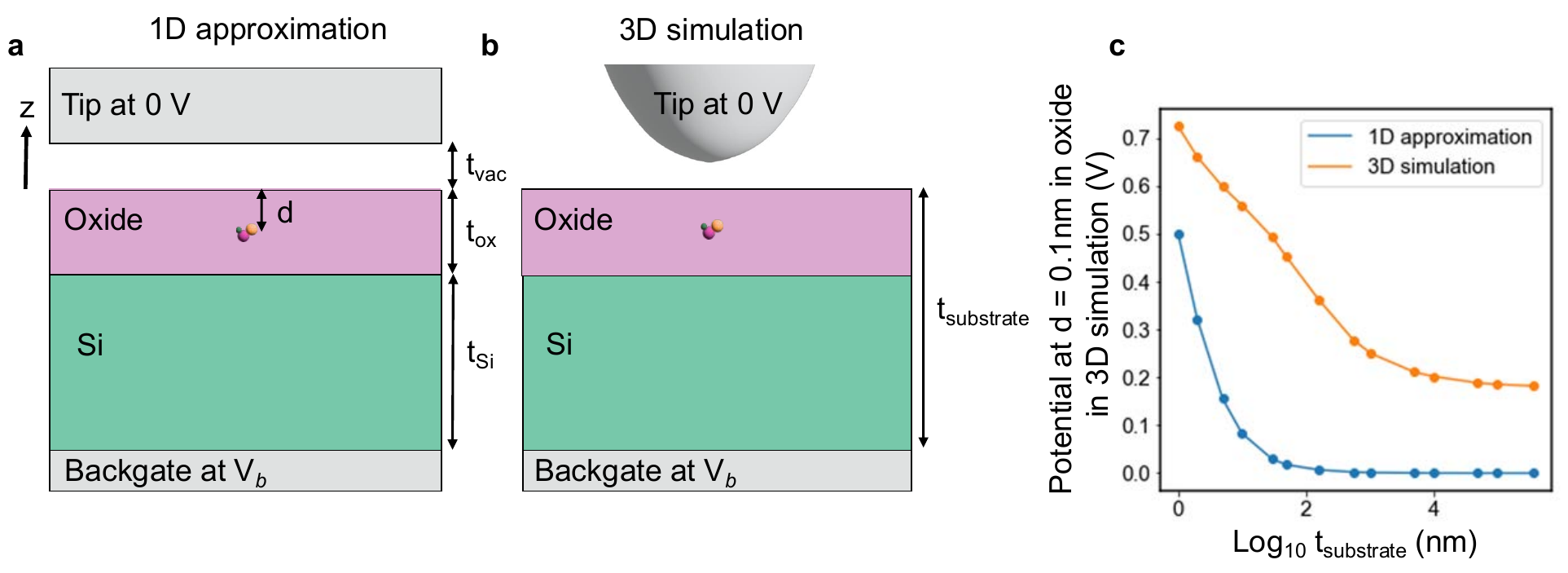}
    \caption{
    Simulations of the electrostatic potential beneath an EFM tip, for 1D or 3D system geometries, sampled at a fixed depth of $d=0.1$~nm below the oxide interface.
    (a) The 1D geometry, modeled as a series of parallel-plate capacitors, as described in Eq.~(\ref{eq:capacitor}).
    (b) The 3D geometry, with a two-section EFM tip, as simulated by COMSOL.
    (c) 1D and 3D simulation results as a function of the substrate thickness $t_\text{substrate}=t_\text{ox}+t_\text{Si}$.
    The results are plotted on a Log$_{10}$ scale in the range $t_\text{substrate}\in(1\text{~nm},381.06~\mu$m).
    We find that the 1D model underestimates the correct (3D) potential by orders of magnitude, for realistically thick substrates.
    Note that the 3D solution shows a small kink at $t_\text{substrate}\approx 30$~nm.
    This marks the crossover point where the wide portion of the tip begins to dominate over the thin portion of the tip.}
    \label{fig:varying_silicon_thickness}
\end{figure*}

\begin{figure*}[t]
    \centering
    \includegraphics[width=0.7\textwidth]{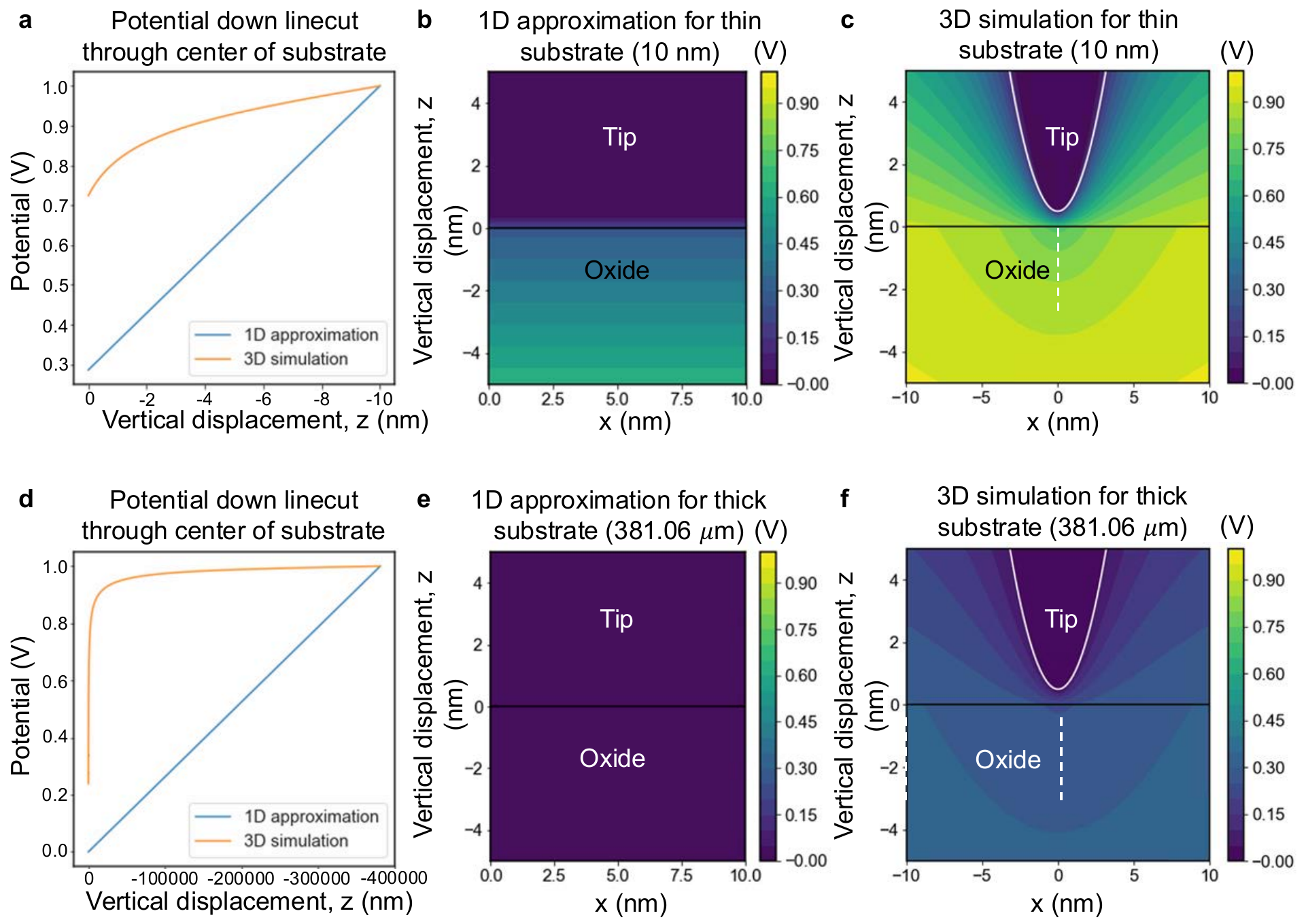}
    \caption{
    Simulations of the electrostatic potential beneath an EFM tip, for 1D or 3D system geometries.
    (a) Vertical linecuts, as a function of position $z$, obtained directly below the tip, for a sample with a thin substrate, $t_\text{substrate}=10$~nm.
    (b),(c) 2D solution cross-sections for the same geometry, as obtained for (b) the 1D model, or (c) the 3D model.
    Here, the dashed line indicates the location of the linecut evaluated in (a).
    (d) Vertical linecuts, as a function of position $z$, obtained directly below the tip, for a sample with a thick substrate, $t_\text{substrate}=381.06~\mu$m.
    (e),(f) 2D potential maps of the same geometry, for (e) the 1D model, or (f) the 3D model.
    Again, the dashed line indicates the location of the linecut evaluated in (d).
    As in Fig.~\ref{fig:varying_silicon_thickness}, the 1D model is seen to greatly underestimate the potential.}
    \label{fig:1d_capacitor_approx}
\end{figure*}

In Fig.~\ref{fig:dft_energy_levels}\emph{C}, we illustrate the experimentally measured transitions for Defects 1-5.
Here, the transition energies are shown as black lines, while the experimental uncertainties are shown as gray boxes, and the experimental measurement range is shown as open black boxes.
To the right, we show our theoretical matches for the corresponding chemical species, following the color coding of Fig.~\ref{fig:dft_energy_levels}\emph{A}.
The first column of colored boxes shows our best match based on the single-defect analysis while the second column shows the best match based on the multi-defect analysis. 
The remaining Defects 8-20 are reported in the same manner in Figs.~\ref{fig:defects8to13}-\ref{fig:defects19to23}.

\section{Comparing a 1D capacitor model to realistic device simulations}

In EFM experimental analyses, it is common to compute electrostatic interactions between the tip, sample, and backgate.
In cases where the dielectric material is thin and the tip is close to the sample surface, a 1D capacitor model provides a reasonable approximation for the potential underneath the tip~\cite{Fatayer2018}.
In this section, we show that such 1D approximations yield very poor results for the thick sample used in our experiments, and that more-realistic 3D device simulations are needed.

For the 1D system, we consider the parallel-plate capacitor model shown in Fig.~\ref{fig:varying_silicon_thickness}(a), where the vacuum, oxide, and Si substrate layers are defined by their dielectric constants $\epsilon_\text{vac}=\epsilon_0$, $\epsilon_\text{ox}=8.0\epsilon_0$, and $\epsilon_\text{Si}=11.4\epsilon_0$, respectively, and their corresponding layer thicknesses $t_\text{vac}$, $t_\text{ox}$, and $t_\text{Si}$.
The resulting electrostatic potential as a function of vertical position is then given by
\begin{equation}
V(z)=\left\{
\begin{array}{lc}
=V_b\frac{(t_\text{vac}-z)\epsilon_\text{ox}\epsilon_\text{Si}}
{t_\text{vac}\epsilon_\text{ox}\epsilon_\text{Si}+t_\text{ox}\epsilon_\text{vac}\epsilon_\text{Si}+t_\text{Si}\epsilon_\text{vac}\epsilon_\text{ox}}, &(z\in \text{vacuum}),  \\[9pt]
=V_b\frac{t_\text{vac}\epsilon_\text{ox}\epsilon_\text{Si}-z\epsilon_\text{vac}\epsilon_\text{Si}}
{t_\text{vac}\epsilon_\text{ox}\epsilon_\text{Si}+t_\text{ox}\epsilon_\text{vac}\epsilon_\text{Si}+t_\text{Si}\epsilon_\text{vac}\epsilon_\text{ox}}, &(z\in \text{oxide}), \\[9pt]
=V_b\frac{t_\text{vac}\epsilon_\text{ox}\epsilon_\text{Si}+t_\text{ox}\epsilon_\text{vac}\epsilon_\text{Si}-(z+t_\text{ox})\epsilon_\text{vac}\epsilon_\text{ox}}
{t_\text{vac}\epsilon_\text{ox}\epsilon_\text{Si}+t_\text{ox}\epsilon_\text{vac}\epsilon_\text{Si}+t_\text{Si}\epsilon_\text{vac}\epsilon_\text{ox}}, &(z\in \text{Si}),
\end{array} \right.
\label{eq:capacitor}
\end{equation}
where $z=0$ corresponds to the vacuum-oxide interface, as illustrated in Fig.~\ref{fig:varying_silicon_thickness}(a).
For the 3D model, we perform COMSOL electrostatic simulations of the two-section tip described above, for the same sample, as illustrated in Fig.~\ref{fig:varying_silicon_thickness}(b).
In both the 1D and 3D models, we ground the tip and apply a backgate voltage of $V_b=1$~V.

We compare the 1D and 3D models in two sets of simulations.
In Fig.~\ref{fig:varying_silicon_thickness}(c), we set $t_\text{vac}=0.1$~nm and solve the potential at a defect site, at location $z=-d=-0.1$~nm, directly below the tip.
We then plot the potential as a function of $t_\text{substrate}=t_\text{ox}+t_\text{Si}$ on a Log$_{10}$ scale, for $t_\text{substrate}\in(1\text{~nm},381.06~\mu$m), where the maximum value corresponds to the actual size of our experimental sample.
When $t_\text{substrate}\leq 60$~nm, we assume the substrate is formed only of oxide material, by setting $t_\text{Si}=0$.
When $t_\text{substrate}>60$~nm, we consider both oxide and Si materials and set $t_\text{ox}=60$~nm, noting that $t_\text{ox}=60$~nm is the correct oxide thickness for our sample.
The results of the simulations show that the 1D model provides a reasonable approximation of the actual potential only for very thin substrates, giving an error of about 30\%.
However for substrate thicknesses $>$100~nm, the 1D approximation is already orders of magnitude too small.

A second set of simuations is shown in Fig.~\ref{fig:1d_capacitor_approx}.
Here, we plot the 1D and 3D electrostatic potentials obtained directly below the tip, as a function of $z$, for two different sample geometries.
In Figs.~\ref{fig:1d_capacitor_approx}(a)-(c), we consider a thin substrate $t_\text{substrate}=10$~nm.
In Figs.~\ref{fig:1d_capacitor_approx}(d)-(f), we consider a thick substrate with $t_\text{substrate}=381.06~\mu$m.
In all cases, we set $t_\text{vac}=0.5$~nm.
Panels (b), (c), (e), and (f) show 2D cross-sections of the potential, while (a) and (d) show vertical linecuts directly below the tip.
Again, the results demonstrate that the 1D approximation vastly underestimates the potential, except for $z$ values very near the backgate. 
The results are especially poor for thick substrates.

\section{Matching Multiple Transitions to a Single Defect}
\label{sec:matching_multiple}

\begin{figure*}[t]
    \centering
    \includegraphics[width=0.8\textwidth]{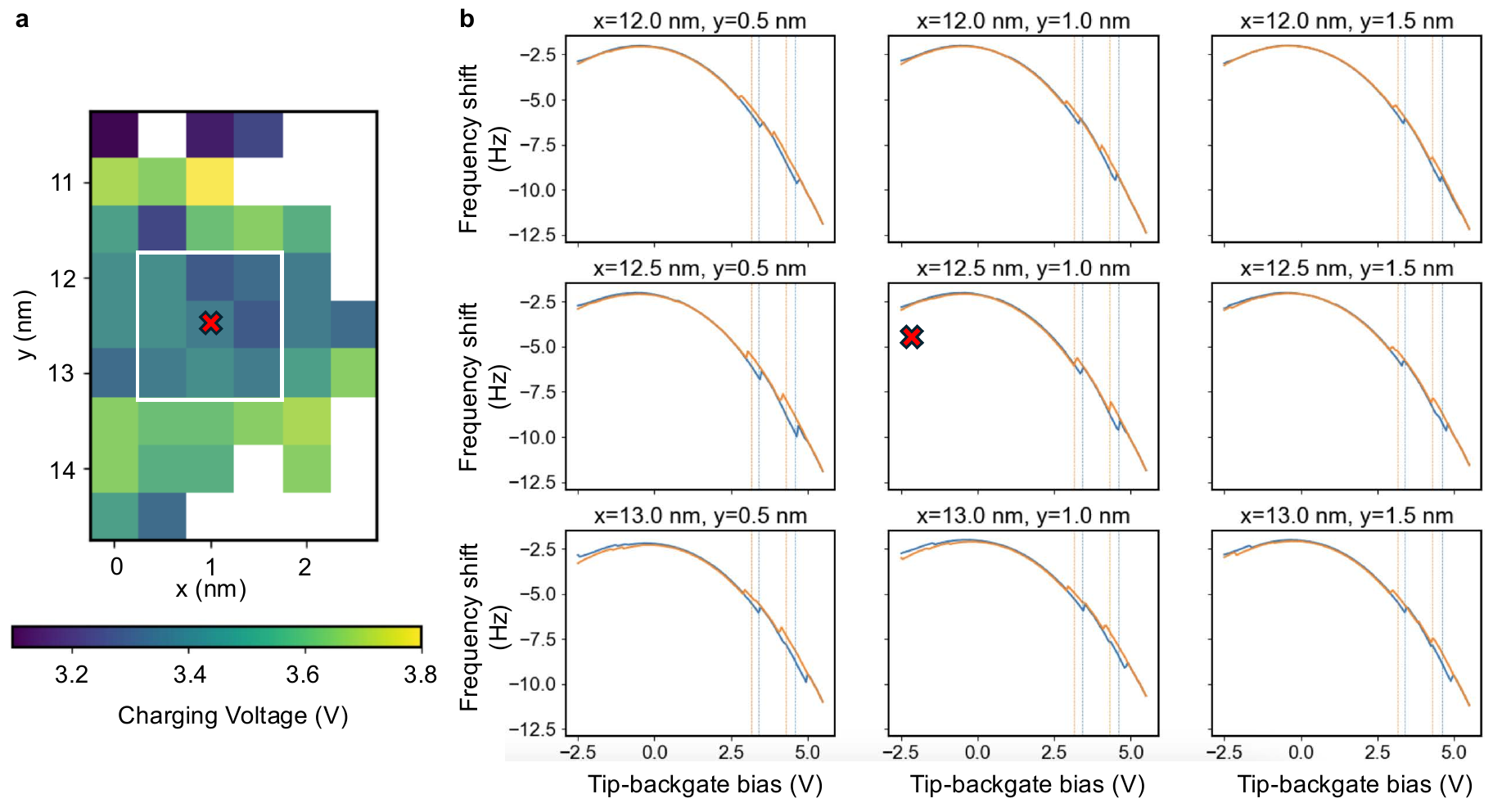}
    \caption{ 
    A study of the hysteresis correlations in the charging transitions on neighboring pixels.
    (a) A map of the lowest-voltage charging transitions observed in the forward $f$-$V$ voltage sweep for Defect 13.
    The red cross shows the center of the defect, as determined in Sec.~\ref{sec:Identifying_Defects}.
    (b) $f$-$V$ forward (blue) and backward (orange) voltage sweeps for Defect 13, measured at the center pixel (red cross) and on the eight neighboring pixels.
    Two pairs of charging transitions are observed at the center pixel, as indicated by vertical dashed lines.
    The same dashed lines (from the center pixel) are shown on all the plots, for reference.
    Greater-or-equal hysteresis is observed for the same pairs of charging transitions in all the surrounding pixels, as compared to the center pixel, indicating that both transitions are centered at the same pixel (red cross), and that, with high likelihood, they originate from the same defect.}
    \label{fig:transitions_correlations}
\end{figure*}

In Sec.~\ref{sec:likelihood}, we described a method for matching one or more charging transitions, observed at a given pixel, to one ore more defects, using a likelihood analysis and theoretical predictions for the charging energies. 
In this section, we describe an alternative method, based on an analysis of the charging map itself.

The method focuses on the correlations between the hysteresis observed in charging transitions on neighboring pixels.
To demonstrate this method, Fig.~\ref{fig:transitions_correlations}(a) shows a map of the lowest-voltage charging transitions for Defect 13, with the predicted center of the defect shown as a red cross.
Figure~\ref{fig:transitions_correlations}(b) shows the $f$-$V$ sweeps for this center pixel, and the eight surrounding pixels.
For reference, in all the plots, the vertical dashed lines show the charging transitions for the center pixel.
As the EFM tip is moved away from the center pixel, changes in the hysteresis provide a probe of the tip-defect separation, due to the exponential decay of tip-defect tunneling with distance.
Thus, if two different charging transitions originate from the same defect, we would expect them both to exhibit a larger hysteresis on the neighboring pixels.
This is precisely the behavior observed in Fig.~\ref{fig:transitions_correlations}(b), indicating that both transitions belong to the same defect.
In this case, this provides strong evidence that the defect should be classified as the defect species V$_\text{O}$, which has two transitions in the experimental voltage range.
In comparison, the second-best classification from our previous likelihood analysis suggests two closely separated C$_\text{Al}$-H defects with charging transitions that are shifted by different local electrostatic environments.
The hysteresis analysis, given above, confirms that the C$_\text{Al}$-H classification is less likely.

Although we do not make further use of this hysteresis-based approach here, we hope to explore it in future work.
This points to untapped information available in the hysteresis data.

\end{document}